\documentclass{emulateapj}

\tabletypesize{\scriptsize}
\usepackage{url}
\usepackage{multirow}

\newcommand{\TauRatioWeightedMean}{0.72}
\newcommand{\TauRatioWeightedError}{0.06}
\newcommand{\TauRatioReducedChiSquared}{11.6}
\newcommand{\TauRatioMedian}{1.73}
\newcommand{\TransitDurationWeightedMean}{7.41}
\newcommand{\TransitDurationWeightedError}{0.15}
\newcommand{\TransitDurationReducedChiSquared}{11.2}
\newcommand{\TauOneMedian}{0.80}
\newcommand{\TauTwoMedian}{1.44}
\newcommand{\TransitDurationUse}{3}
\newcommand{\ROccultREarth}{9.5}
\newcommand{\TravelTime}{1.6}
\newcommand{\TransitDepthAirOverV}{1.04}

\newcommand{\TransitDepthAirOverR}{0.98}
\newcommand{\TransitDepthAirOverRError}{0.23}
\newcommand{\TransitDepthVOverR}{0.96}
\newcommand{\TransitDepthVOverRError}{0.17}
\newcommand{\AngstromVOverR}{0.24}
\newcommand{\AngstromVOverROneSigmaLow}{-0.95}
\newcommand{\AngstromVOverROneSigmaHigh}{1.43}
\newcommand{\AngstromVOverRTwoSigmaLow}{-2.14}
\newcommand{\AngstromVOverRTwoSigmaHigh}{2.62}
\newcommand{\MicronSingleSizeLimit}{0.15}
\newcommand{\MicronSingleSizeLowLimit}{0.06}
\newcommand{\NumSignificantTransits}{9}
\newcommand{\NumSignificantHours}{32}
\newcommand{\FrequencySignificantTransits}{0.28}
\newcommand{\HoursPerSignificantTransits}{3.6}

\newcommand{\TEquilibriumFourPi}{1100}
\newcommand{\ChiSquaredReducedFlatLine}{66}
\newcommand{\FZeroMayThirteenJoint}{0.991}
\newcommand{\FZeroPlusMayThirteenJoint}{0.005}
\newcommand{\FZeroMinusMayThirteenJoint}{0.005}

\newcommand{\ThetaOneMayThirteenJoint}{0.33}
\newcommand{\ThetaOnePlusMayThirteenJoint}{0.06}
\newcommand{\ThetaOneMinusMayThirteenJoint}{0.07}
\newcommand{\ThetaTwoMayThirteenJoint}{1.03}
\newcommand{\ThetaTwoPlusMayThirteenJoint}{0.13}
\newcommand{\ThetaTwoMinusMayThirteenJoint}{0.14}
\newcommand{\ThetaTwoOverThetaOneMayThirteenJoint}{3.1}
\newcommand{\ThetaTwoOverThetaOnePlusMayThirteenJoint}{0.8}
\newcommand{\ThetaTwoOverThetaOneMinusMayThirteenJoint}{0.9}
\newcommand{\FZeroBMayThirteenJoint}{1.007}
\newcommand{\FZeroBPlusMayThirteenJoint}{0.003}
\newcommand{\FZeroBMinusMayThirteenJoint}{0.003}

\newcommand{\FZeroCMayThirteenJoint}{0.996}
\newcommand{\FZeroCPlusMayThirteenJoint}{0.006}
\newcommand{\FZeroCMinusMayThirteenJoint}{0.006}

\newcommand{\CRatioCAMayThirteenJoint}{1.57}
\newcommand{\CRatioCAPlusMayThirteenJoint}{0.28}
\newcommand{\CRatioCAMinusMayThirteenJoint}{0.34}
\newcommand{\CRatioBAMayThirteenJoint}{1.54}
\newcommand{\CRatioBAPlusMayThirteenJoint}{0.22}
\newcommand{\CRatioBAMinusMayThirteenJoint}{0.27}
\newcommand{\CRatioCBMayThirteenJoint}{1.02}
\newcommand{\CRatioCBPlusMayThirteenJoint}{0.13}
\newcommand{\CRatioCBMinusMayThirteenJoint}{0.13}
\newcommand{\TDAPercentMayThirteenJoint}{16.2}
\newcommand{\TDAPercentPlusMayThirteenJoint}{2.6}
\newcommand{\TDAPercentMinusMayThirteenJoint}{2.7}
\newcommand{\TDBPercentMayThirteenJoint}{25.1}
\newcommand{\TDBPercentPlusMayThirteenJoint}{1.4}
\newcommand{\TDBPercentMinusMayThirteenJoint}{1.4}
\newcommand{\TDCPercentMayThirteenJoint}{25.6}
\newcommand{\TDCPercentPlusMayThirteenJoint}{3.2}
\newcommand{\TDCPercentMinusMayThirteenJoint}{3.4}
\newcommand{\TminMayThirteenJoint}{155.79517}
\newcommand{\TminPlusMayThirteenJoint}{0.00006}
\newcommand{\TminMinusMayThirteenJoint}{0.00005}
\newcommand{\TransitDurationMayThirteenJoint}{4.1}
\newcommand{\TransitDurationPlusMayThirteenJoint}{0.4}
\newcommand{\TransitDurationMinusMayThirteenJoint}{0.5}
\newcommand{\FZeroMayNineJoint}{1.005}
\newcommand{\FZeroPlusMayNineJoint}{0.009}
\newcommand{\FZeroMinusMayNineJoint}{0.009}

\newcommand{\ThetaOneMayNineJoint}{0.20}
\newcommand{\ThetaOnePlusMayNineJoint}{0.11}
\newcommand{\ThetaOneMinusMayNineJoint}{0.20}
\newcommand{\ThetaTwoMayNineJoint}{1.96}
\newcommand{\ThetaTwoPlusMayNineJoint}{0.55}
\newcommand{\ThetaTwoMinusMayNineJoint}{0.69}
\newcommand{\ThetaTwoOverThetaOneMayNineJoint}{8.2}
\newcommand{\ThetaTwoOverThetaOnePlusMayNineJoint}{4.9}
\newcommand{\ThetaTwoOverThetaOneMinusMayNineJoint}{6.9}
\newcommand{\FZeroBMayNineJoint}{1.042}
\newcommand{\FZeroBPlusMayNineJoint}{0.007}
\newcommand{\FZeroBMinusMayNineJoint}{0.008}

\newcommand{\FZeroCMayNineJoint}{1.000}
\newcommand{\FZeroCPlusMayNineJoint}{0.009}
\newcommand{\FZeroCMinusMayNineJoint}{0.009}

\newcommand{\CRatioCAMayNineJoint}{0.64}
\newcommand{\CRatioCAPlusMayNineJoint}{0.20}
\newcommand{\CRatioCAMinusMayNineJoint}{0.26}
\newcommand{\CRatioBAMayNineJoint}{1.39}
\newcommand{\CRatioBAPlusMayNineJoint}{0.29}
\newcommand{\CRatioBAMinusMayNineJoint}{0.32}

\newcommand{\TDAPercentMayNineJoint}{14.5}
\newcommand{\TDAPercentPlusMayNineJoint}{3.6}
\newcommand{\TDAPercentMinusMayNineJoint}{3.9}
\newcommand{\TDBPercentMayNineJoint}{20.1}
\newcommand{\TDBPercentPlusMayNineJoint}{2.9}
\newcommand{\TDBPercentMinusMayNineJoint}{3.8}
\newcommand{\TDCPercentMayNineJoint}{9.2}
\newcommand{\TDCPercentPlusMayNineJoint}{3.1}
\newcommand{\TDCPercentMinusMayNineJoint}{3.4}
\newcommand{\TminMayNineJoint}{151.77662}
\newcommand{\TminPlusMayNineJoint}{0.00017}
\newcommand{\TminMinusMayNineJoint}{0.00018}
\newcommand{\TransitDurationMayNineJoint}{6.5}
\newcommand{\TransitDurationPlusMayNineJoint}{1.7}
\newcommand{\TransitDurationMinusMayNineJoint}{2.2}
\newcommand{\FZeroMayTenJoint}{1.028}
\newcommand{\FZeroPlusMayTenJoint}{0.008}
\newcommand{\FZeroMinusMayTenJoint}{0.008}

\newcommand{\ThetaOneMayTenJoint}{0.35}
\newcommand{\ThetaOnePlusMayTenJoint}{0.17}
\newcommand{\ThetaOneMinusMayTenJoint}{0.21}
\newcommand{\ThetaTwoMayTenJoint}{1.08}
\newcommand{\ThetaTwoPlusMayTenJoint}{0.28}
\newcommand{\ThetaTwoMinusMayTenJoint}{0.32}
\newcommand{\ThetaTwoOverThetaOneMayTenJoint}{2.9}
\newcommand{\ThetaTwoOverThetaOnePlusMayTenJoint}{1.6}
\newcommand{\ThetaTwoOverThetaOneMinusMayTenJoint}{2.2}
\newcommand{\FZeroBMayTenJoint}{1.018}
\newcommand{\FZeroBPlusMayTenJoint}{0.005}
\newcommand{\FZeroBMinusMayTenJoint}{0.005}

\newcommand{\FZeroCMayTenJoint}{1.021}
\newcommand{\FZeroCPlusMayTenJoint}{0.012}
\newcommand{\FZeroCMinusMayTenJoint}{0.012}

\newcommand{\CRatioCAMayTenJoint}{1.05}
\newcommand{\CRatioCAPlusMayTenJoint}{0.19}
\newcommand{\CRatioCAMinusMayTenJoint}{0.24}
\newcommand{\CRatioBAMayTenJoint}{1.01}
\newcommand{\CRatioBAPlusMayTenJoint}{0.15}
\newcommand{\CRatioBAMinusMayTenJoint}{0.19}
\newcommand{\CRatioCBMayTenJoint}{1.04}
\newcommand{\CRatioCBPlusMayTenJoint}{0.16}
\newcommand{\CRatioCBMinusMayTenJoint}{0.18}
\newcommand{\TDAPercentMayTenJoint}{27.8}
\newcommand{\TDAPercentPlusMayTenJoint}{4.5}
\newcommand{\TDAPercentMinusMayTenJoint}{5.1}
\newcommand{\TDBPercentMayTenJoint}{28.0}
\newcommand{\TDBPercentPlusMayTenJoint}{3.0}
\newcommand{\TDBPercentMinusMayTenJoint}{3.2}
\newcommand{\TDCPercentMayTenJoint}{29.1}
\newcommand{\TDCPercentPlusMayTenJoint}{4.8}
\newcommand{\TDCPercentMinusMayTenJoint}{5.0}
\newcommand{\TminMayTenJoint}{152.71259}
\newcommand{\TminPlusMayTenJoint}{0.00013}
\newcommand{\TminMinusMayTenJoint}{0.00011}
\newcommand{\TransitDurationMayTenJoint}{4.3}
\newcommand{\TransitDurationPlusMayTenJoint}{1.0}
\newcommand{\TransitDurationMinusMayTenJoint}{1.2}
\newcommand{\FZeroMayElevenJoint}{1.016}
\newcommand{\FZeroPlusMayElevenJoint}{0.005}
\newcommand{\FZeroMinusMayElevenJoint}{0.005}

\newcommand{\ThetaOneMayElevenJoint}{0.36}
\newcommand{\ThetaOnePlusMayElevenJoint}{0.17}
\newcommand{\ThetaOneMinusMayElevenJoint}{0.16}
\newcommand{\ThetaTwoMayElevenJoint}{2.58}
\newcommand{\ThetaTwoPlusMayElevenJoint}{0.40}
\newcommand{\ThetaTwoMinusMayElevenJoint}{0.46}
\newcommand{\ThetaTwoOverThetaOneMayElevenJoint}{7.1}
\newcommand{\ThetaTwoOverThetaOnePlusMayElevenJoint}{2.5}
\newcommand{\ThetaTwoOverThetaOneMinusMayElevenJoint}{3.7}
\newcommand{\FZeroBMayElevenJoint}{1.045}
\newcommand{\FZeroBPlusMayElevenJoint}{0.010}
\newcommand{\FZeroBMinusMayElevenJoint}{0.010}

\newcommand{\FZeroCMayElevenJoint}{1.021}
\newcommand{\FZeroCPlusMayElevenJoint}{0.005}
\newcommand{\FZeroCMinusMayElevenJoint}{0.005}

\newcommand{\CRatioCAMayElevenJoint}{0.97}
\newcommand{\CRatioCAPlusMayElevenJoint}{0.15}
\newcommand{\CRatioCAMinusMayElevenJoint}{0.17}
\newcommand{\CRatioBAMayElevenJoint}{1.75}
\newcommand{\CRatioBAPlusMayElevenJoint}{0.24}
\newcommand{\CRatioBAMinusMayElevenJoint}{0.27}

\newcommand{\TDAPercentMayElevenJoint}{11.0}
\newcommand{\TDAPercentPlusMayElevenJoint}{1.2}
\newcommand{\TDAPercentMinusMayElevenJoint}{1.4}
\newcommand{\TDBPercentMayElevenJoint}{19.3}
\newcommand{\TDBPercentPlusMayElevenJoint}{2.2}
\newcommand{\TDBPercentMinusMayElevenJoint}{2.6}
\newcommand{\TDCPercentMayElevenJoint}{10.6}
\newcommand{\TDCPercentPlusMayElevenJoint}{1.5}
\newcommand{\TDCPercentMinusMayElevenJoint}{1.9}
\newcommand{\TminMayElevenJoint}{153.64838}
\newcommand{\TminPlusMayElevenJoint}{0.00014}
\newcommand{\TminMinusMayElevenJoint}{0.00012}
\newcommand{\TransitDurationMayElevenJoint}{8.8}
\newcommand{\TransitDurationPlusMayElevenJoint}{1.3}
\newcommand{\TransitDurationMinusMayElevenJoint}{1.5}
\newcommand{\FZeroMayElevenSecondJoint}{1.018}
\newcommand{\FZeroPlusMayElevenSecondJoint}{0.005}
\newcommand{\FZeroMinusMayElevenSecondJoint}{0.005}

\newcommand{\ThetaOneMayElevenSecondJoint}{1.90}
\newcommand{\ThetaOnePlusMayElevenSecondJoint}{0.31}
\newcommand{\ThetaOneMinusMayElevenSecondJoint}{0.37}
\newcommand{\ThetaTwoMayElevenSecondJoint}{1.45}
\newcommand{\ThetaTwoPlusMayElevenSecondJoint}{0.32}
\newcommand{\ThetaTwoMinusMayElevenSecondJoint}{0.40}
\newcommand{\ThetaTwoOverThetaOneMayElevenSecondJoint}{0.8}
\newcommand{\ThetaTwoOverThetaOnePlusMayElevenSecondJoint}{0.2}
\newcommand{\ThetaTwoOverThetaOneMinusMayElevenSecondJoint}{0.3}
\newcommand{\FZeroBMayElevenSecondJoint}{1.026}
\newcommand{\FZeroBPlusMayElevenSecondJoint}{0.005}
\newcommand{\FZeroBMinusMayElevenSecondJoint}{0.005}

\newcommand{\FZeroCMayElevenSecondJoint}{0.999}
\newcommand{\FZeroCPlusMayElevenSecondJoint}{0.007}
\newcommand{\FZeroCMinusMayElevenSecondJoint}{0.008}

\newcommand{\CRatioCAMayElevenSecondJoint}{0.91}
\newcommand{\CRatioCAPlusMayElevenSecondJoint}{0.14}
\newcommand{\CRatioCAMinusMayElevenSecondJoint}{0.15}
\newcommand{\CRatioBAMayElevenSecondJoint}{0.95}
\newcommand{\CRatioBAPlusMayElevenSecondJoint}{0.12}
\newcommand{\CRatioBAMinusMayElevenSecondJoint}{0.13}

\newcommand{\TDAPercentMayElevenSecondJoint}{11.0}
\newcommand{\TDAPercentPlusMayElevenSecondJoint}{0.9}
\newcommand{\TDAPercentMinusMayElevenSecondJoint}{0.9}
\newcommand{\TDBPercentMayElevenSecondJoint}{10.5}
\newcommand{\TDBPercentPlusMayElevenSecondJoint}{1.2}
\newcommand{\TDBPercentMinusMayElevenSecondJoint}{1.1}
\newcommand{\TDCPercentMayElevenSecondJoint}{10.0}
\newcommand{\TDCPercentPlusMayElevenSecondJoint}{1.4}
\newcommand{\TDCPercentMinusMayElevenSecondJoint}{1.4}
\newcommand{\TminMayElevenSecondJoint}{153.64015}
\newcommand{\TminPlusMayElevenSecondJoint}{0.00015}
\newcommand{\TminMinusMayElevenSecondJoint}{0.00015}
\newcommand{\TransitDurationMayElevenSecondJoint}{10.0}
\newcommand{\TransitDurationPlusMayElevenSecondJoint}{1.3}
\newcommand{\TransitDurationMinusMayElevenSecondJoint}{1.6}
\newcommand{\FZeroMayElevenThirdJoint}{0.996}
\newcommand{\FZeroPlusMayElevenThirdJoint}{0.000}
\newcommand{\FZeroMinusMayElevenThirdJoint}{0.000}

\newcommand{\ThetaOneMayElevenThirdJoint}{1.44}
\newcommand{\ThetaOnePlusMayElevenThirdJoint}{0.10}
\newcommand{\ThetaOneMinusMayElevenThirdJoint}{0.10}
\newcommand{\ThetaTwoMayElevenThirdJoint}{1.14}
\newcommand{\ThetaTwoPlusMayElevenThirdJoint}{0.09}
\newcommand{\ThetaTwoMinusMayElevenThirdJoint}{0.09}
\newcommand{\ThetaTwoOverThetaOneMayElevenThirdJoint}{0.8}
\newcommand{\ThetaTwoOverThetaOnePlusMayElevenThirdJoint}{0.1}
\newcommand{\ThetaTwoOverThetaOneMinusMayElevenThirdJoint}{0.1}
\newcommand{\FZeroBMayElevenThirdJoint}{0.996}
\newcommand{\FZeroBPlusMayElevenThirdJoint}{0.002}
\newcommand{\FZeroBMinusMayElevenThirdJoint}{0.002}

\newcommand{\FZeroCMayElevenThirdJoint}{0.999}
\newcommand{\FZeroCPlusMayElevenThirdJoint}{0.002}
\newcommand{\FZeroCMinusMayElevenThirdJoint}{0.002}

\newcommand{\CRatioCAMayElevenThirdJoint}{1.04}
\newcommand{\CRatioCAPlusMayElevenThirdJoint}{0.19}
\newcommand{\CRatioCAMinusMayElevenThirdJoint}{0.20}
\newcommand{\CRatioBAMayElevenThirdJoint}{1.16}
\newcommand{\CRatioBAPlusMayElevenThirdJoint}{0.19}
\newcommand{\CRatioBAMinusMayElevenThirdJoint}{0.19}

\newcommand{\TDAPercentMayElevenThirdJoint}{3.8}
\newcommand{\TDAPercentPlusMayElevenThirdJoint}{0.1}
\newcommand{\TDAPercentMinusMayElevenThirdJoint}{0.1}
\newcommand{\TDBPercentMayElevenThirdJoint}{4.4}
\newcommand{\TDBPercentPlusMayElevenThirdJoint}{0.7}
\newcommand{\TDBPercentMinusMayElevenThirdJoint}{0.7}
\newcommand{\TDCPercentMayElevenThirdJoint}{4.0}
\newcommand{\TDCPercentPlusMayElevenThirdJoint}{0.7}
\newcommand{\TDCPercentMinusMayElevenThirdJoint}{0.8}
\newcommand{\TminMayElevenThirdJoint}{153.74647}
\newcommand{\TminPlusMayElevenThirdJoint}{0.00004}
\newcommand{\TminMinusMayElevenThirdJoint}{0.00004}
\newcommand{\TransitDurationMayElevenThirdJoint}{7.8}
\newcommand{\TransitDurationPlusMayElevenThirdJoint}{0.4}
\newcommand{\TransitDurationMinusMayElevenThirdJoint}{0.4}
\newcommand{\FZeroMayTwelveJoint}{1.005}
\newcommand{\FZeroPlusMayTwelveJoint}{0.002}
\newcommand{\FZeroMinusMayTwelveJoint}{0.002}

\newcommand{\ThetaOneMayTwelveJoint}{0.79}
\newcommand{\ThetaOnePlusMayTwelveJoint}{0.13}
\newcommand{\ThetaOneMinusMayTwelveJoint}{0.15}
\newcommand{\ThetaTwoMayTwelveJoint}{1.23}
\newcommand{\ThetaTwoPlusMayTwelveJoint}{0.18}
\newcommand{\ThetaTwoMinusMayTwelveJoint}{0.22}
\newcommand{\ThetaTwoOverThetaOneMayTwelveJoint}{1.6}
\newcommand{\ThetaTwoOverThetaOnePlusMayTwelveJoint}{0.4}
\newcommand{\ThetaTwoOverThetaOneMinusMayTwelveJoint}{0.5}
\newcommand{\FZeroBMayTwelveJoint}{1.000}
\newcommand{\FZeroBPlusMayTwelveJoint}{0.005}
\newcommand{\FZeroBMinusMayTwelveJoint}{0.005}

\newcommand{\FZeroCMayTwelveJoint}{1.000}
\newcommand{\FZeroCPlusMayTwelveJoint}{0.005}
\newcommand{\FZeroCMinusMayTwelveJoint}{0.005}

\newcommand{\CRatioCAMayTwelveJoint}{1.20}
\newcommand{\CRatioCAPlusMayTwelveJoint}{0.14}
\newcommand{\CRatioCAMinusMayTwelveJoint}{0.15}
\newcommand{\CRatioBAMayTwelveJoint}{0.93}
\newcommand{\CRatioBAPlusMayTwelveJoint}{0.12}
\newcommand{\CRatioBAMinusMayTwelveJoint}{0.13}
\newcommand{\CRatioCBMayTwelveJoint}{1.29}
\newcommand{\CRatioCBPlusMayTwelveJoint}{0.19}
\newcommand{\CRatioCBMinusMayTwelveJoint}{0.23}
\newcommand{\TDAPercentMayTwelveJoint}{13.0}
\newcommand{\TDAPercentPlusMayTwelveJoint}{0.8}
\newcommand{\TDAPercentMinusMayTwelveJoint}{0.8}
\newcommand{\TDBPercentMayTwelveJoint}{12.0}
\newcommand{\TDBPercentPlusMayTwelveJoint}{1.5}
\newcommand{\TDBPercentMinusMayTwelveJoint}{1.6}
\newcommand{\TDCPercentMayTwelveJoint}{15.5}
\newcommand{\TDCPercentPlusMayTwelveJoint}{1.8}
\newcommand{\TDCPercentMinusMayTwelveJoint}{1.8}
\newcommand{\TminMayTwelveJoint}{154.76274}
\newcommand{\TminPlusMayTwelveJoint}{0.00007}
\newcommand{\TminMinusMayTwelveJoint}{0.00007}
\newcommand{\TransitDurationMayTwelveJoint}{6.1}
\newcommand{\TransitDurationPlusMayTwelveJoint}{0.7}
\newcommand{\TransitDurationMinusMayTwelveJoint}{0.8}
\newcommand{\FZeroMayTwelveSecondJoint}{1.022}
\newcommand{\FZeroPlusMayTwelveSecondJoint}{0.002}
\newcommand{\FZeroMinusMayTwelveSecondJoint}{0.002}

\newcommand{\ThetaOneMayTwelveSecondJoint}{0.16}
\newcommand{\ThetaOnePlusMayTwelveSecondJoint}{0.08}
\newcommand{\ThetaOneMinusMayTwelveSecondJoint}{0.12}
\newcommand{\ThetaTwoMayTwelveSecondJoint}{2.62}
\newcommand{\ThetaTwoPlusMayTwelveSecondJoint}{0.33}
\newcommand{\ThetaTwoMinusMayTwelveSecondJoint}{0.33}
\newcommand{\ThetaTwoOverThetaOneMayTwelveSecondJoint}{14.1}
\newcommand{\ThetaTwoOverThetaOnePlusMayTwelveSecondJoint}{5.6}
\newcommand{\ThetaTwoOverThetaOneMinusMayTwelveSecondJoint}{7.7}
\newcommand{\FZeroBMayTwelveSecondJoint}{1.010}
\newcommand{\FZeroBPlusMayTwelveSecondJoint}{0.004}
\newcommand{\FZeroBMinusMayTwelveSecondJoint}{0.005}

\newcommand{\FZeroCMayTwelveSecondJoint}{1.006}
\newcommand{\FZeroCPlusMayTwelveSecondJoint}{0.005}
\newcommand{\FZeroCMinusMayTwelveSecondJoint}{0.005}

\newcommand{\CRatioCAMayTwelveSecondJoint}{0.72}
\newcommand{\CRatioCAPlusMayTwelveSecondJoint}{0.13}
\newcommand{\CRatioCAMinusMayTwelveSecondJoint}{0.13}
\newcommand{\CRatioBAMayTwelveSecondJoint}{0.83}
\newcommand{\CRatioBAPlusMayTwelveSecondJoint}{0.12}
\newcommand{\CRatioBAMinusMayTwelveSecondJoint}{0.13}
\newcommand{\CRatioCBMayTwelveSecondJoint}{0.87}
\newcommand{\CRatioCBPlusMayTwelveSecondJoint}{0.17}
\newcommand{\CRatioCBMinusMayTwelveSecondJoint}{0.19}
\newcommand{\TDAPercentMayTwelveSecondJoint}{14.1}
\newcommand{\TDAPercentPlusMayTwelveSecondJoint}{1.1}
\newcommand{\TDAPercentMinusMayTwelveSecondJoint}{1.3}
\newcommand{\TDBPercentMayTwelveSecondJoint}{11.9}
\newcommand{\TDBPercentPlusMayTwelveSecondJoint}{1.5}
\newcommand{\TDBPercentMinusMayTwelveSecondJoint}{1.6}
\newcommand{\TDCPercentMayTwelveSecondJoint}{10.3}
\newcommand{\TDCPercentPlusMayTwelveSecondJoint}{1.7}
\newcommand{\TDCPercentMinusMayTwelveSecondJoint}{1.8}
\newcommand{\TminMayTwelveSecondJoint}{154.77075}
\newcommand{\TminPlusMayTwelveSecondJoint}{0.00016}
\newcommand{\TminMinusMayTwelveSecondJoint}{0.00014}
\newcommand{\TransitDurationMayTwelveSecondJoint}{8.3}
\newcommand{\TransitDurationPlusMayTwelveSecondJoint}{1.0}
\newcommand{\TransitDurationMinusMayTwelveSecondJoint}{1.0}

\newcommand{\FZeroAprilEighteenJoint}{1.028}
\newcommand{\FZeroPlusAprilEighteenJoint}{0.014}
\newcommand{\FZeroMinusAprilEighteenJoint}{0.016}

\newcommand{\ThetaOneAprilEighteenJoint}{2.54}
\newcommand{\ThetaOnePlusAprilEighteenJoint}{1.49}
\newcommand{\ThetaOneMinusAprilEighteenJoint}{2.10}
\newcommand{\ThetaTwoAprilEighteenJoint}{1.80}
\newcommand{\ThetaTwoPlusAprilEighteenJoint}{1.02}
\newcommand{\ThetaTwoMinusAprilEighteenJoint}{1.56}
\newcommand{\ThetaTwoOverThetaOneAprilEighteenJoint}{0.7}
\newcommand{\ThetaTwoOverThetaOnePlusAprilEighteenJoint}{0.4}
\newcommand{\ThetaTwoOverThetaOneMinusAprilEighteenJoint}{0.7}
\newcommand{\FZeroBAprilEighteenJoint}{1.032}
\newcommand{\FZeroBPlusAprilEighteenJoint}{0.019}
\newcommand{\FZeroBMinusAprilEighteenJoint}{0.025}

\newcommand{\TDAPercentAprilEighteenJoint}{14.5}
\newcommand{\TDAPercentPlusAprilEighteenJoint}{6.1}
\newcommand{\TDAPercentMinusAprilEighteenJoint}{7.9}
\newcommand{\TDBPercentAprilEighteenJoint}{12.8}
\newcommand{\TDBPercentPlusAprilEighteenJoint}{5.9}
\newcommand{\TDBPercentMinusAprilEighteenJoint}{7.3}

\newcommand{\TminAprilEighteenJoint}{130.69958}
\newcommand{\TminPlusAprilEighteenJoint}{0.00061}
\newcommand{\TminMinusAprilEighteenJoint}{0.00072}
\newcommand{\TransitDurationAprilEighteenJoint}{13.0}
\newcommand{\TransitDurationPlusAprilEighteenJoint}{5.4}
\newcommand{\TransitDurationMinusAprilEighteenJoint}{7.9}

\newcommand{\FZeroFLWOMarchTwentyTwoVBand}{0.999}
\newcommand{\FZeroPlusFLWOMarchTwentyTwoVBand}{0.002}
\newcommand{\FZeroMinusFLWOMarchTwentyTwoVBand}{0.002}

\newcommand{\ThetaOneFLWOMarchTwentyTwoVBand}{1.61}
\newcommand{\ThetaOnePlusFLWOMarchTwentyTwoVBand}{0.20}
\newcommand{\ThetaOneMinusFLWOMarchTwentyTwoVBand}{0.20}
\newcommand{\ThetaTwoFLWOMarchTwentyTwoVBand}{0.21}
\newcommand{\ThetaTwoPlusFLWOMarchTwentyTwoVBand}{0.11}
\newcommand{\ThetaTwoMinusFLWOMarchTwentyTwoVBand}{0.21}
\newcommand{\ThetaTwoOverThetaOneFLWOMarchTwentyTwoVBand}{0.1}
\newcommand{\ThetaTwoOverThetaOnePlusFLWOMarchTwentyTwoVBand}{0.1}
\newcommand{\ThetaTwoOverThetaOneMinusFLWOMarchTwentyTwoVBand}{0.1}
\newcommand{\TDPercentFLWOMarchTwentyTwoVBand}{19.6}
\newcommand{\TDPercentPlusFLWOMarchTwentyTwoVBand}{2.5}
\newcommand{\TDPercentMinusFLWOMarchTwentyTwoVBand}{3.4}
\newcommand{\TminFLWOMarchTwentyTwoVBand}{104.69519}
\newcommand{\TminPlusFLWOMarchTwentyTwoVBand}{0.00017}
\newcommand{\TminMinusFLWOMarchTwentyTwoVBand}{0.00020}
\newcommand{\TransitDurationFLWOMarchTwentyTwoVBand}{5.5}
\newcommand{\TransitDurationPlusFLWOMarchTwentyTwoVBand}{0.7}
\newcommand{\TransitDurationMinusFLWOMarchTwentyTwoVBand}{0.9}
\newcommand{\FZeroFLWOAprilElevenVBand}{1.023}
\newcommand{\FZeroPlusFLWOAprilElevenVBand}{0.001}
\newcommand{\FZeroMinusFLWOAprilElevenVBand}{0.002}

\newcommand{\ThetaOneFLWOAprilElevenVBand}{0.43}
\newcommand{\ThetaOnePlusFLWOAprilElevenVBand}{0.04}
\newcommand{\ThetaOneMinusFLWOAprilElevenVBand}{0.04}
\newcommand{\ThetaTwoFLWOAprilElevenVBand}{2.41}
\newcommand{\ThetaTwoPlusFLWOAprilElevenVBand}{0.08}
\newcommand{\ThetaTwoMinusFLWOAprilElevenVBand}{0.08}
\newcommand{\ThetaTwoOverThetaOneFLWOAprilElevenVBand}{5.6}
\newcommand{\ThetaTwoOverThetaOnePlusFLWOAprilElevenVBand}{0.5}
\newcommand{\ThetaTwoOverThetaOneMinusFLWOAprilElevenVBand}{0.6}
\newcommand{\TDPercentFLWOAprilElevenVBand}{56.5}
\newcommand{\TDPercentPlusFLWOAprilElevenVBand}{1.0}
\newcommand{\TDPercentMinusFLWOAprilElevenVBand}{1.2}
\newcommand{\TminFLWOAprilElevenVBand}{123.66918}
\newcommand{\TminPlusFLWOAprilElevenVBand}{0.00003}
\newcommand{\TminMinusFLWOAprilElevenVBand}{0.00003}
\newcommand{\TransitDurationFLWOAprilElevenVBand}{8.5}
\newcommand{\TransitDurationPlusFLWOAprilElevenVBand}{0.3}
\newcommand{\TransitDurationMinusFLWOAprilElevenVBand}{0.3}
\newcommand{\FZeroFLWOAprilElevenVBandSecond}{1.028}
\newcommand{\FZeroPlusFLWOAprilElevenVBandSecond}{0.001}
\newcommand{\FZeroMinusFLWOAprilElevenVBandSecond}{0.001}

\newcommand{\ThetaOneFLWOAprilElevenVBandSecond}{1.20}
\newcommand{\ThetaOnePlusFLWOAprilElevenVBandSecond}{0.07}
\newcommand{\ThetaOneMinusFLWOAprilElevenVBandSecond}{0.07}
\newcommand{\ThetaTwoFLWOAprilElevenVBandSecond}{2.07}
\newcommand{\ThetaTwoPlusFLWOAprilElevenVBandSecond}{0.10}
\newcommand{\ThetaTwoMinusFLWOAprilElevenVBandSecond}{0.10}
\newcommand{\ThetaTwoOverThetaOneFLWOAprilElevenVBandSecond}{1.7}
\newcommand{\ThetaTwoOverThetaOnePlusFLWOAprilElevenVBandSecond}{0.2}
\newcommand{\ThetaTwoOverThetaOneMinusFLWOAprilElevenVBandSecond}{0.2}
\newcommand{\TDPercentFLWOAprilElevenVBandSecond}{49.6}
\newcommand{\TDPercentPlusFLWOAprilElevenVBandSecond}{0.7}
\newcommand{\TDPercentMinusFLWOAprilElevenVBandSecond}{0.8}
\newcommand{\TminFLWOAprilElevenVBandSecond}{123.85688}
\newcommand{\TminPlusFLWOAprilElevenVBandSecond}{0.00004}
\newcommand{\TminMinusFLWOAprilElevenVBandSecond}{0.00004}
\newcommand{\TransitDurationFLWOAprilElevenVBandSecond}{9.8}
\newcommand{\TransitDurationPlusFLWOAprilElevenVBandSecond}{0.3}
\newcommand{\TransitDurationMinusFLWOAprilElevenVBandSecond}{0.4}
\newcommand{\FZeroMEarthAprilSixteenMEarthBand}{1.012}
\newcommand{\FZeroPlusMEarthAprilSixteenMEarthBand}{0.016}
\newcommand{\FZeroMinusMEarthAprilSixteenMEarthBand}{0.017}

\newcommand{\ThetaOneMEarthAprilSixteenMEarthBand}{0.98}
\newcommand{\ThetaOnePlusMEarthAprilSixteenMEarthBand}{0.47}
\newcommand{\ThetaOneMinusMEarthAprilSixteenMEarthBand}{0.61}
\newcommand{\ThetaTwoMEarthAprilSixteenMEarthBand}{0.78}
\newcommand{\ThetaTwoPlusMEarthAprilSixteenMEarthBand}{0.41}
\newcommand{\ThetaTwoMinusMEarthAprilSixteenMEarthBand}{0.74}
\newcommand{\ThetaTwoOverThetaOneMEarthAprilSixteenMEarthBand}{0.8}
\newcommand{\ThetaTwoOverThetaOnePlusMEarthAprilSixteenMEarthBand}{0.6}
\newcommand{\ThetaTwoOverThetaOneMinusMEarthAprilSixteenMEarthBand}{0.8}
\newcommand{\TDPercentMEarthAprilSixteenMEarthBand}{42.7}
\newcommand{\TDPercentPlusMEarthAprilSixteenMEarthBand}{13.8}
\newcommand{\TDPercentMinusMEarthAprilSixteenMEarthBand}{16.8}
\newcommand{\TminMEarthAprilSixteenMEarthBand}{129.57600}
\newcommand{\TminPlusMEarthAprilSixteenMEarthBand}{0.00036}
\newcommand{\TminMinusMEarthAprilSixteenMEarthBand}{0.00029}
\newcommand{\TransitDurationMEarthAprilSixteenMEarthBand}{5.3}
\newcommand{\TransitDurationPlusMEarthAprilSixteenMEarthBand}{1.9}
\newcommand{\TransitDurationMinusMEarthAprilSixteenMEarthBand}{2.9}
\newcommand{\FZeroMEarthAprilSixteenMEarthBandSecond}{1.034}
\newcommand{\FZeroPlusMEarthAprilSixteenMEarthBandSecond}{0.019}
\newcommand{\FZeroMinusMEarthAprilSixteenMEarthBandSecond}{0.019}

\newcommand{\ThetaOneMEarthAprilSixteenMEarthBandSecond}{1.39}
\newcommand{\ThetaOnePlusMEarthAprilSixteenMEarthBandSecond}{0.68}
\newcommand{\ThetaOneMinusMEarthAprilSixteenMEarthBandSecond}{1.29}
\newcommand{\ThetaTwoMEarthAprilSixteenMEarthBandSecond}{1.93}
\newcommand{\ThetaTwoPlusMEarthAprilSixteenMEarthBandSecond}{0.95}
\newcommand{\ThetaTwoMinusMEarthAprilSixteenMEarthBandSecond}{1.49}
\newcommand{\ThetaTwoOverThetaOneMEarthAprilSixteenMEarthBandSecond}{1.4}
\newcommand{\ThetaTwoOverThetaOnePlusMEarthAprilSixteenMEarthBandSecond}{1.0}
\newcommand{\ThetaTwoOverThetaOneMinusMEarthAprilSixteenMEarthBandSecond}{1.3}
\newcommand{\TDPercentMEarthAprilSixteenMEarthBandSecond}{27.1}
\newcommand{\TDPercentPlusMEarthAprilSixteenMEarthBandSecond}{9.9}
\newcommand{\TDPercentMinusMEarthAprilSixteenMEarthBandSecond}{10.8}
\newcommand{\TminMEarthAprilSixteenMEarthBandSecond}{129.76341}
\newcommand{\TminPlusMEarthAprilSixteenMEarthBandSecond}{0.00046}
\newcommand{\TminMinusMEarthAprilSixteenMEarthBandSecond}{0.00056}
\newcommand{\TransitDurationMEarthAprilSixteenMEarthBandSecond}{10.0}
\newcommand{\TransitDurationPlusMEarthAprilSixteenMEarthBandSecond}{3.5}
\newcommand{\TransitDurationMinusMEarthAprilSixteenMEarthBandSecond}{5.9}
\newcommand{\FZeroPerkinsMayThirteenVBandSecond}{1.017}
\newcommand{\FZeroPlusPerkinsMayThirteenVBandSecond}{0.005}
\newcommand{\FZeroMinusPerkinsMayThirteenVBandSecond}{0.005}

\newcommand{\ThetaOnePerkinsMayThirteenVBandSecond}{0.36}
\newcommand{\ThetaOnePlusPerkinsMayThirteenVBandSecond}{0.19}
\newcommand{\ThetaOneMinusPerkinsMayThirteenVBandSecond}{0.25}
\newcommand{\ThetaTwoPerkinsMayThirteenVBandSecond}{0.64}
\newcommand{\ThetaTwoPlusPerkinsMayThirteenVBandSecond}{0.19}
\newcommand{\ThetaTwoMinusPerkinsMayThirteenVBandSecond}{0.23}
\newcommand{\ThetaTwoOverThetaOnePerkinsMayThirteenVBandSecond}{1.8}
\newcommand{\ThetaTwoOverThetaOnePlusPerkinsMayThirteenVBandSecond}{1.1}
\newcommand{\ThetaTwoOverThetaOneMinusPerkinsMayThirteenVBandSecond}{1.4}
\newcommand{\TDPercentPerkinsMayThirteenVBandSecond}{15.9}
\newcommand{\TDPercentPlusPerkinsMayThirteenVBandSecond}{3.0}
\newcommand{\TDPercentMinusPerkinsMayThirteenVBandSecond}{4.3}
\newcommand{\TminPerkinsMayThirteenVBandSecond}{155.81033}
\newcommand{\TminPlusPerkinsMayThirteenVBandSecond}{0.00010}
\newcommand{\TminMinusPerkinsMayThirteenVBandSecond}{0.00011}
\newcommand{\TransitDurationPerkinsMayThirteenVBandSecond}{3.0}
\newcommand{\TransitDurationPlusPerkinsMayThirteenVBandSecond}{0.8}
\newcommand{\TransitDurationMinusPerkinsMayThirteenVBandSecond}{1.0}
\newcommand{\FZeroMINERVAMayEighteenAirBandTFour}{0.998}
\newcommand{\FZeroPlusMINERVAMayEighteenAirBandTFour}{0.007}
\newcommand{\FZeroMinusMINERVAMayEighteenAirBandTFour}{0.007}

\newcommand{\ThetaOneMINERVAMayEighteenAirBandTFour}{0.43}
\newcommand{\ThetaOnePlusMINERVAMayEighteenAirBandTFour}{0.16}
\newcommand{\ThetaOneMinusMINERVAMayEighteenAirBandTFour}{0.16}
\newcommand{\ThetaTwoMINERVAMayEighteenAirBandTFour}{0.58}
\newcommand{\ThetaTwoPlusMINERVAMayEighteenAirBandTFour}{0.25}
\newcommand{\ThetaTwoMinusMINERVAMayEighteenAirBandTFour}{0.25}
\newcommand{\ThetaTwoOverThetaOneMINERVAMayEighteenAirBandTFour}{1.4}
\newcommand{\ThetaTwoOverThetaOnePlusMINERVAMayEighteenAirBandTFour}{0.7}
\newcommand{\ThetaTwoOverThetaOneMinusMINERVAMayEighteenAirBandTFour}{1.1}
\newcommand{\TDPercentMINERVAMayEighteenAirBandTFour}{32.9}
\newcommand{\TDPercentPlusMINERVAMayEighteenAirBandTFour}{4.5}
\newcommand{\TDPercentMinusMINERVAMayEighteenAirBandTFour}{6.5}
\newcommand{\TminMINERVAMayEighteenAirBandTFour}{160.66062}
\newcommand{\TminPlusMINERVAMayEighteenAirBandTFour}{0.00010}
\newcommand{\TminMinusMINERVAMayEighteenAirBandTFour}{0.00010}
\newcommand{\TransitDurationMINERVAMayEighteenAirBandTFour}{3.0}
\newcommand{\TransitDurationPlusMINERVAMayEighteenAirBandTFour}{0.9}
\newcommand{\TransitDurationMinusMINERVAMayEighteenAirBandTFour}{0.9}

\shorttitle{Multiwavelength Transit Observations of WD 1145+017} 
\shortauthors{Croll et al.}

\begin{document}

\title{Multiwavelength Transit Observations of the Candidate Disintegrating Planetesimals Orbiting WD 1145+017}

\author{
	Bryce Croll\altaffilmark{1},
	Paul A. Dalba\altaffilmark{2},
	Andrew Vanderburg\altaffilmark{3},
	Jason Eastman\altaffilmark{3},
	Saul Rappaport\altaffilmark{4},
	John DeVore\altaffilmark{5},	
	Allyson Bieryla\altaffilmark{3},
	Philip S. Muirhead\altaffilmark{2} \altaffilmark{1},	
	Eunkyu Han\altaffilmark{2},
	David W. Latham\altaffilmark{3},
	Thomas G. Beatty\altaffilmark{6},
	Robert A. Wittenmyer\altaffilmark{7},
	Jason T. Wright\altaffilmark{6} \altaffilmark{8},
	John Asher Johnson\altaffilmark{3},
	Nate McCrady\altaffilmark{9}
}

\altaffiltext{1}{Institute for Astrophysical Research, Boston University, 725 Commonwealth Ave. Room 506, Boston, MA 02215; croll@bu.edu}

\altaffiltext{2}{Department of Astronomy, Boston University, 725 Commonwealth Ave. Boston, MA 02215, USA}

\altaffiltext{3}{Harvard-Smithsonian Center for Astrophysics, 60 Garden Street, Cambridge, MA 02138 USA}

\altaffiltext{4}{Kavli Institute for Astrophysics and Space Research, Massachusetts Institute of Technology, Cambridge, MA 02139, USA}

\altaffiltext{5}{Visidyne, Inc., Santa Barbara, CA 93105, USA}

\altaffiltext{6}{Department of Astronomy and Astrophysics and Center for Exoplanets and Habitable Worlds, The Pennsylvania State University, University Park, PA 16802}

\altaffiltext{7}{School of Physics and Australian Centre for Astrobiology, UNSW Australia, Sydney, NSW 2052, Australia}

\altaffiltext{8}{NASA Nexus for Exoplanet System Science}

\altaffiltext{9}{Department of Physics and Astronomy, University of Montana, 32 Campus Drive, No. 1080, Missoula, MT 59812 USA}

\begin{abstract}

 We present multiwavelength, multi-telescope, ground-based follow-up photometry of the white dwarf 
WD 1145+017, that has recently been 
suggested to be orbited by up to six or more, short-period, low-mass, disintegrating planetesimals.
We detect \NumSignificantTransits \ significant dips in flux 
of between 10\% and 30\% of the stellar flux
from our ground-based photometry.
We observe transits deeper than 10\% on average every $\sim$\HoursPerSignificantTransits \ $\rm hr$ in our photometry.
This suggests that WD 1145+017 is indeed being orbited by  multiple, short-period objects.
Through fits to the multiple asymmetric transits that we observe, we confirm 
that the transit egress timescale is usually longer than the ingress timescale, and that the transit duration is longer
than expected for a solid body at these short periods, all suggesting that these objects have cometary 
tails streaming behind them.
The precise orbital periods of the planetesimals in this system are unclear from the transit-times,
but at
least one object, and likely more, have orbital periods of $\sim$4.5 hours.
We are otherwise unable to confirm the specific periods that have been reported, 
bringing into question the long-term stability of these periods.
Our high precision photometry also displays low amplitude variations suggesting that dusty material is consistently
passing in front of the white dwarf, either from discarded material from these disintegrating planetesimals or from the 
detected dusty debris disk.
For the significant transits we observe, we compare the transit depths in the V- and R-bands
of our multiwavelength photometry,
and find no significant difference; therefore, for likely compositions 
the radius of single-size particles in the cometary tails streaming behind the planetesimals in this system must be   
$\sim$\MicronSingleSizeLimit \ ${\rm \mu m}$ or larger, or $\sim$\MicronSingleSizeLowLimit \ ${\rm \mu m}$ or smaller, with 2$\sigma$ confidence.

\end{abstract}

\keywords{planetary systems . techniques: photometric -- eclipses . stars: individual: WD 1145+017 }

\section{Introduction}

The white dwarf WD 1145+017 was recently announced to host up to six or more disintegrating candidate planetesimals
in extremely short-period orbits:
\citet{Vanderburg15} presented two-wheeled {\it Kepler} Space Telescope (K2) photometry \citep{Howell14}
of WD 1145+017 with six distinct occultations
with periods from $\sim$4.5 - 4.9 hours and depths up to a few percent.
The depth of the main $\sim$4.5 hour transiting object also evolved
from undetectable to a few percent over the 80 days of long cadence K2 photometry.
As the long cadence ($\sim$29.4 $\rm min$) {\it Kepler} integrations were poorly suited
to resolve what likely should be events with very short transit durations ($\sim$1 $\rm min$)
at these ultra-short periods, follow-up
photometry was performed: \citet{Vanderburg15} detected $\sim$40\% eclipses with an asymmetric transit profile
using the 1.2 $\rm m$ Fred L. Whipple Observatory, 
and the MEarth South array of 0.4 ${\rm m}$ telescopes \citep{NutzmannCharbonneau08,Irwin15}. 
These observed apparent occultations of WD 1145+017
displayed many characteristics in common with other candidate disintegrating, ultra-short 
period planets, including variable transit depths, and an asymmetric transit profile featuring a sharp ingress and gradual
egress; in this case -- and in the case of the three other
ultra short-period, low-mass, disintegrating planet candidates that have been claimed to date 
(KIC 12557548; \citealt{Rappaport12}, KOI-2700b; \citealt{Rappaport14}, and EPIC 201637175b; \citealt{Sanchis15}) --
these characteristics are interpreted as being due to a variable amount of material disintegrating
from the planets/planetesimals 
that condenses at altitude into a 
dusty cometary tail streaming behind the planets/planetesimals.

 That WD 1145+017 might be the best example of a white dwarf orbited by close-in planets/planetesimals is
strengthened by two additional
lines of evidence: the spectrum of WD 1145+017 is significantly polluted, and it displays 
an infrared excess. A visible spectrum of WD 1145+017 revealed spectral lines
of calcium, aluminum, magnesium, silicon, nickel and iron \citep{Vanderburg15};
as the settling time of these elements is rapid ($\sim$1 million years)
compared
with the age of this white dwarf ($\sim$175 million years), these elements must have recently accreted onto the
white dwarf.  \citet{Vanderburg15} also 
found evidence for an infrared excess likely arising from a $\sim$1150 K warm dusty debris disk;
such a debris disk could be the source of the planetesimals that have been observed to pass in front of this white dwarf.

WD 1145+017 is arguably the most compelling example 
of the many white dwarfs that have been observed to be significantly polluted as a result of what has been claimed to be 
the accretion of rocky bodies.
Approximately 1/3 of all white dwarfs cooler than 
20,000 K display the presence of elements heavier than hydrogen/helium \citep{Zuckerman03,Zuckerman10,Koester14};
for white dwarfs of these temperatures, elements heavier than hydrogen and helium sink beneath the outer layers quickly
compared to the cooling time.
Although it was originally suggested that this material originated from the interstellar medium (e.g. \citealt{Dupuis93}),
the currently accepted, canonical origin for these elements is that they resulted from material
from asteroids or more massive rocky
bodies that have been orbitally perturbed \citep{DebesSigurdsson02}, are tidally disrupted -- often into a debris disk --
and then
material from these bodies gradually or quickly
accretes onto the white dwarf \citep{Jura03,Zuckerman10}.
The reason that the origin of these polluting elements is thought to result from rocky bodies
is that analyses of high resolution
spectra of polluted white dwarfs have allowed the elemental abundances of these polluting materials to be measured,
and they are broadly consistent with rocky, terrestrial solar system bodies with refractory-rich and
volatile-poor material\footnote{Although there have now been a few polluted white dwarfs with spectra that are believed to result from volatile/water-rich asteroids (e.g. \citealt{Raddi15}).}
\citep{Zuckerman07}.


 One method of lending credence to the disintegrating planet/planetesimal candidate scenario is
through multiwavelength observations to constrain 
the particle size of the dusty material in the cometary tails trailing these objects. 
This has already been attempted for the candidate disintegrating planet KIC 12557548b; optical and
near-infrared observations of two transits of KIC 12557548b suggested that the 
grain sizes trailing KIC 12557548b were $\sim$0.5 ${\rm \mu m}$ in radius or larger \citep{Croll14}, while
optical multiwavelength observations suggested the particle sizes were 0.25 - 1.0 ${\rm \mu m}$ 
\citep{Bochinski15}.
Multiwavelength optical photometry of one transit of the disintegrating low-mass candidate exoplanet EPIC 201637175b suggests approximate
particle sizes of 0.2 - 0.4 ${\rm \mu m}$ in the cometary tail trailing that body \citep{Sanchis15}.


 For the candidate planetesimals orbiting WD 1145+017, multiwavelength observations may help determine the mechanism 
that generates the dust that is believed to be trailing these objects.
The preferred explanation of \citet{Vanderburg15} for the apparent dusty tails was that the high temperatures of these planetesimals in
these short-period orbits would result in material sublimating off the planetesimals' surfaces
with sufficient thermal speed to overcome the escape
speed on these low surface gravity objects; at altitude these vapours would condense into dust.
For the other 
disintegrating, planet-mass candidates, 
vapour is believed to be driven 
from the higher surface gravity of these planets
by a Parker-wind, before similarly condensing into dust at altitude \citep{Rappaport12,PerezBeckerChiang13,Rappaport14,Sanchis15}.
In these other disintegrating systems the presumed higher planet masses of these candidates, and therefore the higher surface gravities,
require a Parker-wind, compared
to the assumed Ceres-mass planetesimals of the WD 1145+017 system, 
where the lower surface gravities allow material to freely stream from the planetesimals.
Alternative possibilities to explain the dusty material in the cometary tails of the planetesimals in this system 
include (i) that these bodies could be similar to comets in our own solar system with low enough surface
gravities that their dust tails are carried off by the disintegration of volatiles, (ii) that these planetesimals and their cometary tails
are the result of collisions with other planetesimals in the system or the observed debris disk, (iii) or that  
tidal forces from the white dwarf have ripped larger bodies apart, or are in the process of ripping apart such bodies, and we are observing
the tidally disrupted bodies that have possibly formed the observed debris disk. 
The rigid-body Roche limit for a Ceres density asteroid around this white dwarf is at an orbital period of $\sim$3.4 hours,
suggesting that if these planetesimals are similar to the asteroids in our own solar system they
should already
be subjected to considerable tidal forces that may be threatening to rip them apart. In the latter two cases of a collision or a tidally disrupted body, 
shear would likely quickly result in material trailing behind the planetesimals \citep{Veras14}, forming cometary tails.
Naively, a tidally disrupted body would suggest larger particle sizes in the trailing tails,
if they are similar to disrupted bodies in our own solar
system (e.g. \citealt{Michikami08,Jewitt10,Jewitt13}). 

 Here we report a wealth of multiwavelength follow-up photometry of WD 1145+017 that considerably strengthens the conclusion
that this star is being orbited by a number of low-mass bodies with dusty trails trailing behind them.
We present multiwavelength ground-based photometry from a variety of telescopes in Section \ref{SecObs}; we display a number of significant
decrements in flux of up to $\sim$30\% of the stellar flux, likely the result of planetesimals with dusty tails passing in front of  the white dwarf host
and scattering light out of the line of sight. We analyze the depths, duration and timing of 
these eclipses in Section \ref{SecAnalysis}; the egress timescale of these transits is usually longer
than the ingress timescale, and the transit duration is longer than we would expect for a circular orbit of an Earth-sized body 
at these short periods, findings that are both consistent with the hypothesis that this system contains short-period
planetesimals trailed by dusty, cometary tails.
The exact periods of these plantesimals are uncertain, but several objects appear to have periods of approximately $\sim$4.5 hours; with this many
objects with nearly identical periods it is unclear whether the orbits of these objects are stable. 
Lastly, the ratio of the transit depths 
from
our multiwavelength V and R-band observations allow us to conclude
that if the dust grains trailing the planetesimals in the WD 1145+017 system are all of a single size then 
they have a radius of 
$\sim$\MicronSingleSizeLimit \ ${\rm \mu m}$ or larger, or $\sim$\MicronSingleSizeLowLimit \ ${\rm \mu m}$ or smaller, 
with 2$\sigma$ confidence.

\section{Observations}
\label{SecObs}

\begin{deluxetable*}{ccccccccc}
\tablecaption{Observing Log}
\tablehead{
\colhead{Date} 		& \colhead{Telescope}		& \colhead{Observing}	& \colhead{Duration}	& \colhead{Exposure} 	& \colhead{Overhead\tablenotemark{a}}	& \colhead{Airmass}				& \colhead{Conditions}	& \colhead{Aperture \tablenotemark{b}}\\
\colhead{(UTC)}		& \colhead{\& Instrument}	& \colhead{Band}	& \colhead{(hours)}	& \colhead{Time (sec)}	& \colhead{(sec)}			& \colhead{}					& \colhead{}		& \colhead{(pixels)}\\
}
\centering
\startdata
2015/05/08		& Perkins/PRISM			& V			& 3.85 			& 45			& 6.3	& 1.21 $\rightarrow$ 1.20 $\rightarrow$ 1.98	& Occasional clouds	& 8, 16, 24  \\	%
2015/05/08		& FLWO/KeplerCam		& V			& 1.77 			& 60			& 16.0	& 1.16 $\rightarrow$ 1.30			& Occasional clouds	& 8, 20, 32  \\	%
2015/05/09		& MINERVA T1			& R			& 3.88 			& 60			& 6.0	& 1.16 $\rightarrow$ 2.19			& Clear			& 8, 20, 32  \\	%
2015/05/09		& MINERVA T3			& air			& 3.77 			& 60			& 6.0	& 1.16 $\rightarrow$ 2.22			& Clear			& 8, 20, 32  \\	%
2015/05/09		& MINERVA T4			& air			& 4.50 			& 60			& 6.0	& 1.18 $\rightarrow$ 1.16 $\rightarrow$ 1.99	& Clear			& 8, 20, 32  \\	%
2015/05/10		& MINERVA T1			& R			& 5.15 			& 60			& 6.0	& 1.16 $\rightarrow$ 5.81			& Clear			& 8, 20, 32  \\	%
2015/05/10		& MINERVA T2			& V			& 4.91			& 60			& 6.0	& 1.16 $\rightarrow$ 4.86			& Clear			& 8, 20, 32  \\	%
2015/05/10		& MINERVA T3			& air			& 4.46 			& 60			& 6.0	& 1.20 $\rightarrow$ 7.91			& Clear			& 8, 20, 32  \\	%
2015/05/10		& MINERVA T4			& air			& 2.79 			& 60			& 6.0	& 1.52 $\rightarrow$ 7.94			& Clear			& 8, 20, 32  \\	%
2015/05/11		& DCT/LMI			& V			& 3.40 			& 30			& 8.6	& 1.48 $\rightarrow$ 1.03 			& Clear			& 13, 25, 35 \\	%
2015/05/11		& FLWO/KeplerCam		& V			& 4.53 			& 60			& 16.0	& 1.17 $\rightarrow$ 1.16 $\rightarrow$ 2.17 	& Clear			& 8, 20, 32  \\	%
2015/05/11		& MINERVA T1			& R			& 2.17 			& 60			& 6.0	& 1.24 $\rightarrow$ 1.92			& Clear			& 8, 20, 32  \\	%
2015/05/11		& MINERVA T2			& V			& 4.17			& 60			& 6.0	& 1.17 $\rightarrow$ 1.16 $\rightarrow$ 1.91	& Clear			& 8, 20, 32  \\	%
2015/05/11		& MINERVA T3			& air			& 4.24 			& 60			& 6.0	& 1.18 $\rightarrow$ 1.16  $\rightarrow$ 1.92	& Clear			& 8, 20, 32  \\	%
2015/05/11		& MINERVA T4			& air			& 4.21 			& 60			& 6.0	& 1.17 $\rightarrow$ 1.16  $\rightarrow$ 1.92	& Clear			& 8, 20, 32  \\	%
2015/05/12		& Perkins/PRISM			& R			& 3.96 			& 45			& 6.0	& 1.21 $\rightarrow$ 1.20 $\rightarrow$ 1.97	& Clear			& 8, 16, 24  \\	%
2015/05/12		& MINERVA T1			& V			& 3.99 			& 60			& 6.0	& 1.16 $\rightarrow$ 1.91			& Occasional clouds	& 8, 20, 32  \\	%
2015/05/12		& MINERVA T2			& V			& 4.15 			& 60			& 6.0	& 1.17 $\rightarrow$ 1.16 $\rightarrow$ 1.91	& Occasional clouds	& 8, 20, 32  \\	%
2015/05/12		& MINERVA T3			& air			& 3.99 			& 60			& 6.0	& 1.16 $\rightarrow$ 1.90			& Occasional clouds	& 8, 20, 32  \\	%
2015/05/12		& MINERVA T4			& air			& 4.15 			& 60			& 6.0	& 1.17 $\rightarrow$ 1.16 $\rightarrow$ 1.91	& Occasional clouds	& 8, 20, 32  \\	%
2015/05/13		& Perkins/PRISM			& V			& 2.34			& 45			& 6.5	& 1.26 $\rightarrow$ 1.97 			& Thin clouds to clear	& 8, 16, 24  \\	%
2015/05/13		& MINERVA T1			& R			& 1.56 			& 60			& 6.0	& 1.34 $\rightarrow$ 1.94 			& Occasional clouds	& 8, 20, 32  \\	%
2015/05/13		& MINERVA T2			& B			& 4.13 			& 60			& 6.0	& 1.17 $\rightarrow$ 1.16 $\rightarrow$ 1.95	& Occasional clouds	& 8, 20, 32  \\	%
2015/05/13		& MINERVA T4			& air			& 4.13 			& 60			& 6.0	& 1.16 $\rightarrow$ 1.95 			& Occasional clouds	& 8, 20, 32  \\	%
2015/05/18		& DCT/LMI			& R 			& 0.96			& 30			& 8.5	& 1.27 $\rightarrow$ 1.43			& Clear 		& 10, 20, 30 \\	%
2015/05/18		& MINERVA T1			& R			& 3.20 			& 60			& 6.0	& 1.16 $\rightarrow$ 1.88 			& Clear			& 8, 20, 32  \\	%
2015/05/18		& MINERVA T2			& B			& 3.72 			& 60			& 6.0	& 1.16 $\rightarrow$ 1.93			& Clear			& 8, 20, 32  \\	%
2015/05/18		& MINERVA T4			& air			& 3.71 			& 60			& 6.0	& 1.16 $\rightarrow$ 1.93 			& Clear			& 8, 20, 32  \\	%
2015/05/22		& Perkins/PRISM			& R 			& 2.79			& 45			& 6.0	& 1.22 $\rightarrow$ 1.90			& Clear 		& 8, 16, 24  \\	%
2015/05/23		& Perkins/PRISM			& R 			& 2.15			& 45			& 7.0	& 1.22 $\rightarrow$ 1.59			& Occasional clouds 	& 8, 16, 24  \\	%
\enddata
\tablenotetext{a}{The overhead includes time for read-out, and any other applicable overheads.}
\tablenotetext{b}{We give the radius of the aperture, the radius of the inner annulus and the radius of the outer annulus that we use for sky subtraction in pixels.}
\label{TableObs}
\end{deluxetable*}

\begin{figure*}
\includegraphics[scale=0.37, angle = 270]{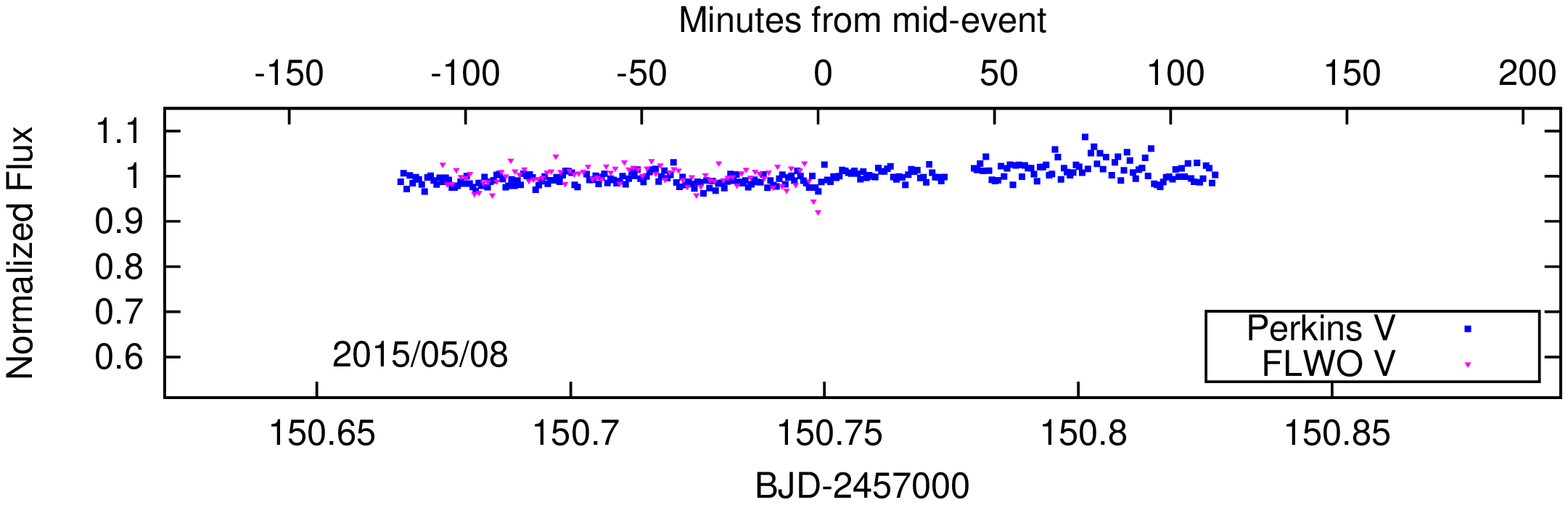}
\includegraphics[scale=0.37, angle = 270]{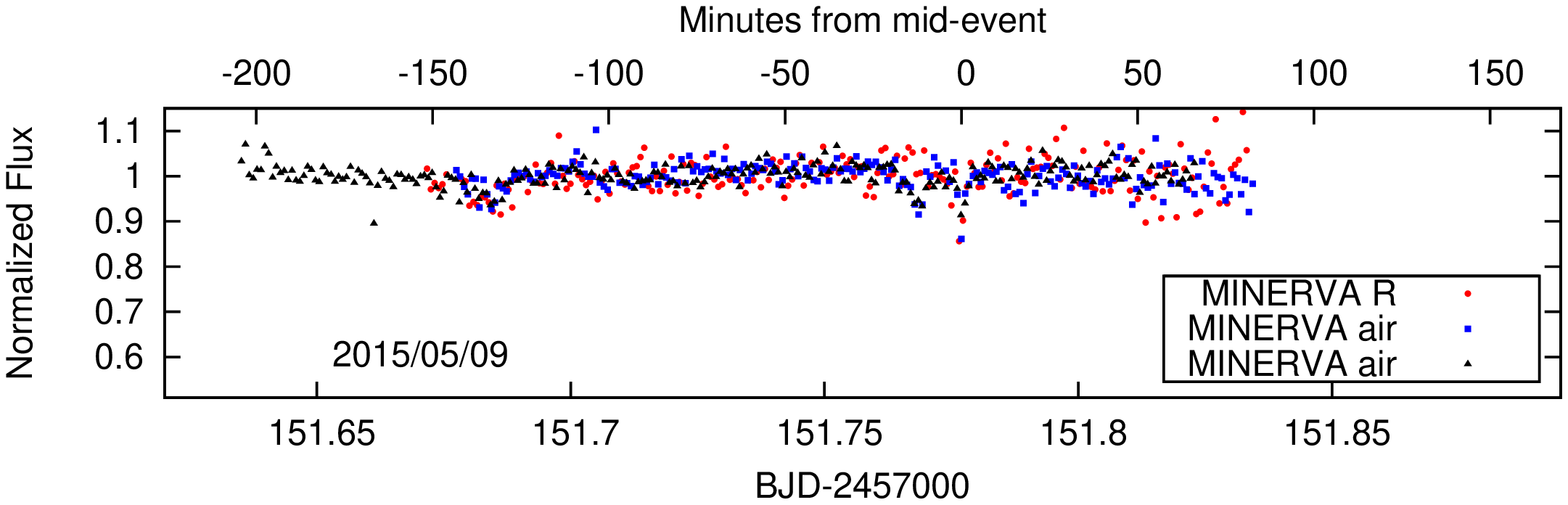}
\includegraphics[scale=0.37, angle = 270]{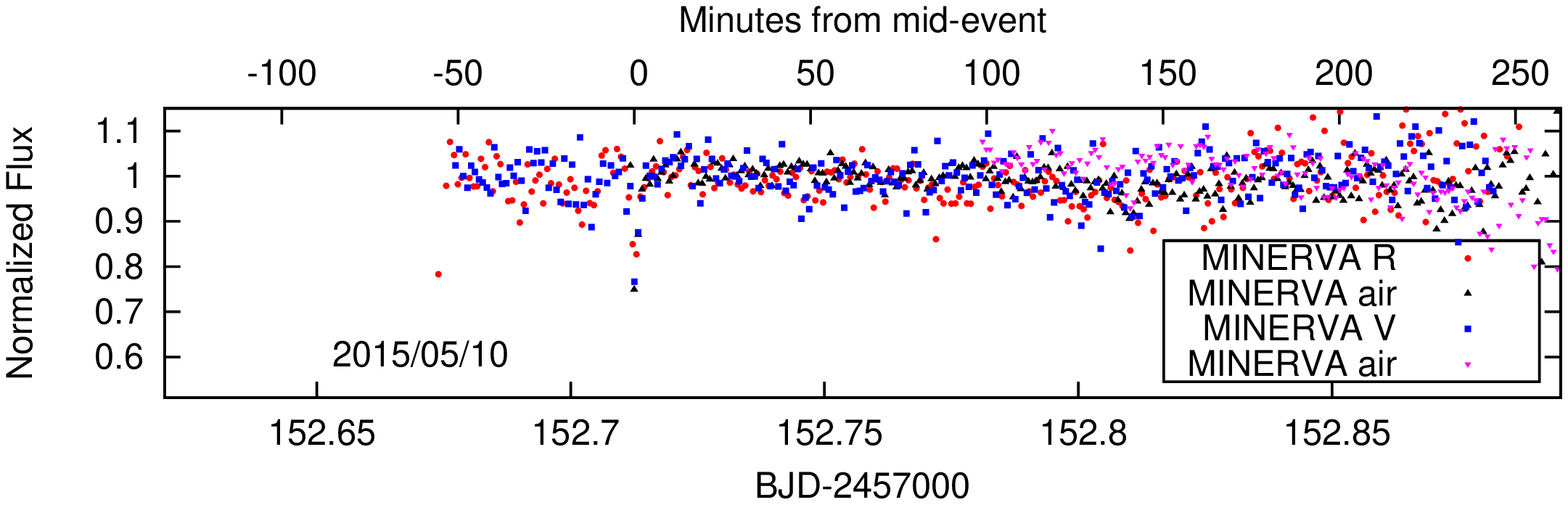}
\includegraphics[scale=0.37, angle = 270]{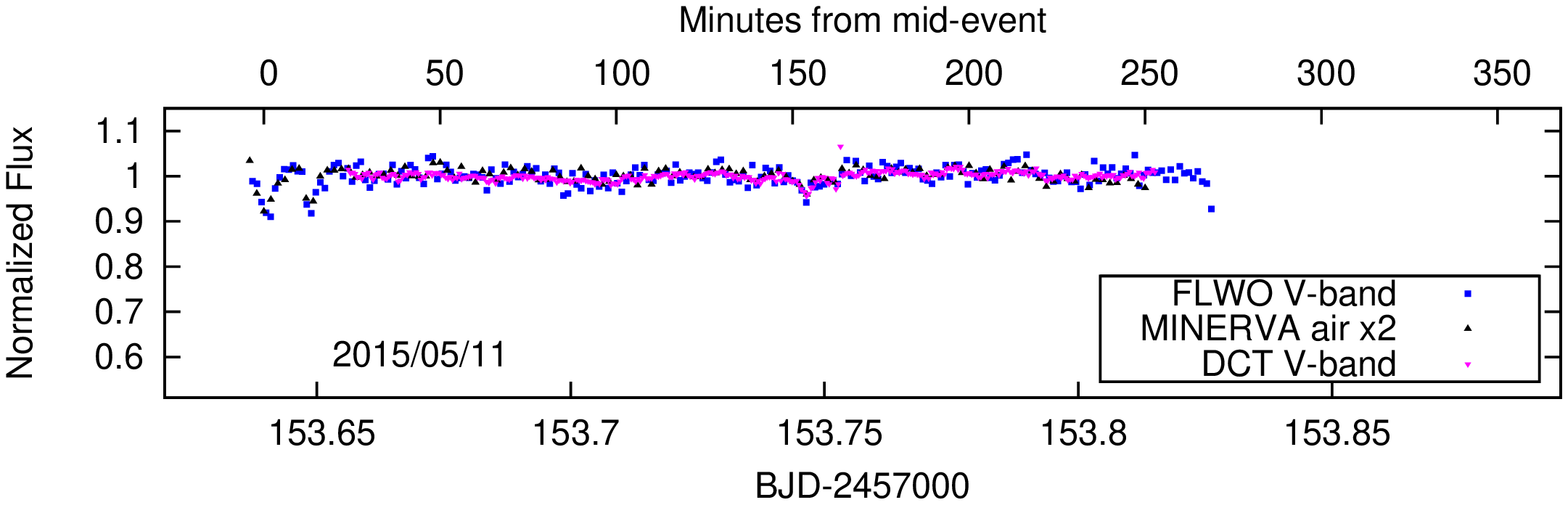}
\includegraphics[scale=0.37, angle = 270]{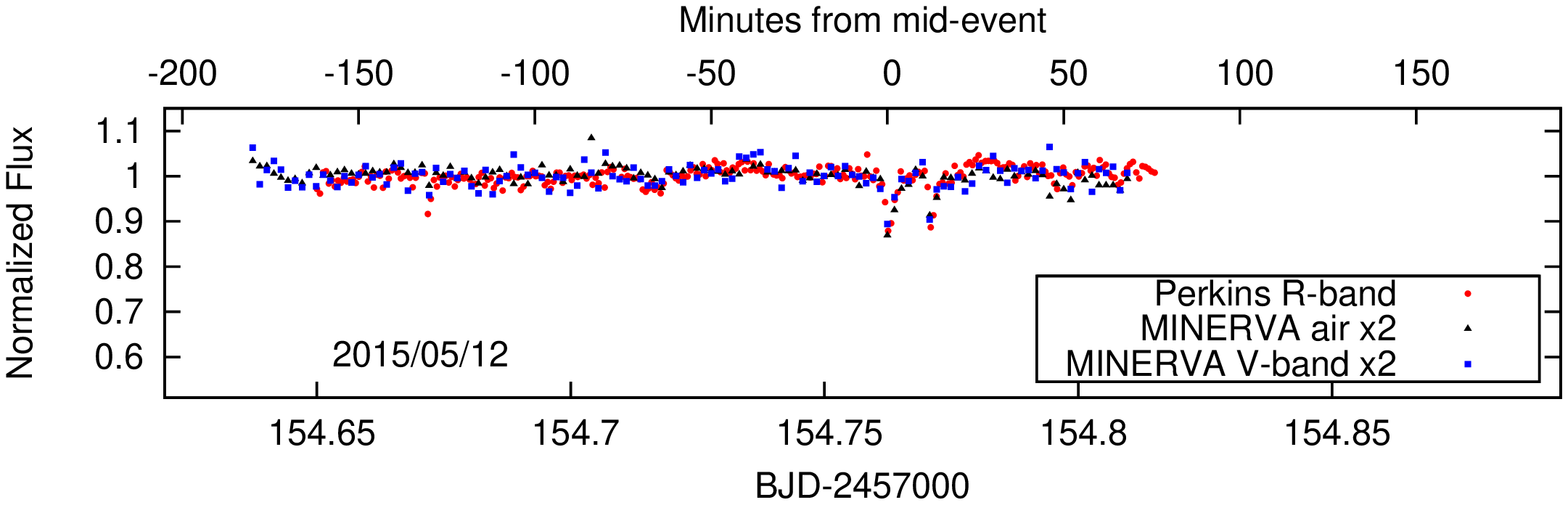}
\includegraphics[scale=0.37, angle = 270]{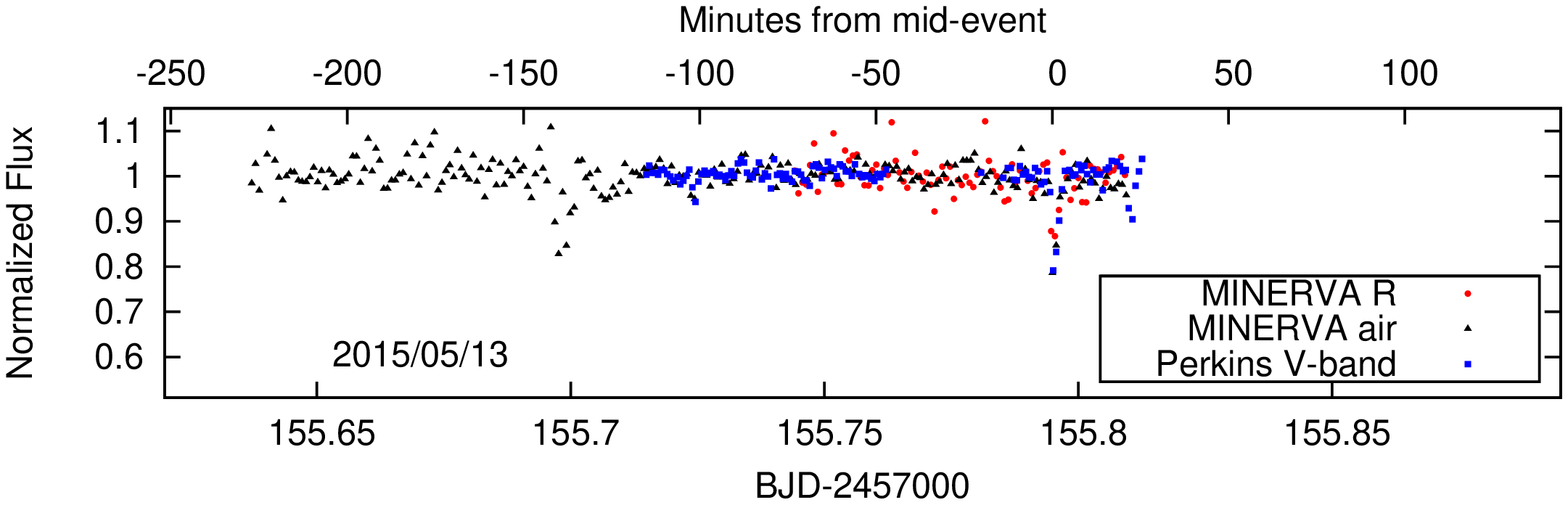}
\includegraphics[scale=0.37, angle = 270]{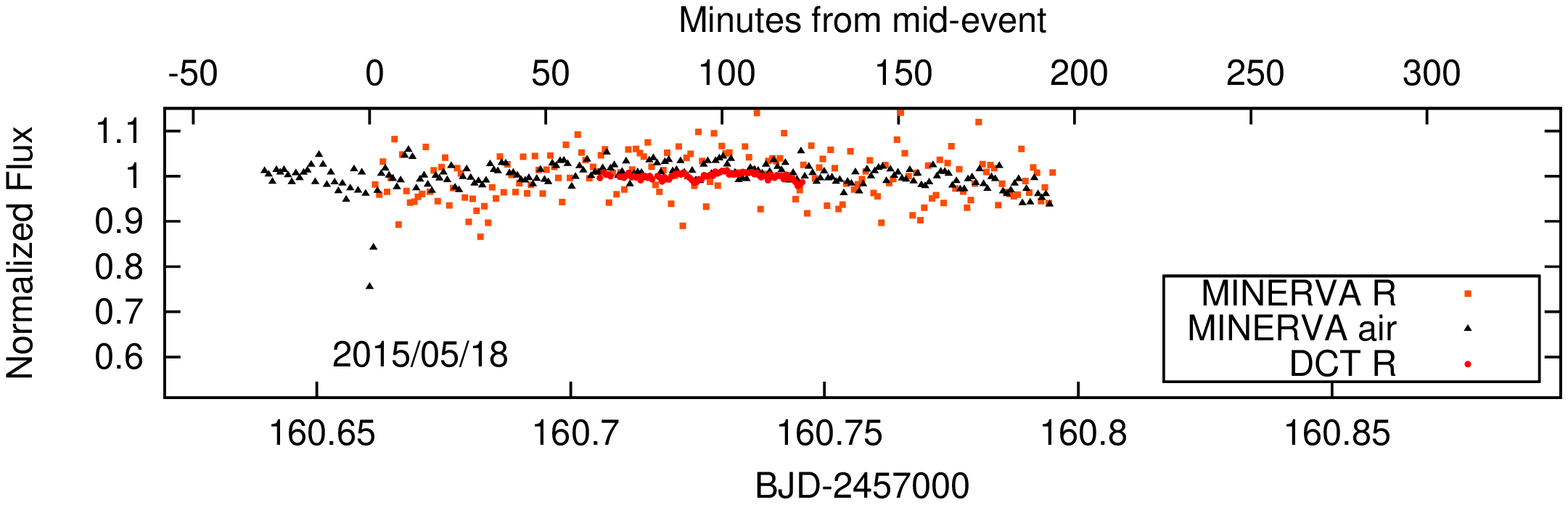}
\includegraphics[scale=0.37, angle = 270]{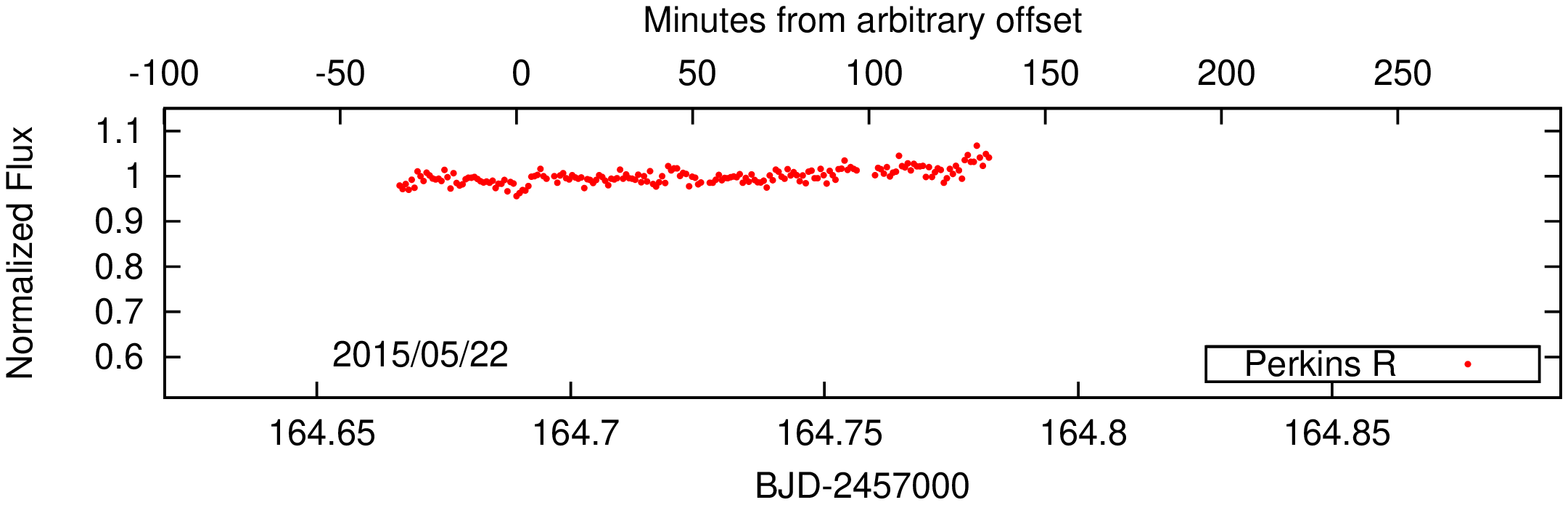}
\includegraphics[scale=0.37, angle = 270]{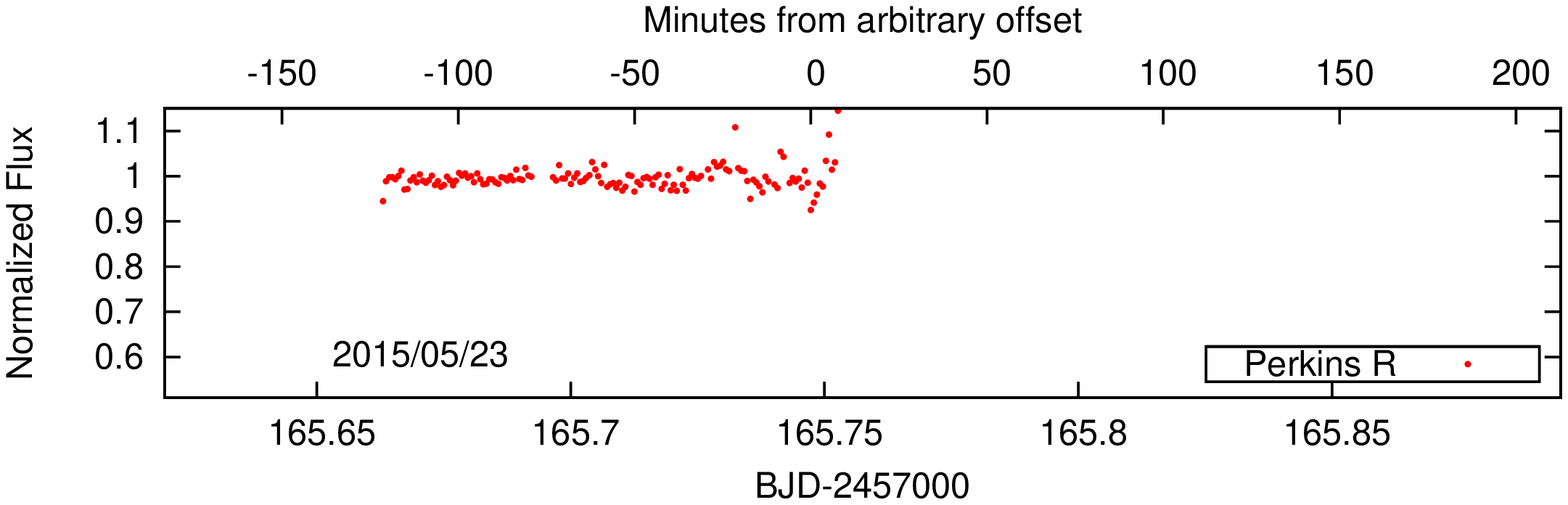}
\caption[]
	{	Perkins/PRISM, DCT/LMI, FLWO/KeplerCam \& MINERVA photometric observations of WD 1145+017.
		The UTC date of observations are given in the lower-left of each panel, while the telescope and band of observations
		are given in the legend at the lower-right.
		The minutes from mid-event for each
		night of observations are given from the deepest decrement in flux observed in each night, if this decrement
		in flux is believed to be statistically significant. 
		For the ``MINERVA air x2'' and ``MINERVA V-band x2'' data, observations of two MINERVA telescopes
		in the ``air'' and V-bands respectively, have been combined using their weighted mean.
	}
\label{FigLC}
\end{figure*}

\begin{figure*}
\includegraphics[scale=0.37, angle = 270]{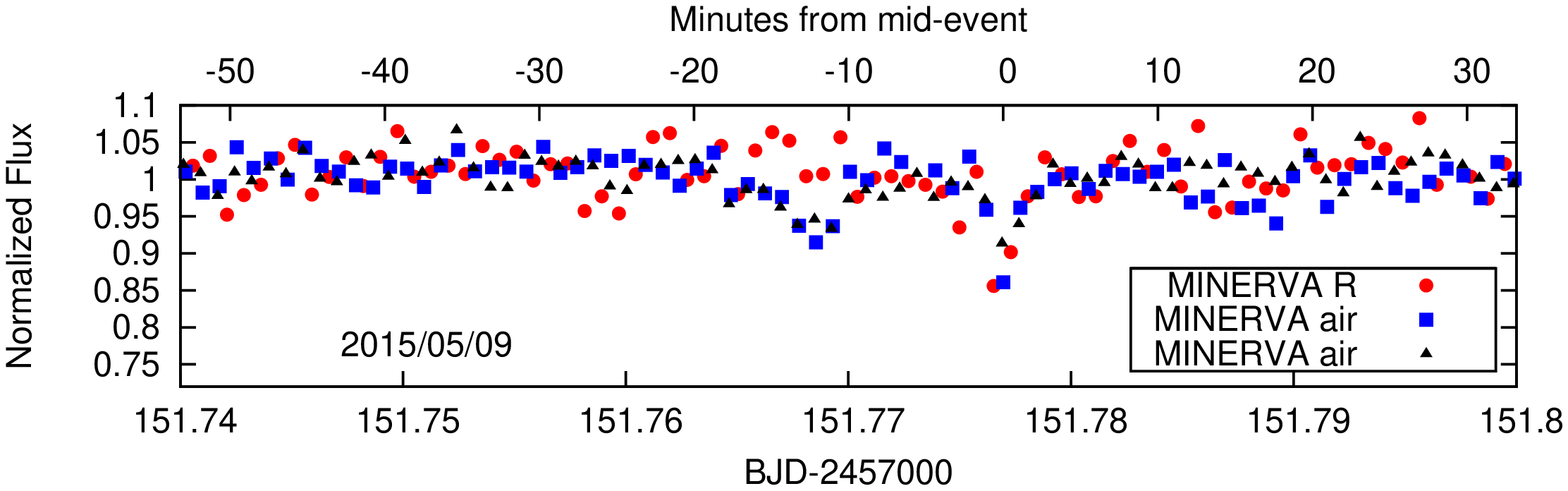} 
\includegraphics[scale=0.37, angle = 270]{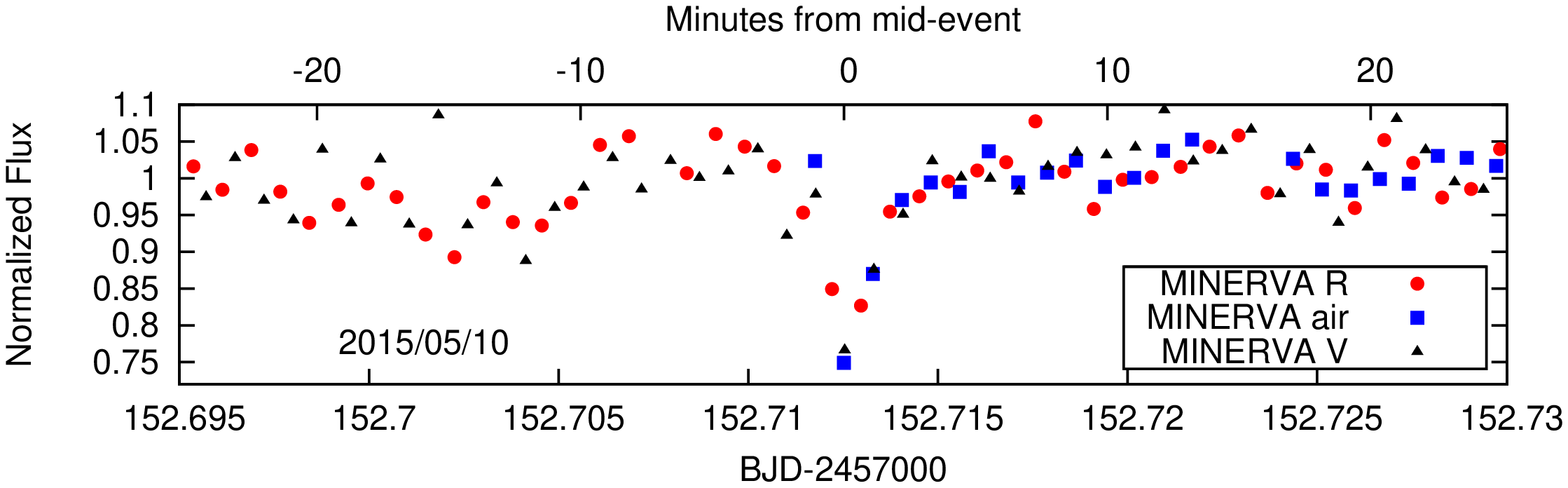}
\includegraphics[scale=0.37, angle = 270]{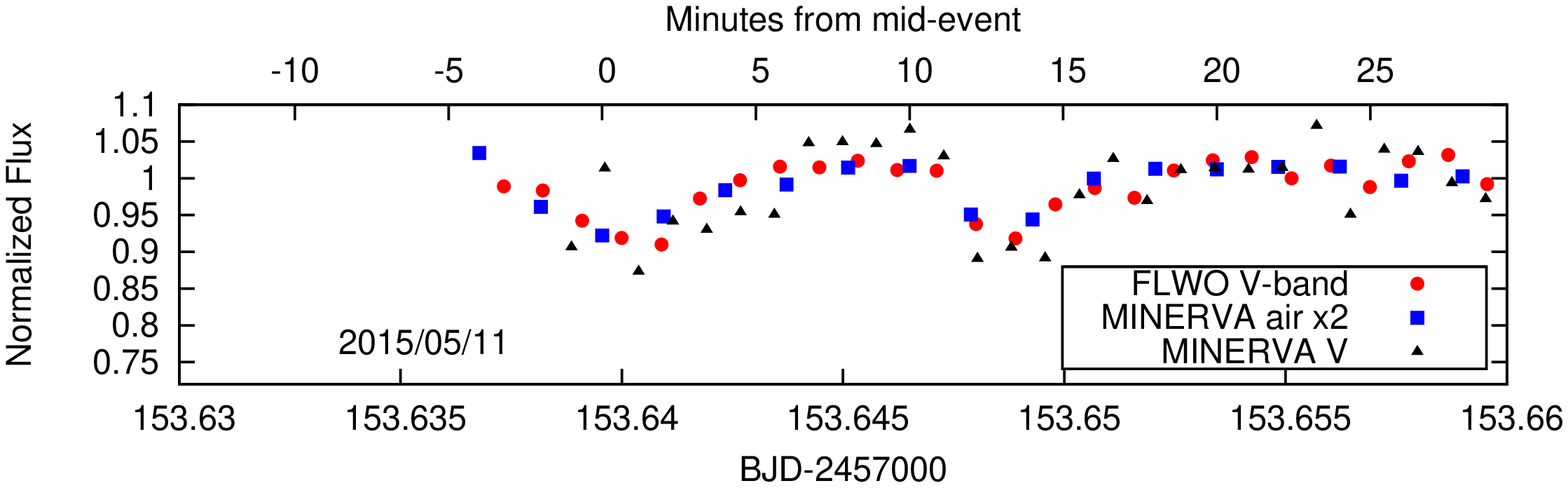}
\includegraphics[scale=0.37, angle = 270]{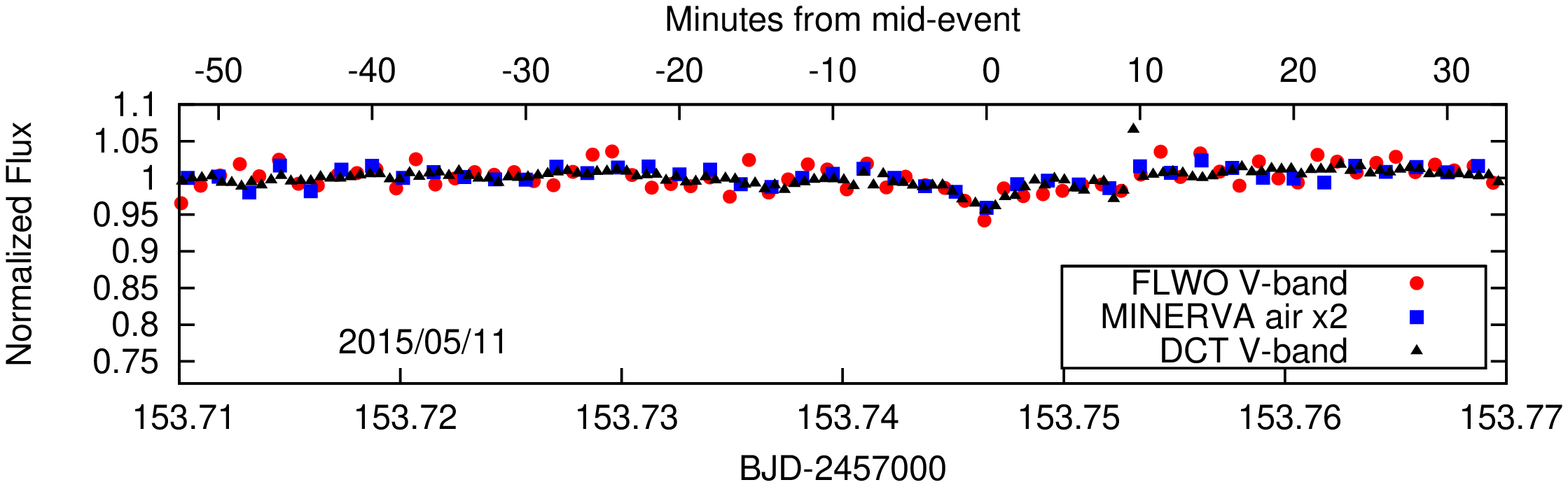}
\includegraphics[scale=0.37, angle = 270]{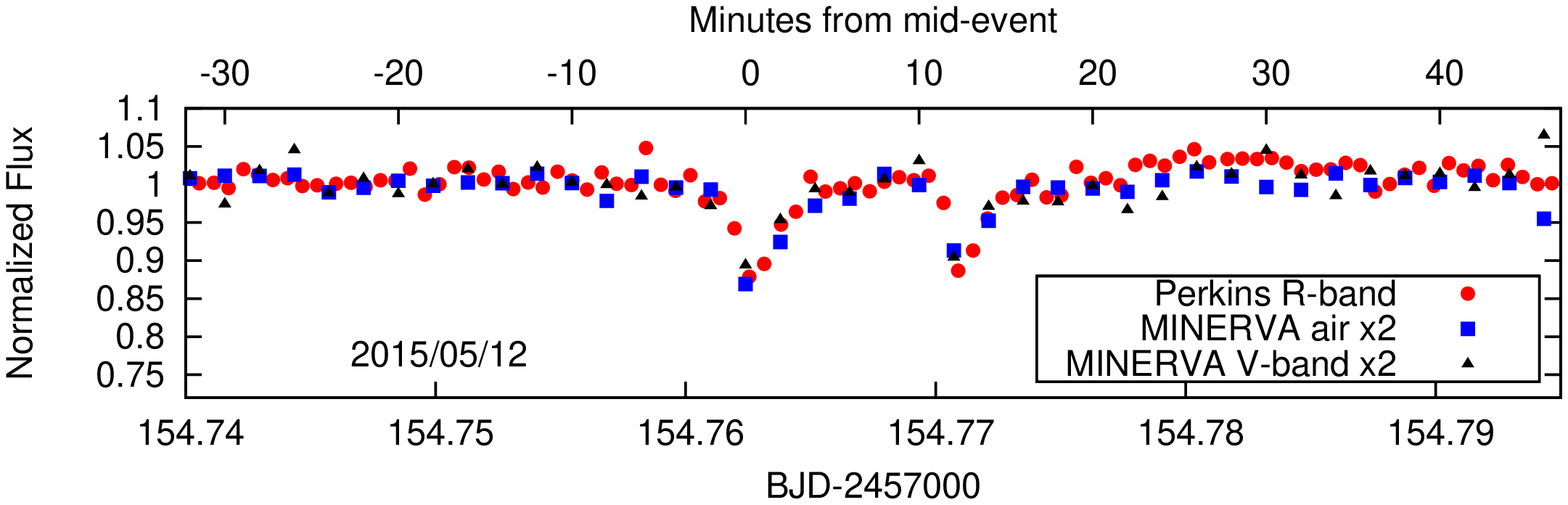}
\includegraphics[scale=0.37, angle = 270]{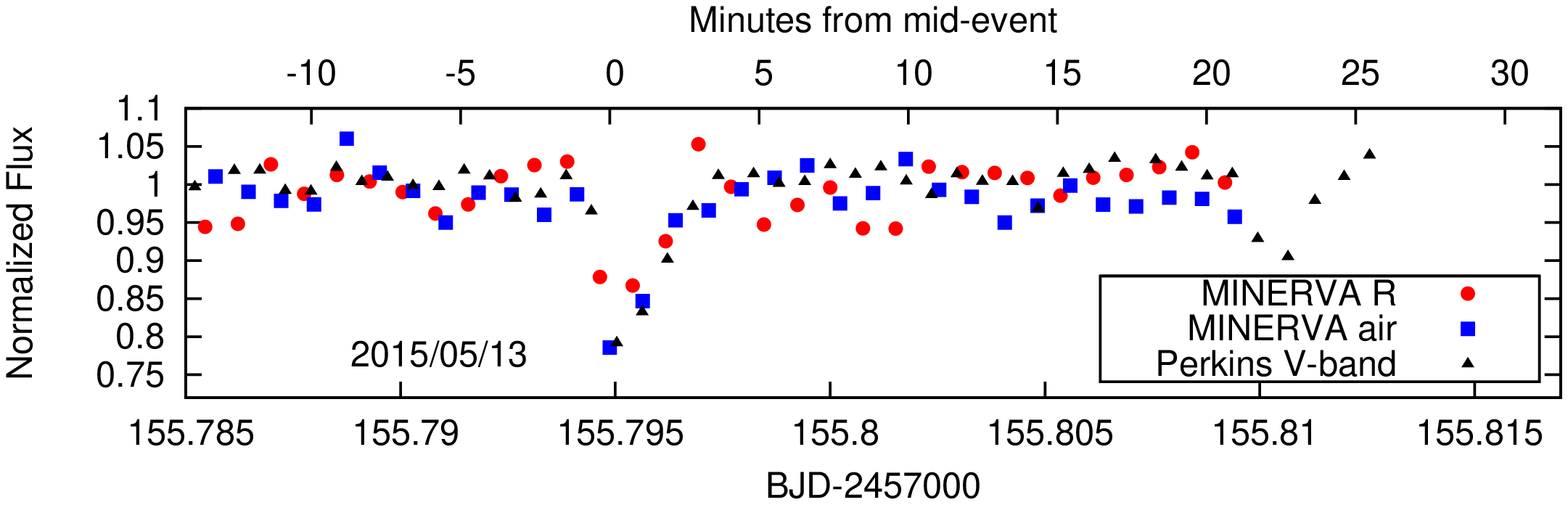}
\includegraphics[scale=0.37, angle = 270]{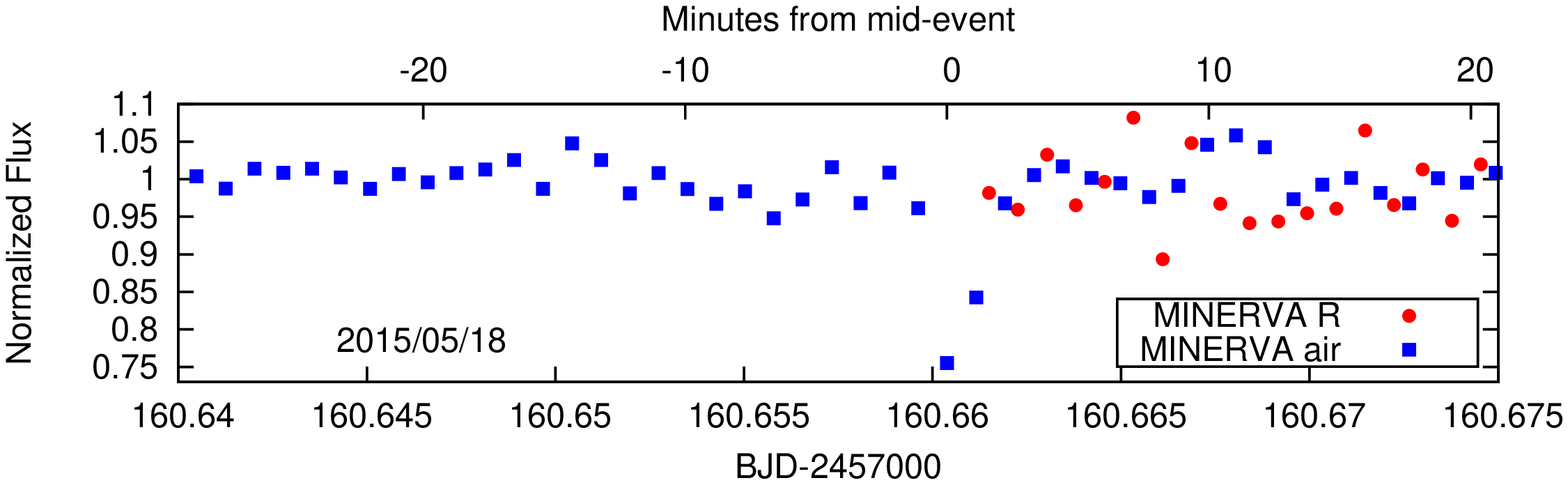}
\caption[]
	{	DCT, Perkins, FLWO, \& MINERVA multiwavelength photometry of WD 1145+017, zoomed-in on durations
		when significant decrements in flux are observed.
		The UTC date of observations are given in the lower-left of each panel, while the telescope and band of observations
		are given in the legend at the lower-right.
		For the ``MINERVA air x2'' and ``MINERVA V-band x2'' data, observations of two MINERVA telescopes
		in the ``air'' and V-bands respectively, have been combined using their weighted mean.
		The minutes from mid-event are given from the deepest decrement in flux observed in each panel.
	}
\label{FigLCMINERVA}
\end{figure*}

We observed WD 1145+017 on a variety of nights in 2015 May with a number of different ground-based
telescopes. These include the Discovery Channel Telescope (DCT) and its Large Monolithic Imager (LMI; \citealt{Massey13}),
the Perkins 1.8-${\rm m}$ telescope and its PRISM imager \citep{Janes04},
the Fred L. Whipple Observatory 1.2-${\rm m}$ telescope and its KeplerCam imager \citep{Szentgyorgyi05},
and the four MINERVA 0.7-${\rm m}$ robotic telescopes, labelled T1-T4 \citep{Swift15}.
We summarize these observations in Table \ref{TableObs}.
The DCT/LMI, Perkins/PRISM and MINERVA data are processed by dark/bias subtracting the data, and then we divide through by a sky-flat.
The FLWO/KeplerCam data are processed using the techniques discussed in \citet{Carter11}.
Aperture
photometry is performed using the techniques discussed in \citet{Croll15} and references therein;
the aperture radii, and the radii of the inner and outer annuli we use to subtract the sky are given in Table \ref{TableObs}.
The filters used on the various telescopes were:
the DCT/LMI observations utilized an Andover V-band filter and a Kron Cousins R-band filter, 
the FLWO/KeplerCam observations utilized a Harris V-filter, 
the MINERVA observations utilized Johnson B, V \& R-band filters, 
while the the Perkins/PRISM observations utilized Johnson V \& R-band filters.
The MINERVA ``air''-band observations are simply those conducted without a filter, and therefore the wavelength dependence
is given by the quantum efficiency of the MINERVA CCDs \citep{Swift15};
the MINERVA CCDs are quantum efficient across a wide wavelength range -- from near-ultraviolet to near-infrared wavelengths --
with central wavelengths of: $\sim$0.64 ${\rm \mu m}$ for the MINERVA T1 \& T4 telescopes,
and $\sim$0.71 ${\rm \mu m}$ for the MINERVA T3 telescope\footnote{The MINERVA T2 telescope observed only in the B \& V-bands.}.
 
As our observations indicate that WD 1145+017 displays low level variability, likely due to dusty material passing consistently
in front of the star (see Section \ref{SecLLV}), to estimate the errors on our photometry we cannot take the root mean square of the 
differential photometry of WD 1145+017; instead we take the mean of the root mean square of the differential photometry of nearby
reference stars that have similar aperture flux values as WD 1145+017 (we take the mean of the root mean square of the differential photometry
of all stars, that are not obvious variables, with aperture flux values within 20\% of WD 1145+017). 
All the Julian dates have been converted to and presented as barycentric Julian dates using the terrestrial time standard \citep{Eastman10}.

We present our DCT, Perkins, FLWO and MINERVA observations in Figure \ref{FigLC}.
We observe occasional significant decrements in flux of up
to $\sim$30\% that we interpret as objects, likely with dusty tails streaming behind them,
passing in front of the white dwarf along our line of sight.
We present our 
multi-telescope, and often multiwavelength,
observations of these significant flux decrements in Figure \ref{FigLCMINERVA}.

\section{Analysis}
\label{SecAnalysis}

\begin{deluxetable*}{cccccccccc}
\tablecaption{Hyperbolic Secant Fits}
\tablehead{
\colhead{Date (UTC)}	& \colhead{Telescope}	& \colhead{Band}	& \colhead{$F_0$} 		& \colhead{$D$ (\%)}	& \colhead{$T_{\rm min}$}		& \colhead{$\tau_1$} 			& \colhead{$\tau_2$}	& \colhead{$\tau_2$/$\tau_1$}	& \colhead{\TransitDurationUse$\times$($\tau_2 + \tau_1$)}	\\
\colhead{\& Transit \#}	& \colhead{}		& \colhead{}		& \colhead{}	 		& \colhead{}		& \colhead{(JD-2457000)}	& \colhead{($\rm min$)} 			& \colhead{($\rm min$)}			& \colhead{}			& \colhead{($\rm min$)} \\
}
\centering
\startdata
\multicolumn{10}{c}{Joint Fits} \\
2015/05/09	& MINERVA/T1		& R			& \FZeroMayNineJoint$^{+\FZeroPlusMayNineJoint}_{-\FZeroMinusMayNineJoint}$									&	\TDAPercentMayNineJoint$^{+\TDAPercentPlusMayNineJoint}_{-\TDAPercentMinusMayNineJoint}$ 								& \multirow{3}{*}{\TminMayNineJoint$^{+\TminPlusMayNineJoint}_{-\TminMinusMayNineJoint}$}								& \multirow{3}{*}{\ThetaOneMayNineJoint$^{+\ThetaOnePlusMayNineJoint}_{-\ThetaOneMinusMayNineJoint}$}							& \multirow{3}{*}{\ThetaTwoMayNineJoint$^{+\ThetaTwoPlusMayNineJoint}_{-\ThetaTwoMinusMayNineJoint}$}							& \multirow{3}{*}{\ThetaTwoOverThetaOneMayNineJoint$^{+\ThetaTwoOverThetaOnePlusMayNineJoint}_{-\ThetaTwoOverThetaOneMinusMayNineJoint}$} 							& \multirow{3}{*}{\TransitDurationMayNineJoint$^{+\TransitDurationPlusMayNineJoint}_{-\TransitDurationMinusMayNineJoint}$}\\
2015/05/09	& MINERVA/T3		& Air			& \FZeroBMayNineJoint$^{+\FZeroBPlusMayNineJoint}_{-\FZeroBMinusMayNineJoint}$									&	\TDBPercentMayNineJoint$^{+\TDBPercentPlusMayNineJoint}_{-\TDBPercentMinusMayNineJoint}$ 								& \\
2015/05/09	& MINERVA/T4		& Air			& \FZeroCMayNineJoint$^{+\FZeroCPlusMayNineJoint}_{-\FZeroCMinusMayNineJoint}$									&	\TDCPercentMayNineJoint$^{+\TDCPercentPlusMayNineJoint}_{-\TDCPercentMinusMayNineJoint}$ 								& \\
																																																																																																																																											
2015/05/10	& MINERVA/T1		& R			& \FZeroMayTenJoint$^{+\FZeroPlusMayTenJoint}_{-\FZeroMinusMayTenJoint}$									&	\TDAPercentMayTenJoint$^{+\TDAPercentPlusMayTenJoint}_{-\TDAPercentMinusMayTenJoint}$ 									& \multirow{3}{*}{\TminMayTenJoint$^{+\TminPlusMayTenJoint}_{-\TminMinusMayTenJoint}$} 									& \multirow{3}{*}{\ThetaOneMayTenJoint$^{+\ThetaOnePlusMayTenJoint}_{-\ThetaOneMinusMayTenJoint}$}							& \multirow{3}{*}{\ThetaTwoMayTenJoint$^{+\ThetaTwoPlusMayTenJoint}_{-\ThetaTwoMinusMayTenJoint}$}							& \multirow{3}{*}{\ThetaTwoOverThetaOneMayTenJoint$^{+\ThetaTwoOverThetaOnePlusMayTenJoint}_{-\ThetaTwoOverThetaOneMinusMayTenJoint}$}								& \multirow{3}{*}{\TransitDurationMayTenJoint$^{+\TransitDurationPlusMayTenJoint}_{-\TransitDurationMinusMayTenJoint}$} \\
2015/05/10	& MINERVA/T2		& V			& \FZeroBMayTenJoint$^{+\FZeroBPlusMayTenJoint}_{-\FZeroBMinusMayTenJoint}$									&	\TDBPercentMayTenJoint$^{+\TDBPercentPlusMayTenJoint}_{-\TDBPercentMinusMayTenJoint}$ 									& \\
2015/05/10	& MINERVA/T3		& Air 			& \FZeroCMayTenJoint$^{+\FZeroCPlusMayTenJoint}_{-\FZeroCMinusMayTenJoint}$									&	\TDCPercentMayTenJoint$^{+\TDCPercentPlusMayTenJoint}_{-\TDCPercentMinusMayTenJoint}$ 									& \\
																																																																																																																																											
2015/05/11-A	& FLWO			& V			& \FZeroMayElevenSecondJoint$^{+\FZeroPlusMayElevenSecondJoint}_{-\FZeroMinusMayElevenSecondJoint}$						&	\TDAPercentMayElevenSecondJoint$^{+\TDAPercentPlusMayElevenSecondJoint}_{-\TDAPercentMinusMayElevenSecondJoint}$ 					& \multirow{3}{*}{\TminMayElevenSecondJoint$^{+\TminPlusMayElevenSecondJoint}_{-\TminMinusMayElevenSecondJoint}$}					& \multirow{3}{*}{\ThetaOneMayElevenSecondJoint$^{+\ThetaOnePlusMayElevenSecondJoint}_{-\ThetaOneMinusMayElevenSecondJoint}$}				& \multirow{3}{*}{\ThetaTwoMayElevenSecondJoint$^{+\ThetaTwoPlusMayElevenSecondJoint}_{-\ThetaTwoMinusMayElevenSecondJoint}$}				& \multirow{3}{*}{\ThetaTwoOverThetaOneMayElevenSecondJoint$^{+\ThetaTwoOverThetaOnePlusMayElevenSecondJoint}_{-\ThetaTwoOverThetaOneMinusMayElevenSecondJoint}$} 				& \multirow{3}{*}{\TransitDurationMayElevenSecondJoint$^{+\TransitDurationPlusMayElevenSecondJoint}_{-\TransitDurationMinusMayElevenSecondJoint}$}\\
2015/05/11-A	& MINERVA/T3		& Air			& \FZeroBMayElevenSecondJoint$^{+\FZeroBPlusMayElevenSecondJoint}_{-\FZeroBMinusMayElevenSecondJoint}$						&	\TDBPercentMayElevenSecondJoint$^{+\TDBPercentPlusMayElevenSecondJoint}_{-\TDBPercentMinusMayElevenSecondJoint}$					& \\
2015/05/11-A	& MINERVA/T4		& Air			& \FZeroCMayElevenSecondJoint$^{+\FZeroCPlusMayElevenSecondJoint}_{-\FZeroCMinusMayElevenSecondJoint}$						&	\TDCPercentMayElevenSecondJoint$^{+\TDCPercentPlusMayElevenSecondJoint}_{-\TDCPercentMinusMayElevenSecondJoint}$					& \\
																																																																																																																																											
																																																																																																																																											
2015/05/11-B	& FLWO			& V			& \FZeroMayElevenJoint$^{+\FZeroPlusMayElevenJoint}_{-\FZeroMinusMayElevenJoint}$								&	\TDAPercentMayElevenJoint$^{+\TDAPercentPlusMayElevenJoint}_{-\TDAPercentMinusMayElevenJoint}$ 								& \multirow{3}{*}{\TminMayElevenJoint$^{+\TminPlusMayElevenJoint}_{-\TminMinusMayElevenJoint}$}								& \multirow{3}{*}{\ThetaOneMayElevenJoint$^{+\ThetaOnePlusMayElevenJoint}_{-\ThetaOneMinusMayElevenJoint}$}						& \multirow{3}{*}{\ThetaTwoMayElevenJoint$^{+\ThetaTwoPlusMayElevenJoint}_{-\ThetaTwoMinusMayElevenJoint}$}						& \multirow{3}{*}{\ThetaTwoOverThetaOneMayElevenJoint$^{+\ThetaTwoOverThetaOnePlusMayElevenJoint}_{-\ThetaTwoOverThetaOneMinusMayElevenJoint}$} 						& \multirow{3}{*}{\TransitDurationMayElevenJoint$^{+\TransitDurationPlusMayElevenJoint}_{-\TransitDurationMinusMayElevenJoint}$}\\
2015/05/11-B	& MINERVA/T2		& V			& \FZeroBMayElevenJoint$^{+\FZeroBPlusMayElevenJoint}_{-\FZeroBMinusMayElevenJoint}$								&	\TDBPercentMayElevenJoint$^{+\TDBPercentPlusMayElevenJoint}_{-\TDBPercentMinusMayElevenJoint}$ 								& \\
2015/05/11-B	& MINERVA/T3 \& T4	& Air			& \FZeroCMayElevenJoint$^{+\FZeroCPlusMayElevenJoint}_{-\FZeroCMinusMayElevenJoint}$								&	\TDCPercentMayElevenJoint$^{+\TDCPercentPlusMayElevenJoint}_{-\TDCPercentMinusMayElevenJoint}$ 								& \\
																																																																																																																																											
2015/05/11-C	& DCT			& V			& \FZeroMayElevenThirdJoint$^{+\FZeroPlusMayElevenThirdJoint}_{-\FZeroMinusMayElevenThirdJoint}$						&	\TDAPercentMayElevenThirdJoint$^{+\TDAPercentPlusMayElevenThirdJoint}_{-\TDAPercentMinusMayElevenThirdJoint}$						& \multirow{3}{*}{\TminMayElevenThirdJoint$^{+\TminPlusMayElevenThirdJoint}_{-\TminMinusMayElevenThirdJoint}$}						& \multirow{3}{*}{\ThetaOneMayElevenThirdJoint$^{+\ThetaOnePlusMayElevenThirdJoint}_{-\ThetaOneMinusMayElevenThirdJoint}$}				& \multirow{3}{*}{\ThetaTwoMayElevenThirdJoint$^{+\ThetaTwoPlusMayElevenThirdJoint}_{-\ThetaTwoMinusMayElevenThirdJoint}$}				& \multirow{3}{*}{\ThetaTwoOverThetaOneMayElevenThirdJoint$^{+\ThetaTwoOverThetaOnePlusMayElevenThirdJoint}_{-\ThetaTwoOverThetaOneMinusMayElevenThirdJoint}$} 					& \multirow{3}{*}{\TransitDurationMayElevenThirdJoint$^{+\TransitDurationPlusMayElevenThirdJoint}_{-\TransitDurationMinusMayElevenThirdJoint}$} \\
2015/05/11-C	& FLWO			& V			& \FZeroBMayElevenThirdJoint$^{+\FZeroBPlusMayElevenThirdJoint}_{-\FZeroBMinusMayElevenThirdJoint}$						&	\TDBPercentMayElevenThirdJoint$^{+\TDBPercentPlusMayElevenThirdJoint}_{-\TDBPercentMinusMayElevenThirdJoint}$						& \\
2015/05/11-C	& MINERVA/T3 \& T4	& Air			& \FZeroCMayElevenThirdJoint$^{+\FZeroCPlusMayElevenThirdJoint}_{-\FZeroCMinusMayElevenThirdJoint}$						&	\TDCPercentMayElevenThirdJoint$^{+\TDCPercentPlusMayElevenThirdJoint}_{-\TDCPercentMinusMayElevenThirdJoint}$ 						& \\
																																																																																																																																											
2015/05/12-A	& Perkins		& R			& \FZeroMayTwelveJoint$^{+\FZeroPlusMayTwelveJoint}_{-\FZeroMinusMayTwelveJoint}$								&	\TDAPercentMayTwelveJoint$^{+\TDAPercentPlusMayTwelveJoint}_{-\TDAPercentMinusMayTwelveJoint}$ 								& \multirow{3}{*}{\TminMayTwelveJoint$^{+\TminPlusMayTwelveJoint}_{-\TminMinusMayTwelveJoint}$}								& \multirow{3}{*}{\ThetaOneMayTwelveJoint$^{+\ThetaOnePlusMayTwelveJoint}_{-\ThetaOneMinusMayTwelveJoint}$}						& \multirow{3}{*}{\ThetaTwoMayTwelveJoint$^{+\ThetaTwoPlusMayTwelveJoint}_{-\ThetaTwoMinusMayTwelveJoint}$}						& \multirow{3}{*}{\ThetaTwoOverThetaOneMayTwelveJoint$^{+\ThetaTwoOverThetaOnePlusMayTwelveJoint}_{-\ThetaTwoOverThetaOneMinusMayTwelveJoint}$} 						& \multirow{3}{*}{\TransitDurationMayTwelveJoint$^{+\TransitDurationPlusMayTwelveJoint}_{-\TransitDurationMinusMayTwelveJoint}$} \\
2015/05/12-A	& MINERVA/T1 \& T2	& V			& \FZeroBMayTwelveJoint$^{+\FZeroBPlusMayTwelveJoint}_{-\FZeroBMinusMayTwelveJoint}$								&	\TDBPercentMayTwelveJoint$^{+\TDBPercentPlusMayTwelveJoint}_{-\TDBPercentMinusMayTwelveJoint}$ 								& \\
2015/05/12-A	& MINERVA/T3 \& T4	& Air			& \FZeroCMayTwelveJoint$^{+\FZeroCPlusMayTwelveJoint}_{-\FZeroCMinusMayTwelveJoint}$								&	\TDCPercentMayTwelveJoint$^{+\TDCPercentPlusMayTwelveJoint}_{-\TDCPercentMinusMayTwelveJoint}$ 								& \\
																																																																																																																																											
2015/05/12-B	& Perkins		& R			& \FZeroMayTwelveSecondJoint$^{+\FZeroPlusMayTwelveSecondJoint}_{-\FZeroMinusMayTwelveSecondJoint}$						&	\TDAPercentMayTwelveSecondJoint$^{+\TDAPercentPlusMayTwelveSecondJoint}_{-\TDAPercentMinusMayTwelveSecondJoint}$					& \multirow{3}{*}{\TminMayTwelveSecondJoint$^{+\TminPlusMayTwelveSecondJoint}_{-\TminMinusMayTwelveSecondJoint}$}					& \multirow{3}{*}{\ThetaOneMayTwelveSecondJoint$^{+\ThetaOnePlusMayTwelveSecondJoint}_{-\ThetaOneMinusMayTwelveSecondJoint}$}				& \multirow{3}{*}{\ThetaTwoMayTwelveSecondJoint$^{+\ThetaTwoPlusMayTwelveSecondJoint}_{-\ThetaTwoMinusMayTwelveSecondJoint}$}				& \multirow{3}{*}{\ThetaTwoOverThetaOneMayTwelveSecondJoint$^{+\ThetaTwoOverThetaOnePlusMayTwelveSecondJoint}_{-\ThetaTwoOverThetaOneMinusMayTwelveSecondJoint}$}				& \multirow{3}{*}{\TransitDurationMayTwelveSecondJoint$^{+\TransitDurationPlusMayTwelveSecondJoint}_{-\TransitDurationMinusMayTwelveSecondJoint}$}\\
2015/05/12-B	& MINERVA/T1 \& T2	& V			& \FZeroBMayTwelveSecondJoint$^{+\FZeroBPlusMayTwelveSecondJoint}_{-\FZeroBMinusMayTwelveSecondJoint}$						&	\TDBPercentMayTwelveSecondJoint$^{+\TDBPercentPlusMayTwelveSecondJoint}_{-\TDBPercentMinusMayTwelveSecondJoint}$					& \\
2015/05/12-B	& MINERVA/T3 \& T4	& Air			& \FZeroCMayTwelveSecondJoint$^{+\FZeroCPlusMayTwelveSecondJoint}_{-\FZeroCMinusMayTwelveSecondJoint}$						&	\TDCPercentMayTwelveSecondJoint$^{+\TDCPercentPlusMayTwelveSecondJoint}_{-\TDCPercentMinusMayTwelveSecondJoint}$					& \\
																																																																																																																																											
2015/05/13-A	& MINERVA/T1		& R			& \FZeroMayThirteenJoint$^{+\FZeroPlusMayThirteenJoint}_{-\FZeroMinusMayThirteenJoint}$								&	\TDAPercentMayThirteenJoint$^{+\TDAPercentPlusMayThirteenJoint}_{-\TDAPercentMinusMayThirteenJoint}$ 							& \multirow{3}{*}{\TminMayThirteenJoint$^{+\TminPlusMayThirteenJoint}_{-\TminMinusMayThirteenJoint}$}							& \multirow{3}{*}{\ThetaOneMayThirteenJoint$^{+\ThetaOnePlusMayThirteenJoint}_{-\ThetaOneMinusMayThirteenJoint}$}					& \multirow{3}{*}{\ThetaTwoMayThirteenJoint$^{+\ThetaTwoPlusMayThirteenJoint}_{-\ThetaTwoMinusMayThirteenJoint}$}					& \multirow{3}{*}{\ThetaTwoOverThetaOneMayThirteenJoint$^{+\ThetaTwoOverThetaOnePlusMayThirteenJoint}_{-\ThetaTwoOverThetaOneMinusMayThirteenJoint}$}						& \multirow{3}{*}{\TransitDurationMayThirteenJoint$^{+\TransitDurationPlusMayThirteenJoint}_{-\TransitDurationMinusMayThirteenJoint}$} \\
2015/05/13-A	& Perkins		& V			& \FZeroBMayThirteenJoint$^{+\FZeroBPlusMayThirteenJoint}_{-\FZeroBMinusMayThirteenJoint}$							&	\TDBPercentMayThirteenJoint$^{+\TDBPercentPlusMayThirteenJoint}_{-\TDBPercentMinusMayThirteenJoint}$ 							& \\
2015/05/13-A	& MINERVA/T4		& Air			& \FZeroCMayThirteenJoint$^{+\FZeroCPlusMayThirteenJoint}_{-\FZeroCMinusMayThirteenJoint}$							&	\TDCPercentMayThirteenJoint$^{+\TDCPercentPlusMayThirteenJoint}_{-\TDCPercentMinusMayThirteenJoint}$ 							& \\
\hline
\multicolumn{10}{c}{Individual Fits} \\
2015/05/13-B	& Perkins		& V			& \FZeroPerkinsMayThirteenVBandSecond$^{+\FZeroPlusPerkinsMayThirteenVBandSecond}_{-\FZeroMinusPerkinsMayThirteenVBandSecond}$			&	\TDPercentPerkinsMayThirteenVBandSecond$^{+\TDPercentPlusPerkinsMayThirteenVBandSecond}_{-\TDPercentMinusPerkinsMayThirteenVBandSecond}$ 		&	\TminPerkinsMayThirteenVBandSecond$^{+\TminPlusPerkinsMayThirteenVBandSecond}_{-\TminMinusPerkinsMayThirteenVBandSecond}$			& \ThetaOnePerkinsMayThirteenVBandSecond$^{+\ThetaOnePlusPerkinsMayThirteenVBandSecond}_{-\ThetaOneMinusPerkinsMayThirteenVBandSecond}$			& \ThetaTwoPerkinsMayThirteenVBandSecond$^{+\ThetaTwoPlusPerkinsMayThirteenVBandSecond}_{-\ThetaTwoMinusPerkinsMayThirteenVBandSecond}$			& \ThetaTwoOverThetaOnePerkinsMayThirteenVBandSecond$^{+\ThetaTwoOverThetaOnePlusPerkinsMayThirteenVBandSecond}_{-\ThetaTwoOverThetaOneMinusPerkinsMayThirteenVBandSecond}$ 			& \TransitDurationPerkinsMayThirteenVBandSecond$^{+\TransitDurationPlusPerkinsMayThirteenVBandSecond}_{-\TransitDurationMinusPerkinsMayThirteenVBandSecond}$\\

2015/05/18	& MINERVA/T4		& Air			& \FZeroMINERVAMayEighteenAirBandTFour$^{+\FZeroPlusMINERVAMayEighteenAirBandTFour}_{-\FZeroMinusMINERVAMayEighteenAirBandTFour}$		&	\TDPercentMINERVAMayEighteenAirBandTFour$^{+\TDPercentPlusMINERVAMayEighteenAirBandTFour}_{-\TDPercentMinusMINERVAMayEighteenAirBandTFour}$ 		&	\TminMINERVAMayEighteenAirBandTFour$^{+\TminPlusMINERVAMayEighteenAirBandTFour}_{-\TminMinusMINERVAMayEighteenAirBandTFour}$			& \ThetaOneMINERVAMayEighteenAirBandTFour$^{+\ThetaOnePlusMINERVAMayEighteenAirBandTFour}_{-\ThetaOneMinusMINERVAMayEighteenAirBandTFour}$		& \ThetaTwoMINERVAMayEighteenAirBandTFour$^{+\ThetaTwoPlusMINERVAMayEighteenAirBandTFour}_{-\ThetaTwoMinusMINERVAMayEighteenAirBandTFour}$		& \ThetaTwoOverThetaOneMINERVAMayEighteenAirBandTFour$^{+\ThetaTwoOverThetaOnePlusMINERVAMayEighteenAirBandTFour}_{-\ThetaTwoOverThetaOneMinusMINERVAMayEighteenAirBandTFour}$			& \TransitDurationMINERVAMayEighteenAirBandTFour$^{+\TransitDurationPlusMINERVAMayEighteenAirBandTFour}_{-\TransitDurationMinusMINERVAMayEighteenAirBandTFour}$ \\
\hline
\multicolumn{10}{c}{Individual Fits to ground-based eclipses presented by \citet{Vanderburg15}} \\
2015/03/22	& FLWO			& V			& \FZeroFLWOMarchTwentyTwoVBand$^{+\FZeroPlusFLWOMarchTwentyTwoVBand}_{-\FZeroMinusFLWOMarchTwentyTwoVBand}$					&	\TDPercentFLWOMarchTwentyTwoVBand$^{+\TDPercentPlusFLWOMarchTwentyTwoVBand}_{-\TDPercentMinusFLWOMarchTwentyTwoVBand}$ 					&	\TminFLWOMarchTwentyTwoVBand$^{+\TminPlusFLWOMarchTwentyTwoVBand}_{-\TminMinusFLWOMarchTwentyTwoVBand}$						& \ThetaOneFLWOMarchTwentyTwoVBand$^{+\ThetaOnePlusFLWOMarchTwentyTwoVBand}_{-\ThetaOneMinusFLWOMarchTwentyTwoVBand}$					& \ThetaTwoFLWOMarchTwentyTwoVBand$^{+\ThetaTwoPlusFLWOMarchTwentyTwoVBand}_{-\ThetaTwoMinusFLWOMarchTwentyTwoVBand}$					& \ThetaTwoOverThetaOneFLWOMarchTwentyTwoVBand$^{+\ThetaTwoOverThetaOnePlusFLWOMarchTwentyTwoVBand}_{-\ThetaTwoOverThetaOneMinusFLWOMarchTwentyTwoVBand}$ 					& \TransitDurationFLWOMarchTwentyTwoVBand$^{+\TransitDurationPlusFLWOMarchTwentyTwoVBand}_{-\TransitDurationMinusFLWOMarchTwentyTwoVBand}$\\

2015/04/11-A	& FLWO			& V			& \FZeroFLWOAprilElevenVBand$^{+\FZeroPlusFLWOAprilElevenVBand}_{-\FZeroMinusFLWOAprilElevenVBand}$						&	\TDPercentFLWOAprilElevenVBand$^{+\TDPercentPlusFLWOAprilElevenVBand}_{-\TDPercentMinusFLWOAprilElevenVBand}$ 						&	\TminFLWOAprilElevenVBand$^{+\TminPlusFLWOAprilElevenVBand}_{-\TminMinusFLWOAprilElevenVBand}$							& \ThetaOneFLWOAprilElevenVBand$^{+\ThetaOnePlusFLWOAprilElevenVBand}_{-\ThetaOneMinusFLWOAprilElevenVBand}$						& \ThetaTwoFLWOAprilElevenVBand$^{+\ThetaTwoPlusFLWOAprilElevenVBand}_{-\ThetaTwoMinusFLWOAprilElevenVBand}$						& \ThetaTwoOverThetaOneFLWOAprilElevenVBand$^{+\ThetaTwoOverThetaOnePlusFLWOAprilElevenVBand}_{-\ThetaTwoOverThetaOneMinusFLWOAprilElevenVBand}$ 						& \TransitDurationFLWOAprilElevenVBand$^{+\TransitDurationPlusFLWOAprilElevenVBand}_{-\TransitDurationMinusFLWOAprilElevenVBand}$\\
																																																																																																																																											
2015/04/11-B	& FLWO			& V			& \FZeroFLWOAprilElevenVBandSecond$^{+\FZeroPlusFLWOAprilElevenVBandSecond}_{-\FZeroMinusFLWOAprilElevenVBandSecond}$				&	\TDPercentFLWOAprilElevenVBandSecond$^{+\TDPercentPlusFLWOAprilElevenVBandSecond}_{-\TDPercentMinusFLWOAprilElevenVBandSecond}$ 			&	\TminFLWOAprilElevenVBandSecond$^{+\TminPlusFLWOAprilElevenVBandSecond}_{-\TminMinusFLWOAprilElevenVBandSecond}$				& \ThetaOneFLWOAprilElevenVBandSecond$^{+\ThetaOnePlusFLWOAprilElevenVBandSecond}_{-\ThetaOneMinusFLWOAprilElevenVBandSecond}$				& \ThetaTwoFLWOAprilElevenVBandSecond$^{+\ThetaTwoPlusFLWOAprilElevenVBandSecond}_{-\ThetaTwoMinusFLWOAprilElevenVBandSecond}$				& \ThetaTwoOverThetaOneFLWOAprilElevenVBandSecond$^{+\ThetaTwoOverThetaOnePlusFLWOAprilElevenVBandSecond}_{-\ThetaTwoOverThetaOneMinusFLWOAprilElevenVBandSecond}$ 				& \TransitDurationFLWOAprilElevenVBandSecond$^{+\TransitDurationPlusFLWOAprilElevenVBandSecond}_{-\TransitDurationMinusFLWOAprilElevenVBandSecond}$\\
2015/04/17-A	& MEarth		& MEarth		& \FZeroMEarthAprilSixteenMEarthBand$^{+\FZeroPlusMEarthAprilSixteenMEarthBand}_{-\FZeroMinusMEarthAprilSixteenMEarthBand}$			&	\TDPercentMEarthAprilSixteenMEarthBand$^{+\TDPercentPlusMEarthAprilSixteenMEarthBand}_{-\TDPercentMinusMEarthAprilSixteenMEarthBand}$ 			&	\TminMEarthAprilSixteenMEarthBand$^{+\TminPlusMEarthAprilSixteenMEarthBand}_{-\TminMinusMEarthAprilSixteenMEarthBand}$				& \ThetaOneMEarthAprilSixteenMEarthBand$^{+\ThetaOnePlusMEarthAprilSixteenMEarthBand}_{-\ThetaOneMinusMEarthAprilSixteenMEarthBand}$			& \ThetaTwoMEarthAprilSixteenMEarthBand$^{+\ThetaTwoPlusMEarthAprilSixteenMEarthBand}_{-\ThetaTwoMinusMEarthAprilSixteenMEarthBand}$			& \ThetaTwoOverThetaOneMEarthAprilSixteenMEarthBand$^{+\ThetaTwoOverThetaOnePlusMEarthAprilSixteenMEarthBand}_{-\ThetaTwoOverThetaOneMinusMEarthAprilSixteenMEarthBand}$ 			& \TransitDurationMEarthAprilSixteenMEarthBand$^{+\TransitDurationPlusMEarthAprilSixteenMEarthBand}_{-\TransitDurationMinusMEarthAprilSixteenMEarthBand}$\\
																																																																																																																																											
2015/04/17-B	& MEarth		& MEarth		& \FZeroMEarthAprilSixteenMEarthBandSecond$^{+\FZeroPlusMEarthAprilSixteenMEarthBandSecond}_{-\FZeroMinusMEarthAprilSixteenMEarthBandSecond}$	&	\TDPercentMEarthAprilSixteenMEarthBandSecond$^{+\TDPercentPlusMEarthAprilSixteenMEarthBandSecond}_{-\TDPercentMinusMEarthAprilSixteenMEarthBandSecond}$ &	\TminMEarthAprilSixteenMEarthBandSecond$^{+\TminPlusMEarthAprilSixteenMEarthBandSecond}_{-\TminMinusMEarthAprilSixteenMEarthBandSecond}$	& \ThetaOneMEarthAprilSixteenMEarthBandSecond$^{+\ThetaOnePlusMEarthAprilSixteenMEarthBandSecond}_{-\ThetaOneMinusMEarthAprilSixteenMEarthBandSecond}$	& \ThetaTwoMEarthAprilSixteenMEarthBandSecond$^{+\ThetaTwoPlusMEarthAprilSixteenMEarthBandSecond}_{-\ThetaTwoMinusMEarthAprilSixteenMEarthBandSecond}$	& \ThetaTwoOverThetaOneMEarthAprilSixteenMEarthBandSecond$^{+\ThetaTwoOverThetaOnePlusMEarthAprilSixteenMEarthBandSecond}_{-\ThetaTwoOverThetaOneMinusMEarthAprilSixteenMEarthBandSecond}$ 	& \TransitDurationMEarthAprilSixteenMEarthBandSecond$^{+\TransitDurationPlusMEarthAprilSixteenMEarthBandSecond}_{-\TransitDurationMinusMEarthAprilSixteenMEarthBandSecond}$\\
\hline
\multicolumn{10}{c}{Joint Fits to ground-based eclipses presented by \citet{Vanderburg15}} \\
2015/04/18	& MEarth		& MEarth		& \FZeroAprilEighteenJoint$^{+\FZeroPlusAprilEighteenJoint}_{-\FZeroMinusAprilEighteenJoint}$								&	\TDAPercentAprilEighteenJoint$^{+\TDAPercentPlusAprilEighteenJoint}_{-\TDAPercentMinusAprilEighteenJoint}$ 							& \multirow{2}{*}{\TminAprilEighteenJoint$^{+\TminPlusAprilEighteenJoint}_{-\TminMinusAprilEighteenJoint}$}							& \multirow{3}{*}{\ThetaOneAprilEighteenJoint$^{+\ThetaOnePlusAprilEighteenJoint}_{-\ThetaOneMinusAprilEighteenJoint}$}					& \multirow{3}{*}{\ThetaTwoAprilEighteenJoint$^{+\ThetaTwoPlusAprilEighteenJoint}_{-\ThetaTwoMinusAprilEighteenJoint}$}					& \multirow{3}{*}{\ThetaTwoOverThetaOneAprilEighteenJoint$^{+\ThetaTwoOverThetaOnePlusAprilEighteenJoint}_{-\ThetaTwoOverThetaOneMinusAprilEighteenJoint}$}						& \multirow{3}{*}{\TransitDurationAprilEighteenJoint$^{+\TransitDurationPlusAprilEighteenJoint}_{-\TransitDurationMinusAprilEighteenJoint}$} \\
2015/04/18	& MINERVA/T3		& Air			& \FZeroBAprilEighteenJoint$^{+\FZeroBPlusAprilEighteenJoint}_{-\FZeroBMinusAprilEighteenJoint}$							&	\TDBPercentAprilEighteenJoint$^{+\TDBPercentPlusAprilEighteenJoint}_{-\TDBPercentMinusAprilEighteenJoint}$ 							& \\

\enddata
\label{TableHSFitsJoint}
\end{deluxetable*}

\begin{figure}
 \centering

\includegraphics[scale=0.5, angle = 270]{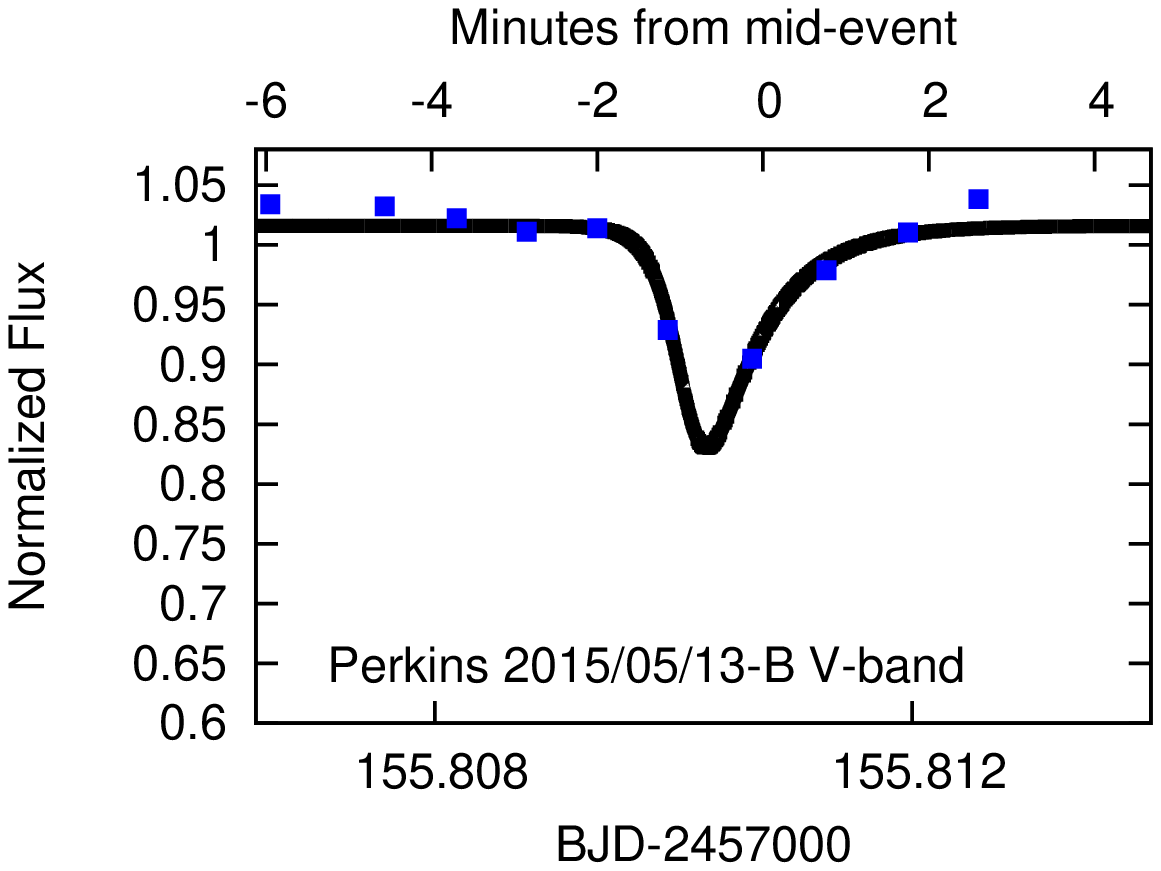} %

\includegraphics[scale=0.5, angle = 270]{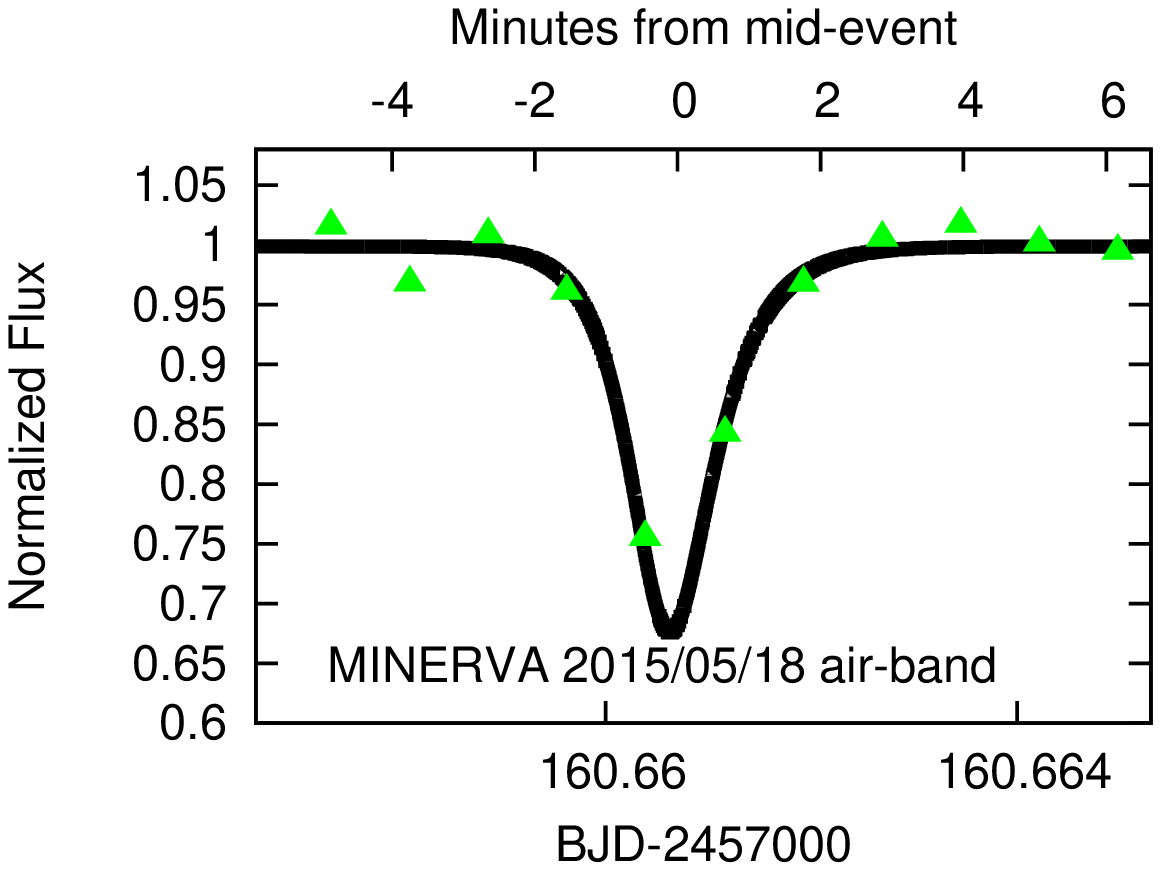} %
\caption[]
	{	Asymmetric hyperbolic secant function fits to various significant flux drops of WD 1145+017,
		observed with a single telescope (from top to bottom, the Perkins and MINERVA telescopes, respectively).
	}
\label{FigHSFitsIndividual}
\end{figure}

\begin{figure*}
\centering
\includegraphics[scale=0.33, angle = 270]{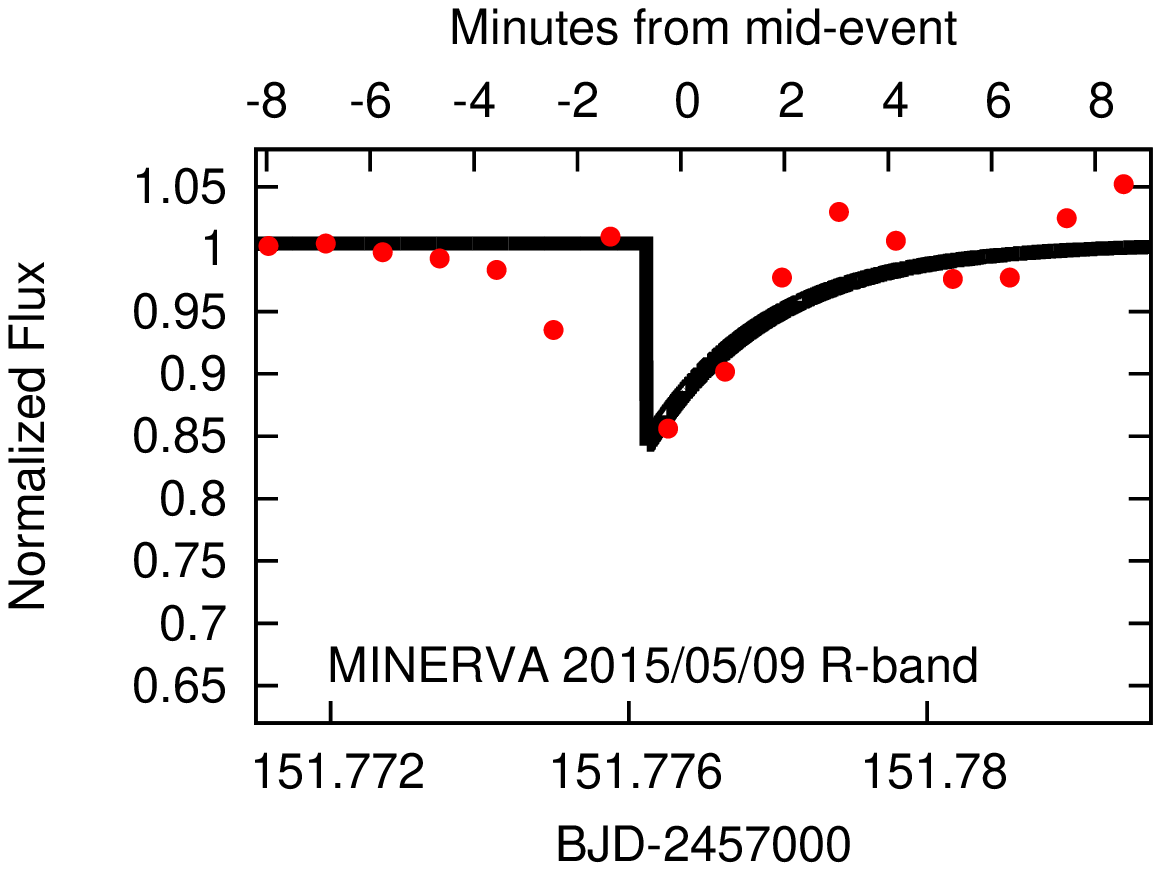} %
\includegraphics[scale=0.33, angle = 270]{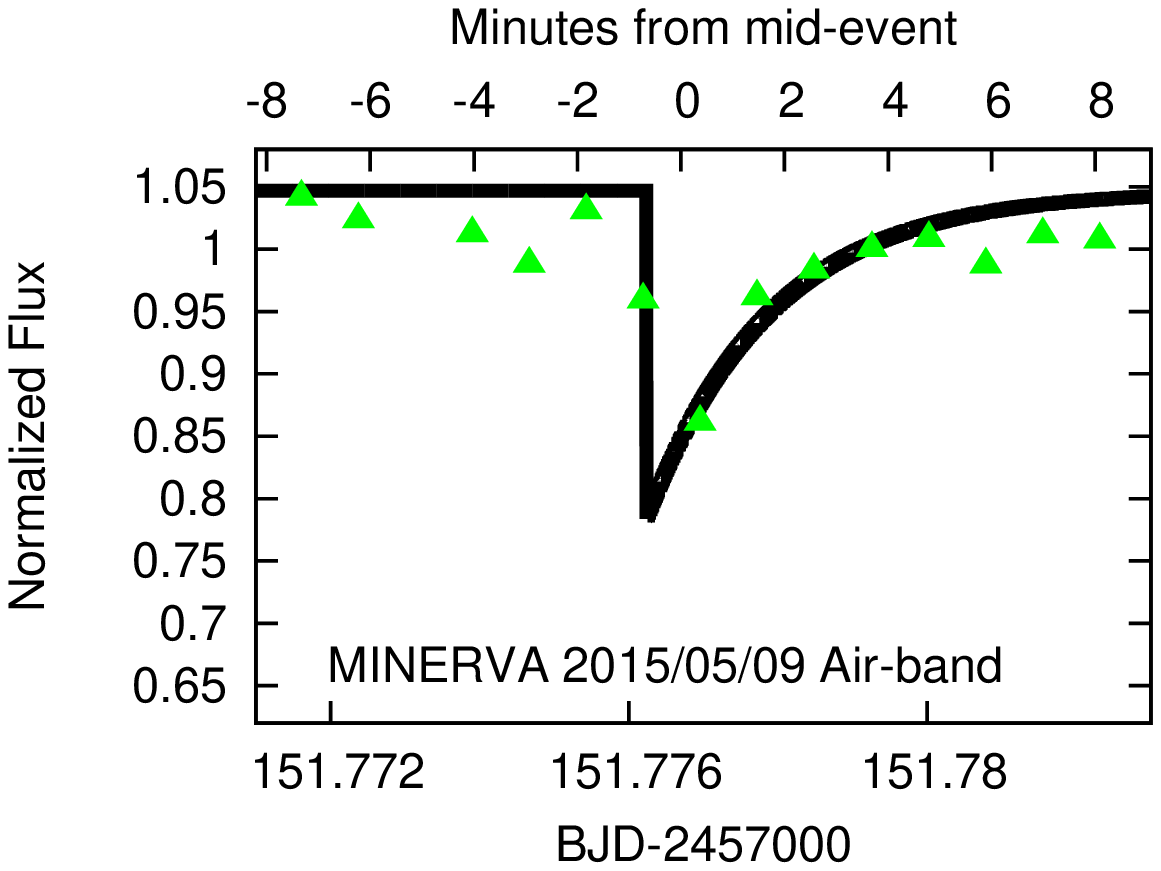} %
\includegraphics[scale=0.33, angle = 270]{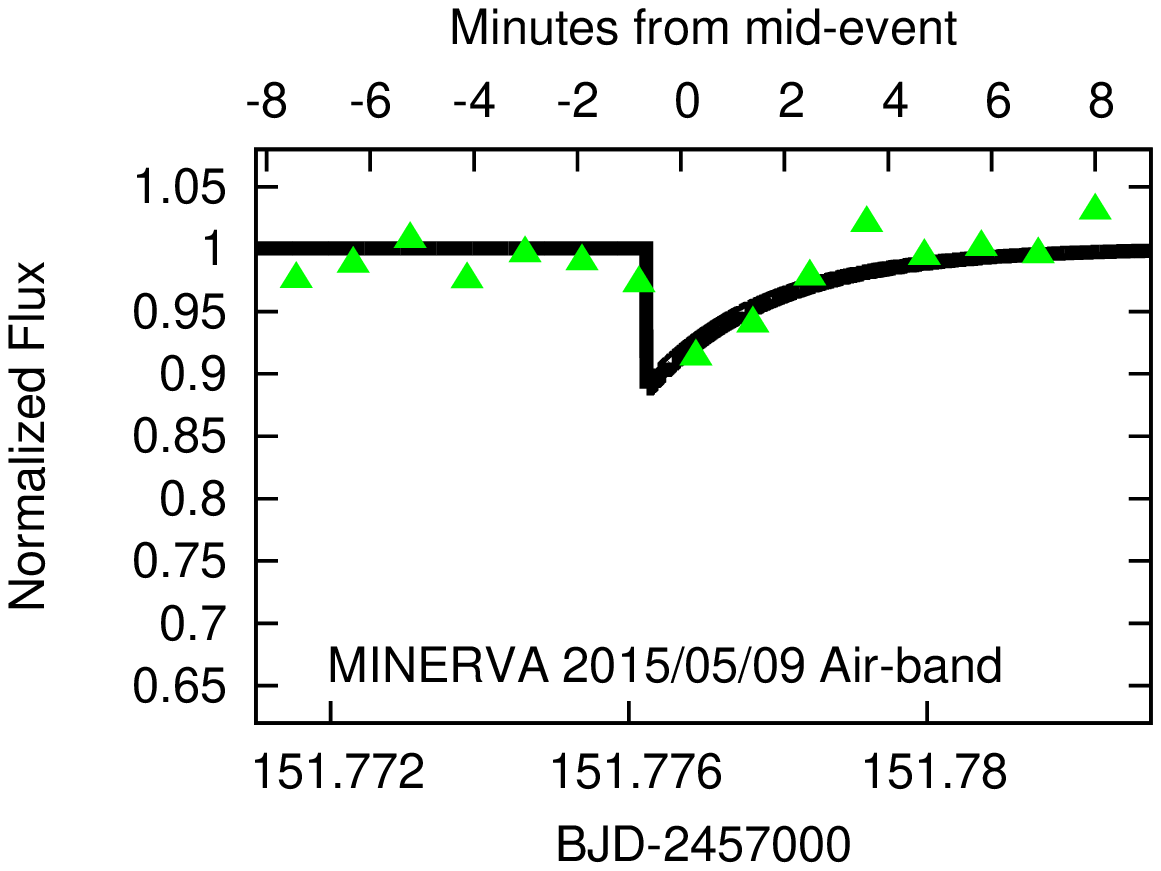} %

\includegraphics[scale=0.33, angle = 270]{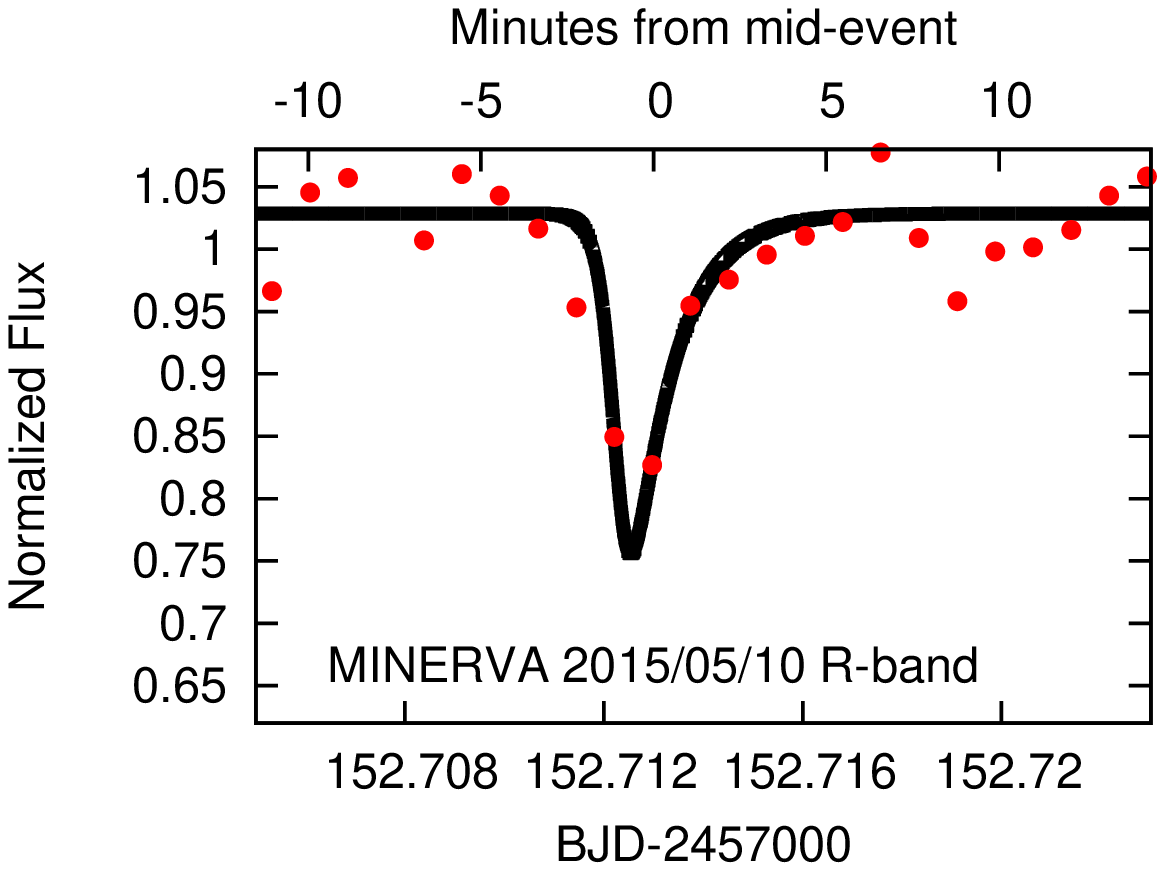} %
\includegraphics[scale=0.33, angle = 270]{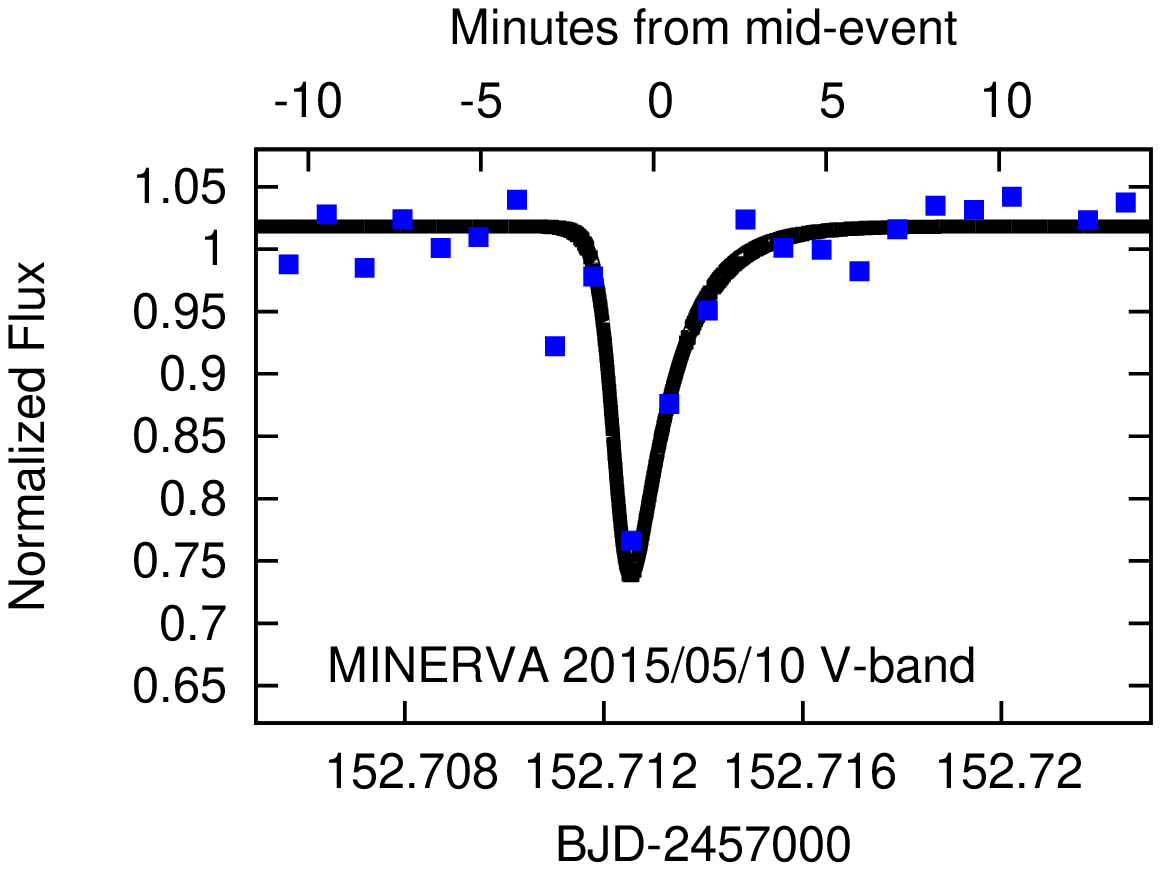} %
\includegraphics[scale=0.33, angle = 270]{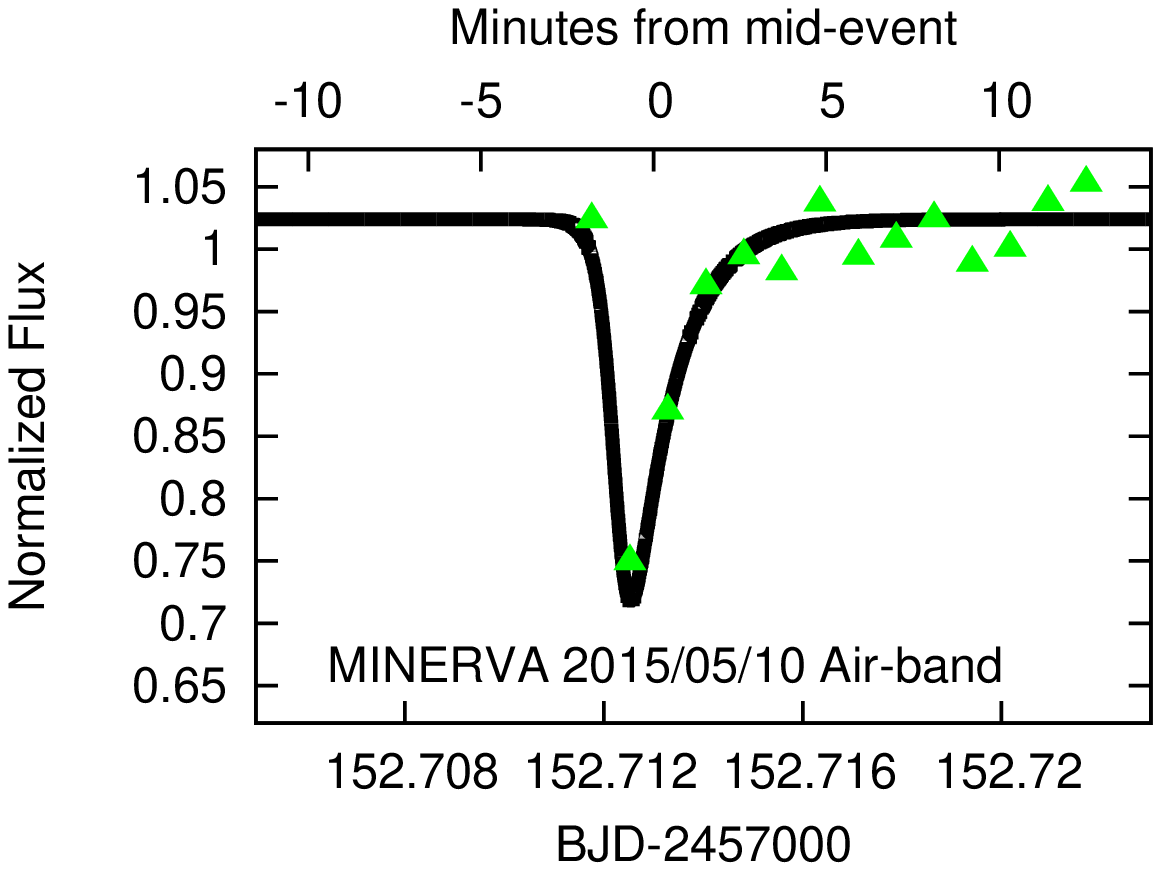} %

\includegraphics[scale=0.33, angle = 270]{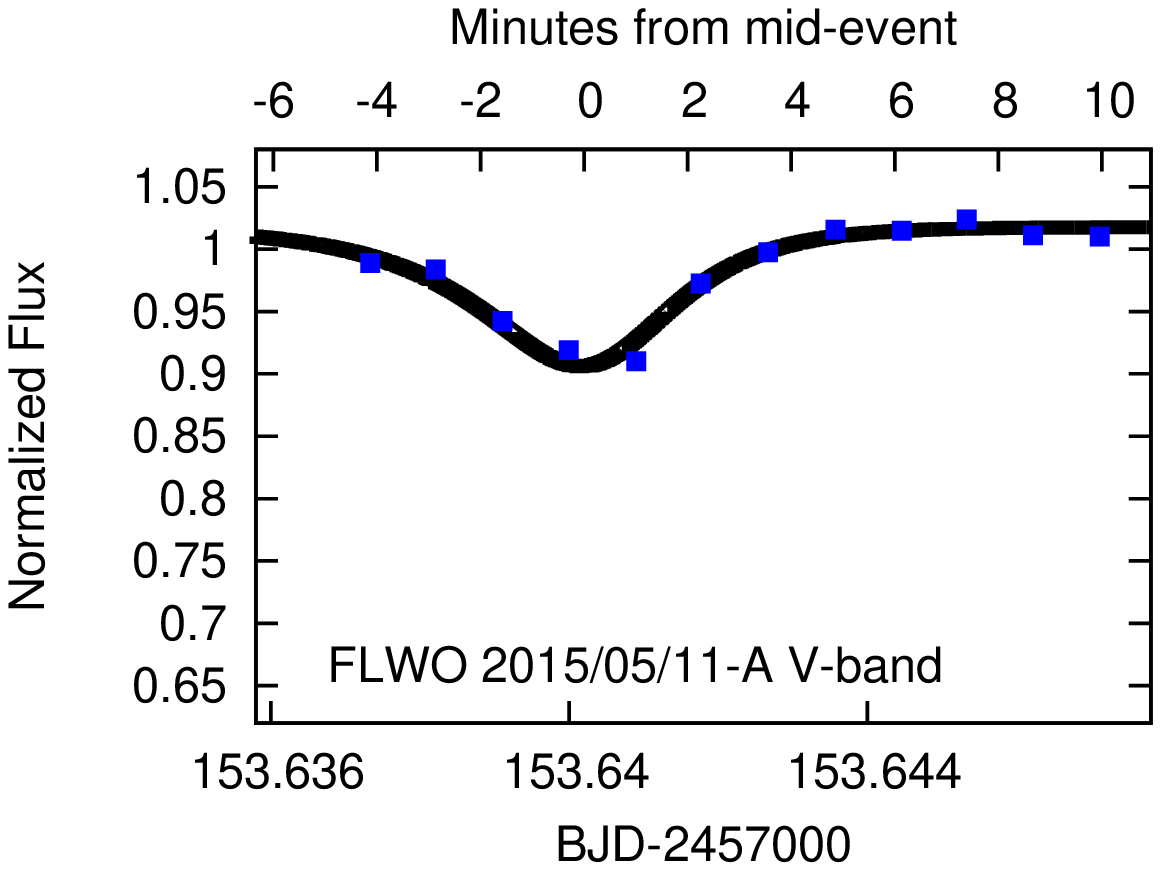} %
\includegraphics[scale=0.33, angle = 270]{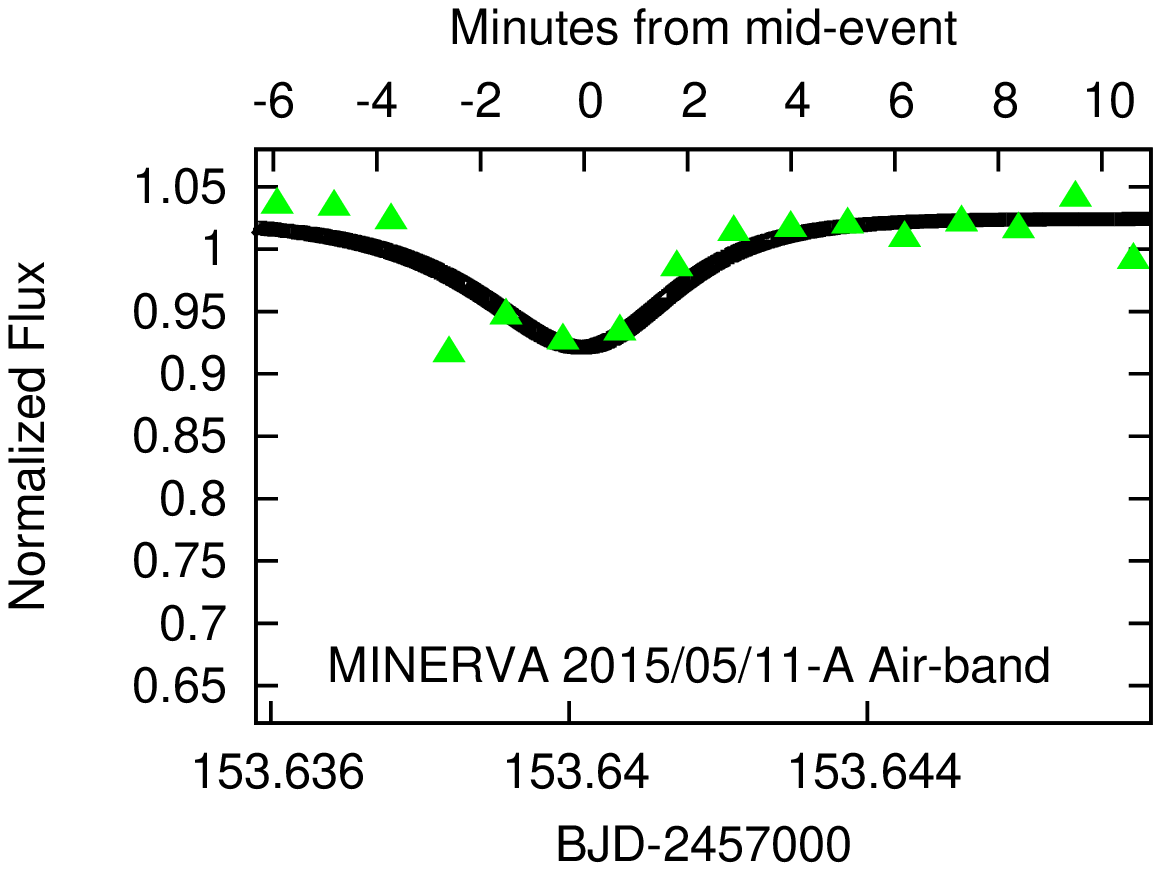} %
\includegraphics[scale=0.33, angle = 270]{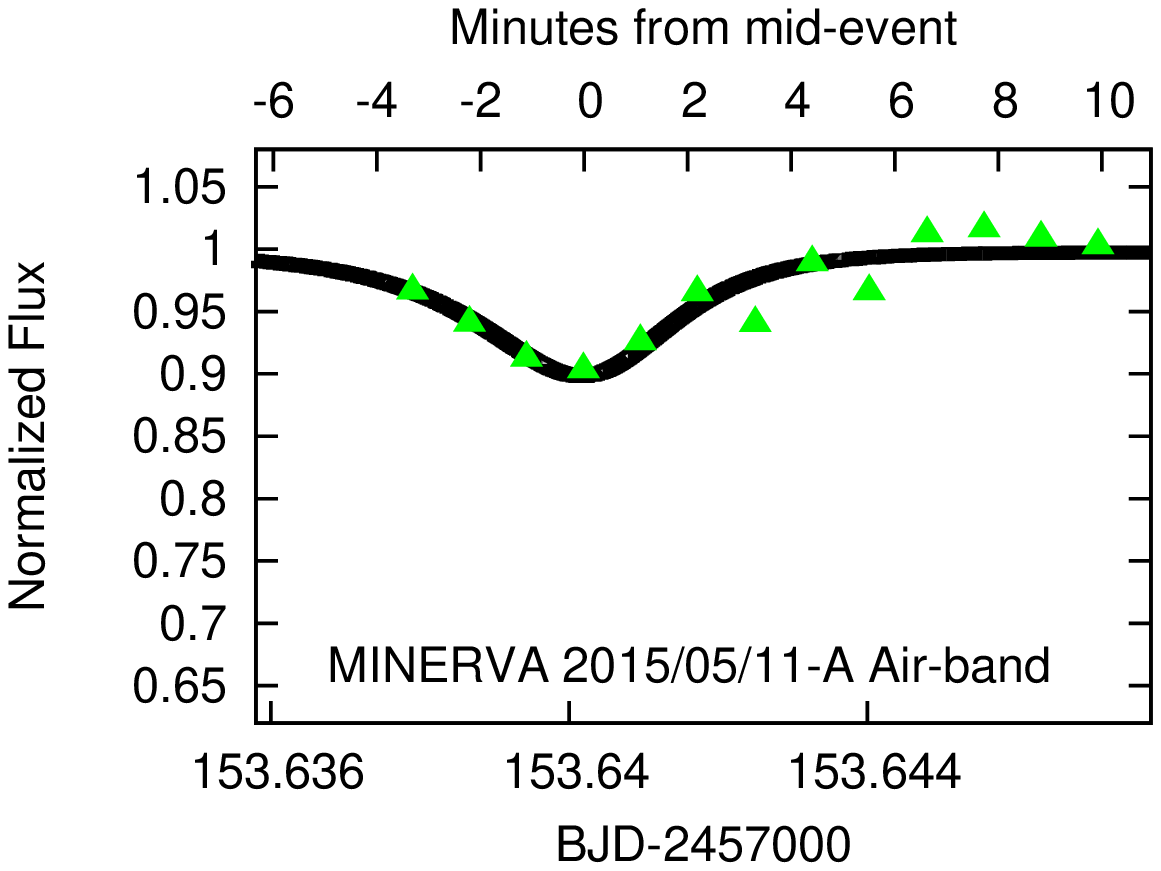} %

\includegraphics[scale=0.33, angle = 270]{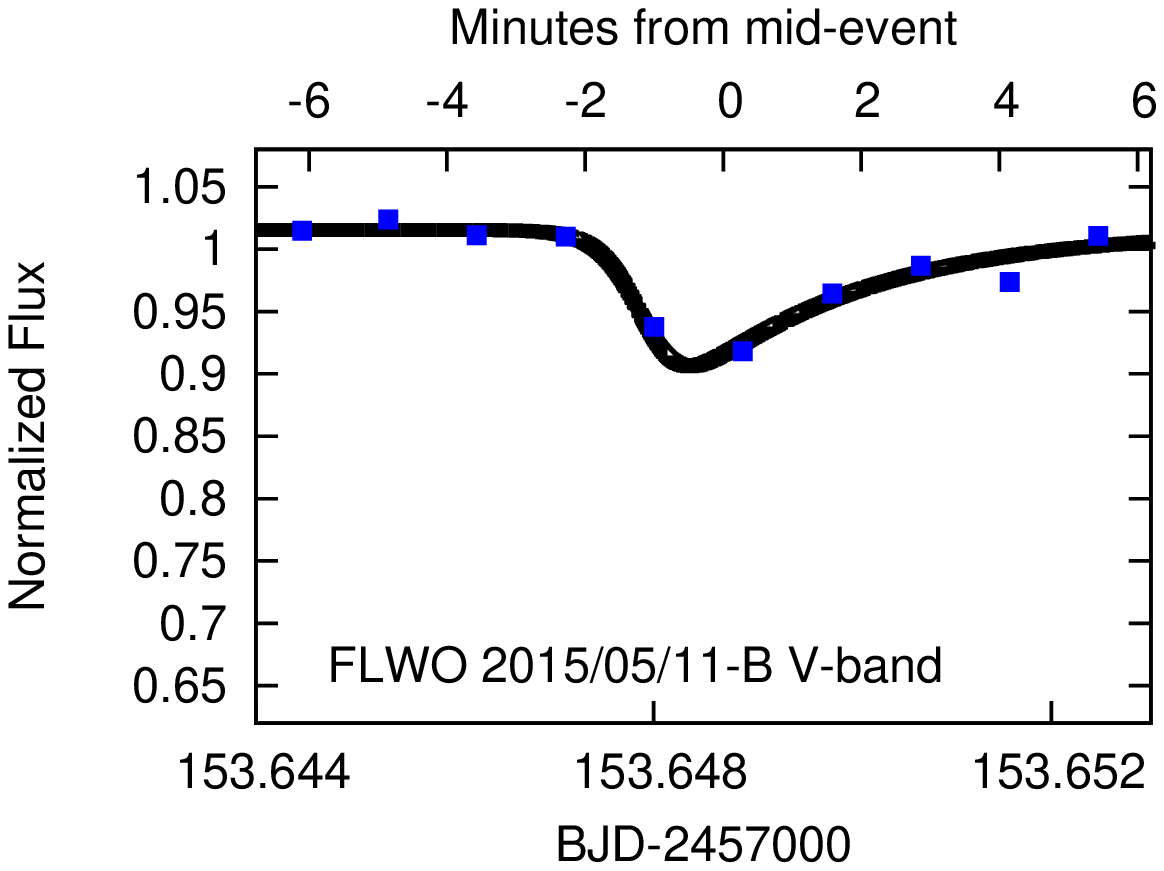} %
\includegraphics[scale=0.33, angle = 270]{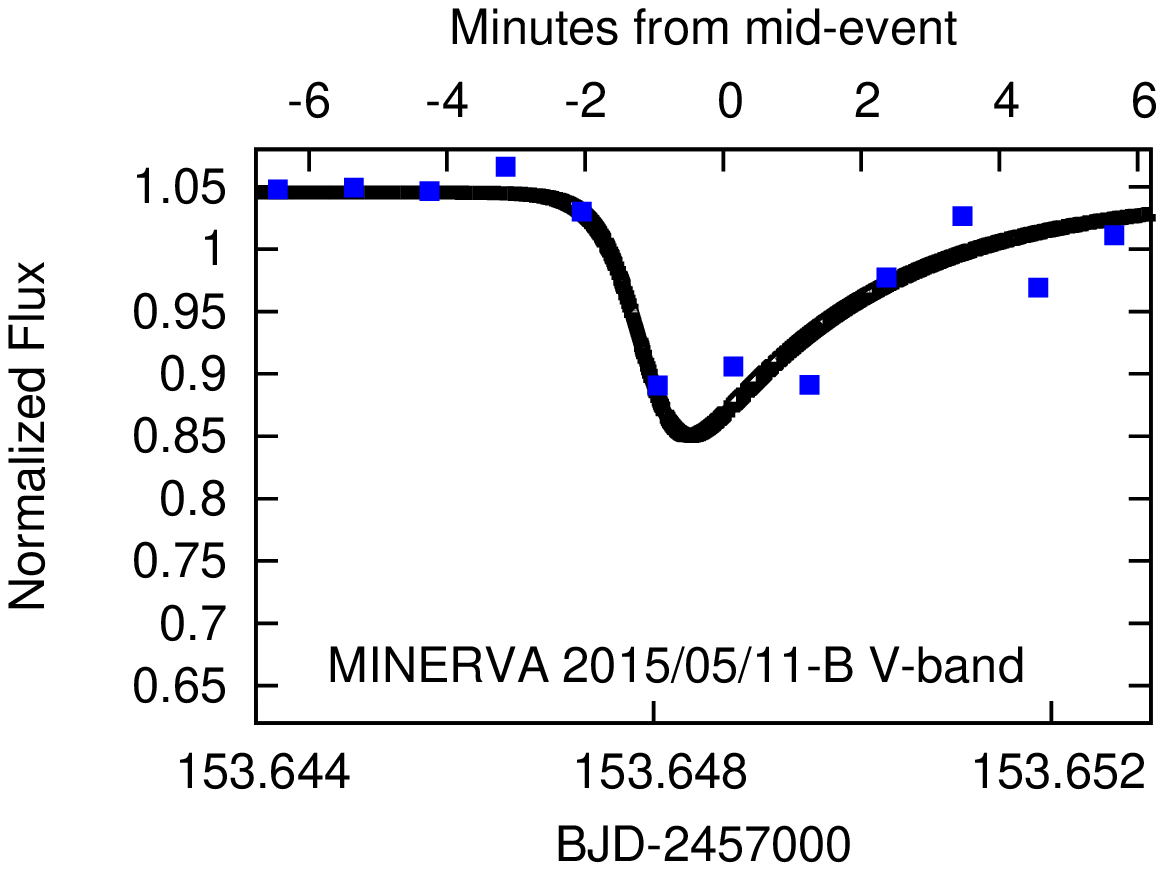} %
\includegraphics[scale=0.33, angle = 270]{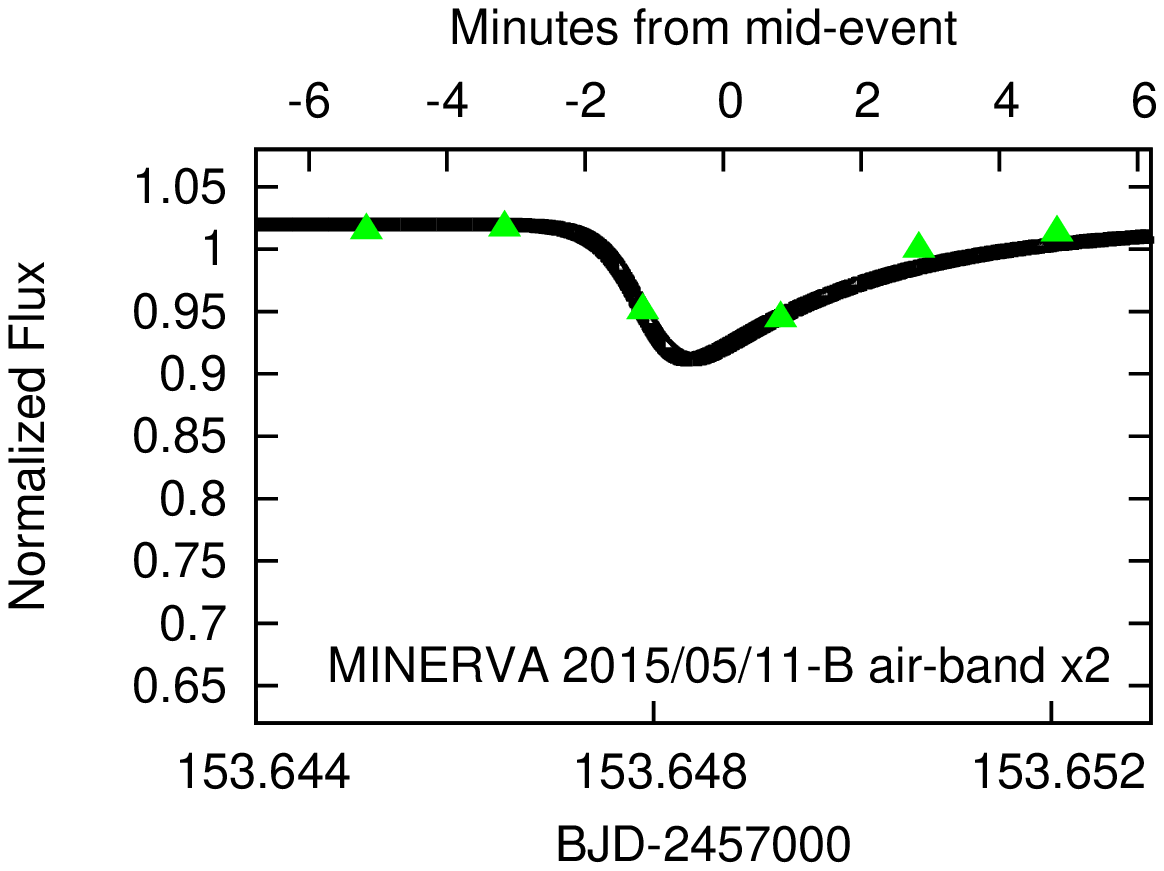} %

\includegraphics[scale=0.33, angle = 270]{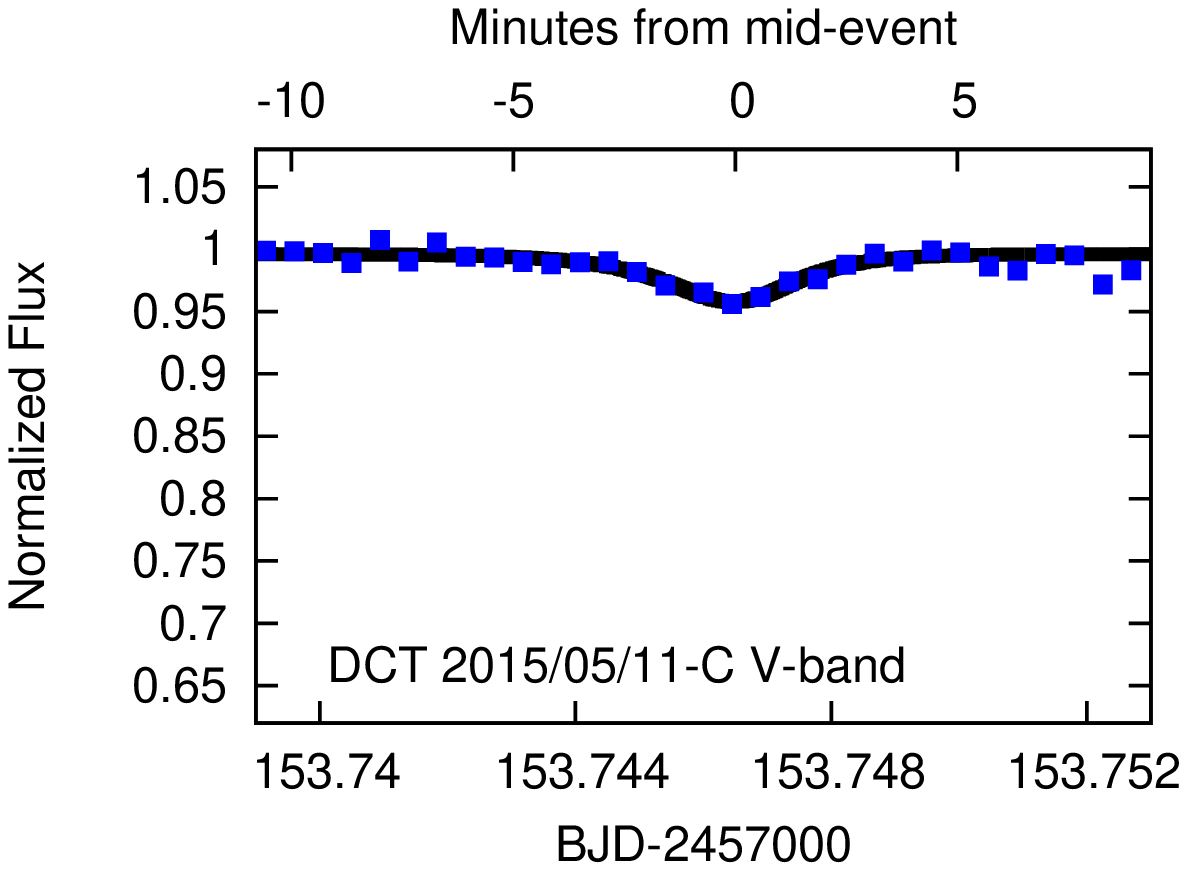} %
\includegraphics[scale=0.33, angle = 270]{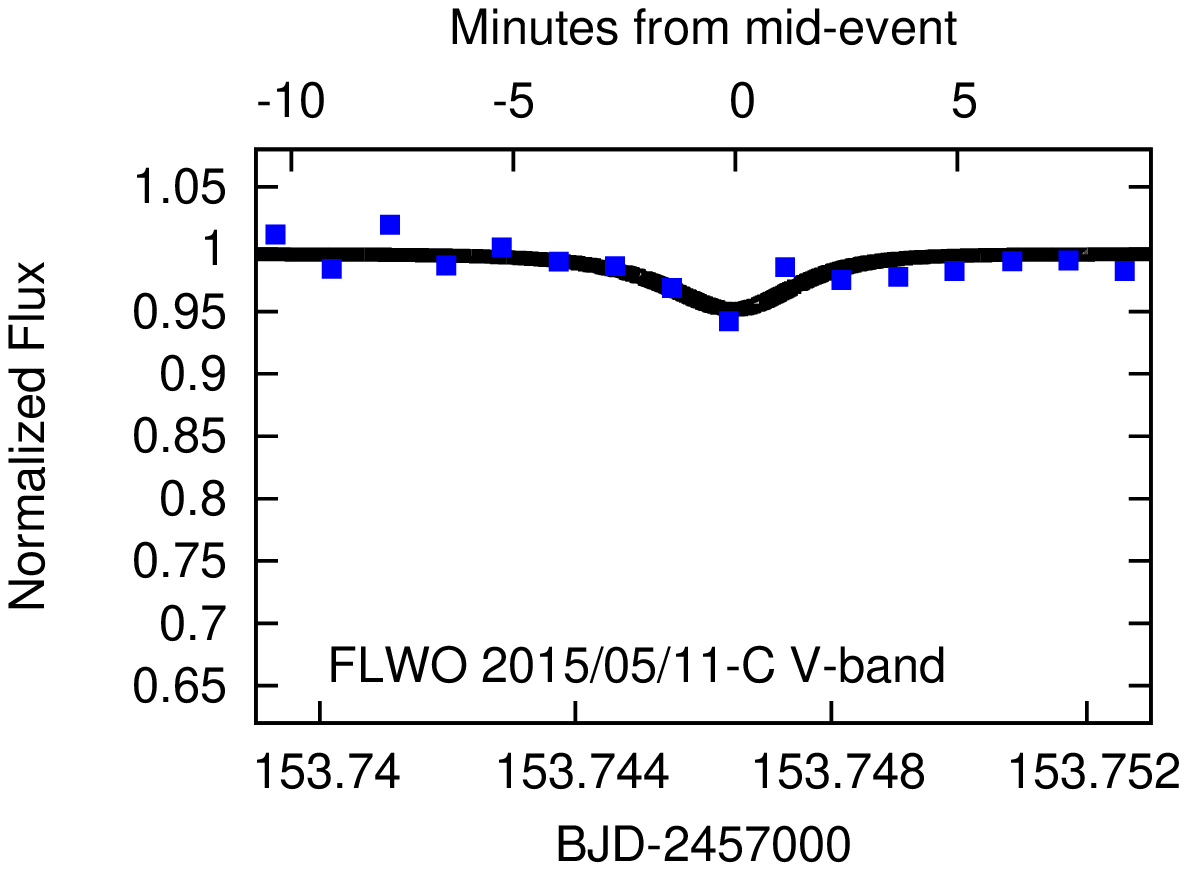} %
\includegraphics[scale=0.33, angle = 270]{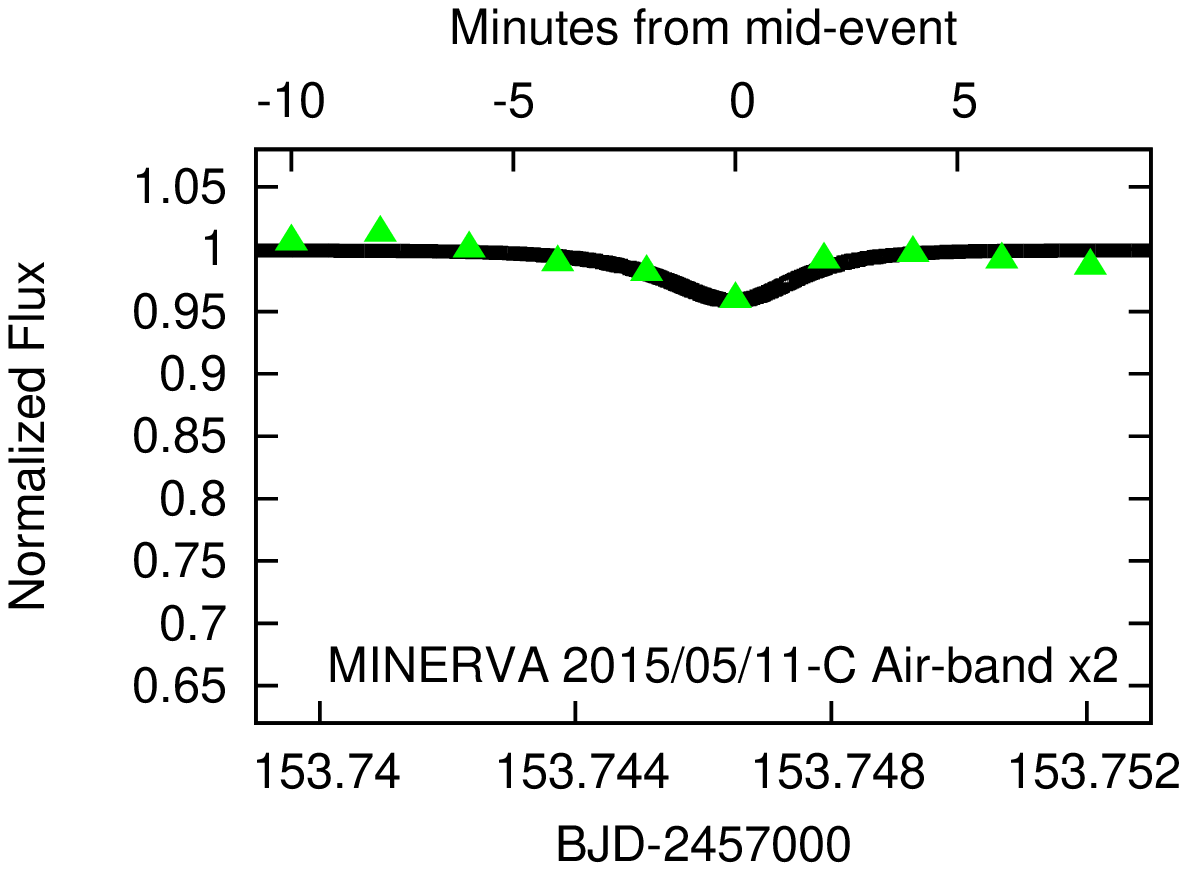} %

\includegraphics[scale=0.33, angle = 270]{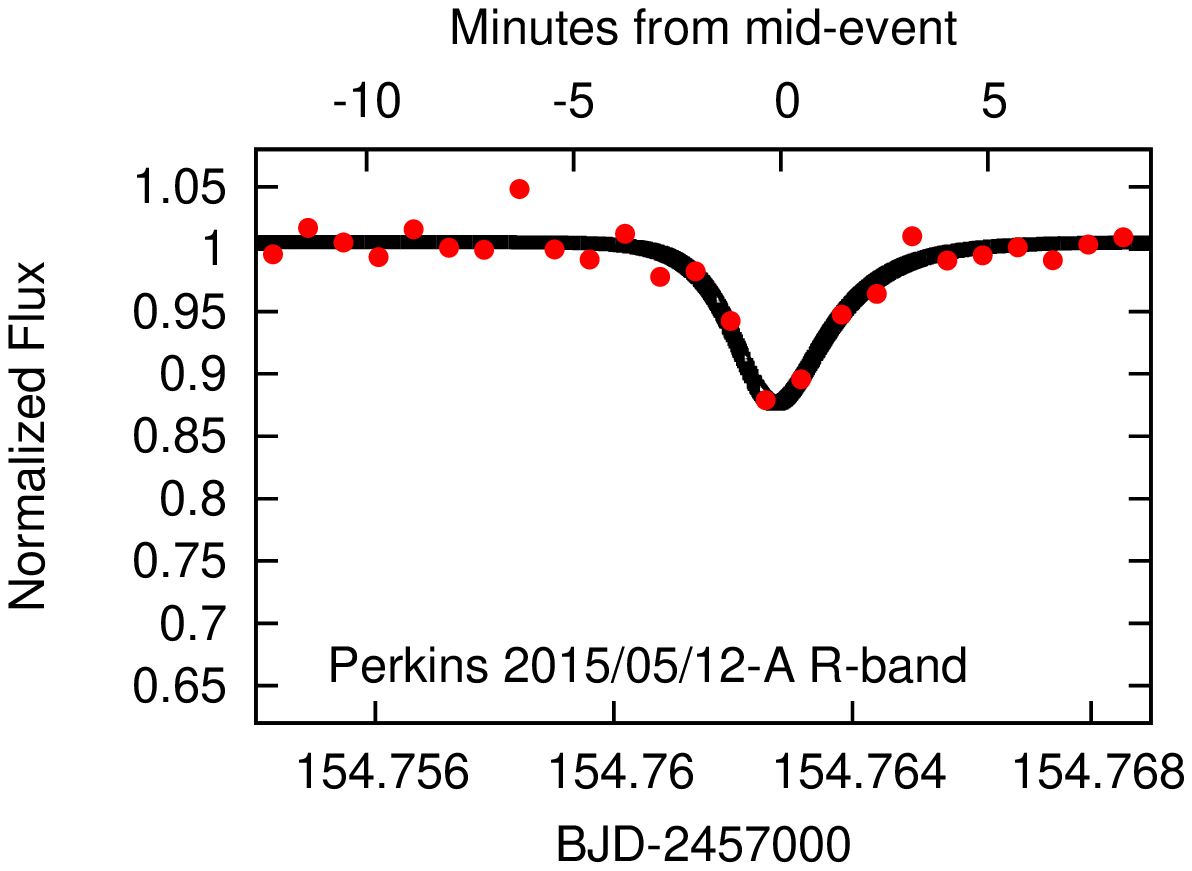} %
\includegraphics[scale=0.33, angle = 270]{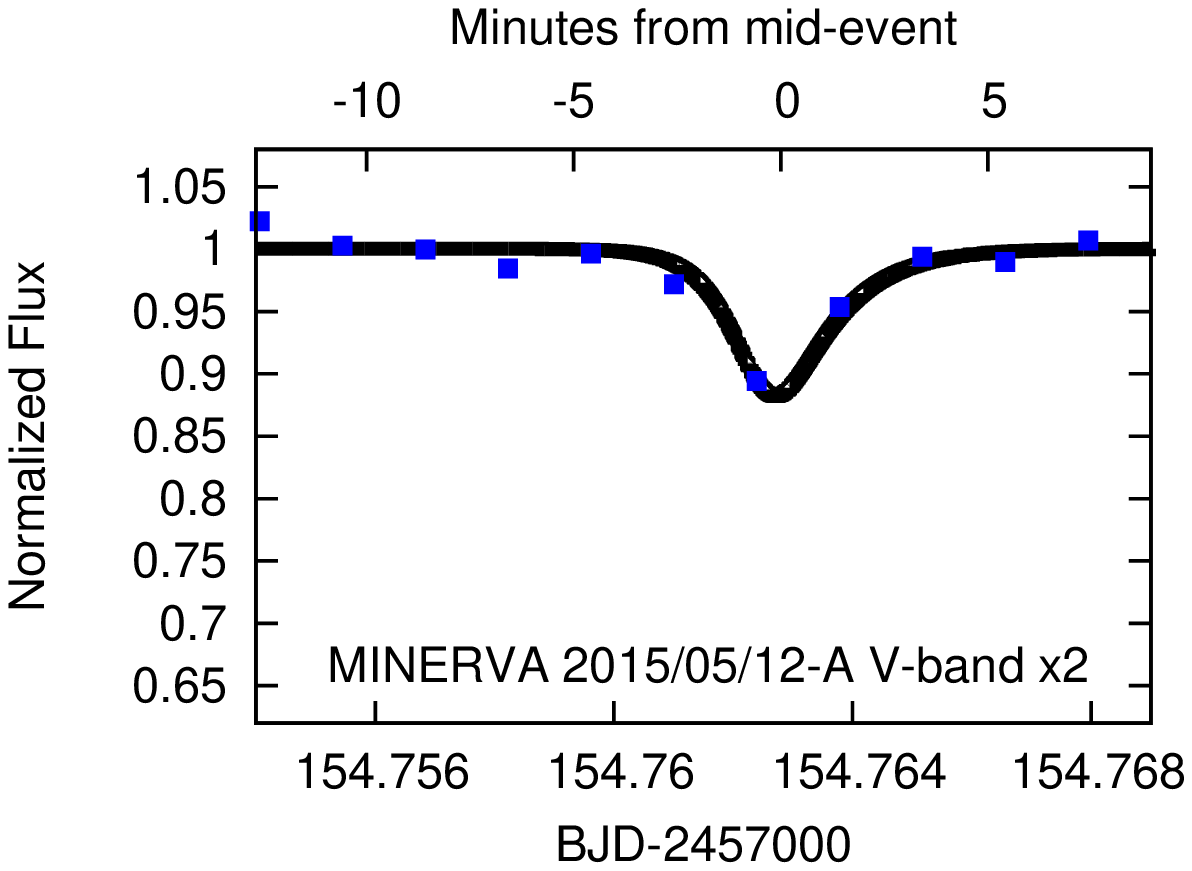} %
\includegraphics[scale=0.33, angle = 270]{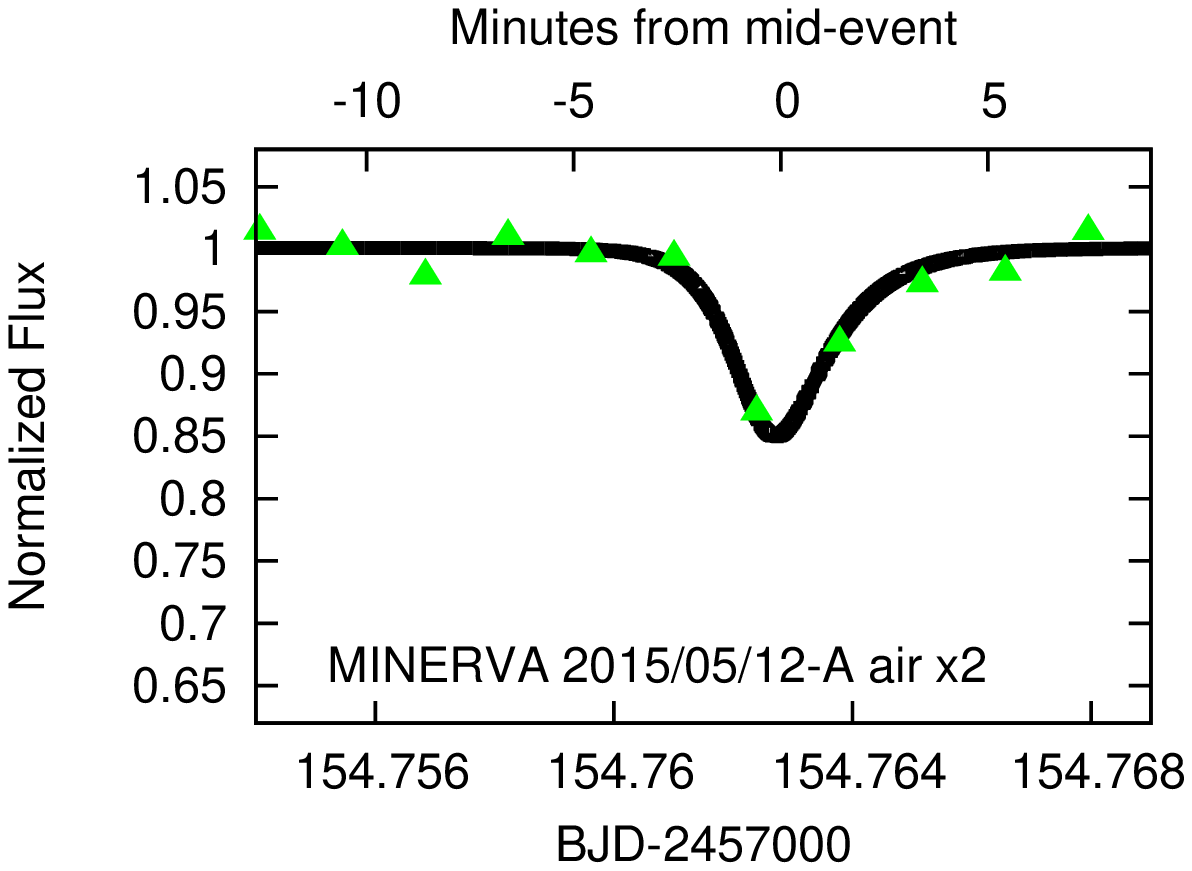} %

\includegraphics[scale=0.33, angle = 270]{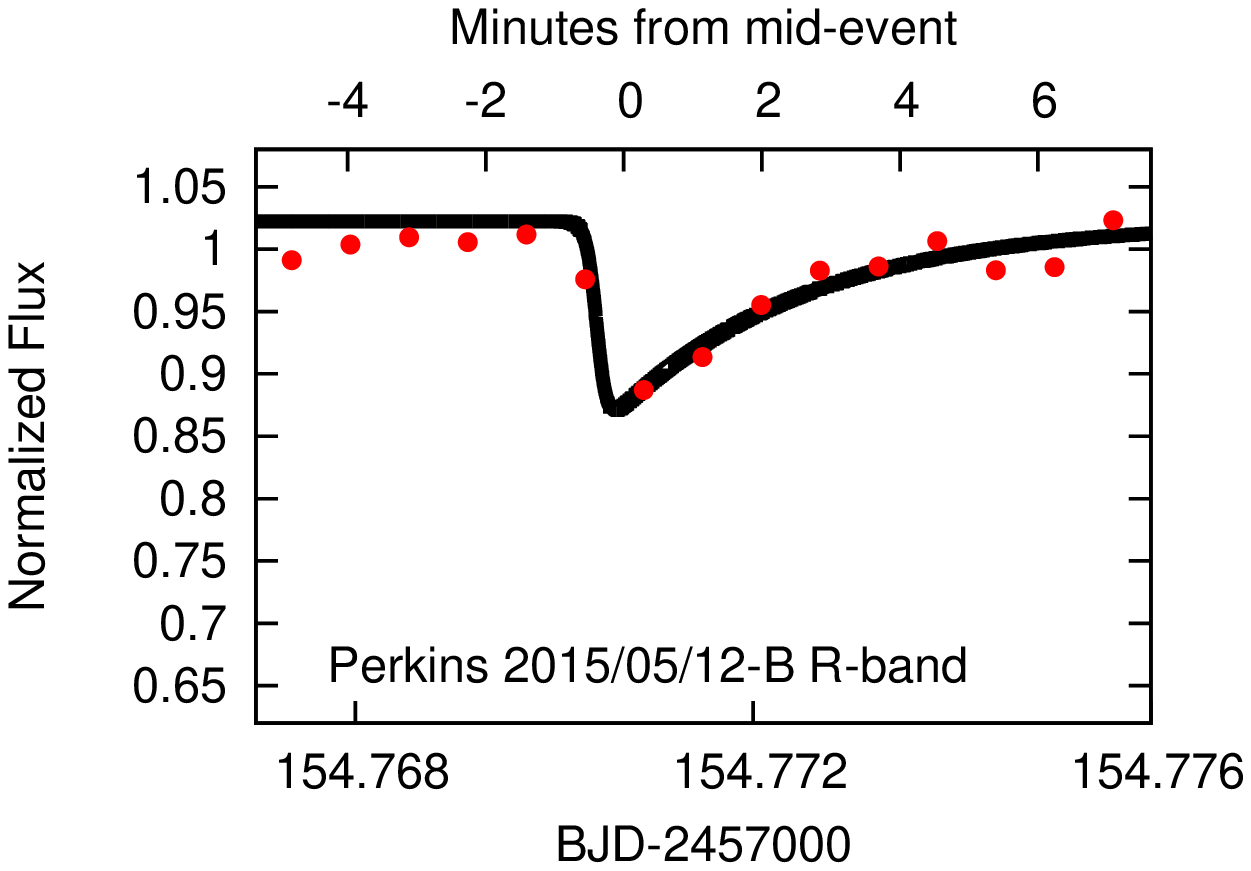} %
\includegraphics[scale=0.33, angle = 270]{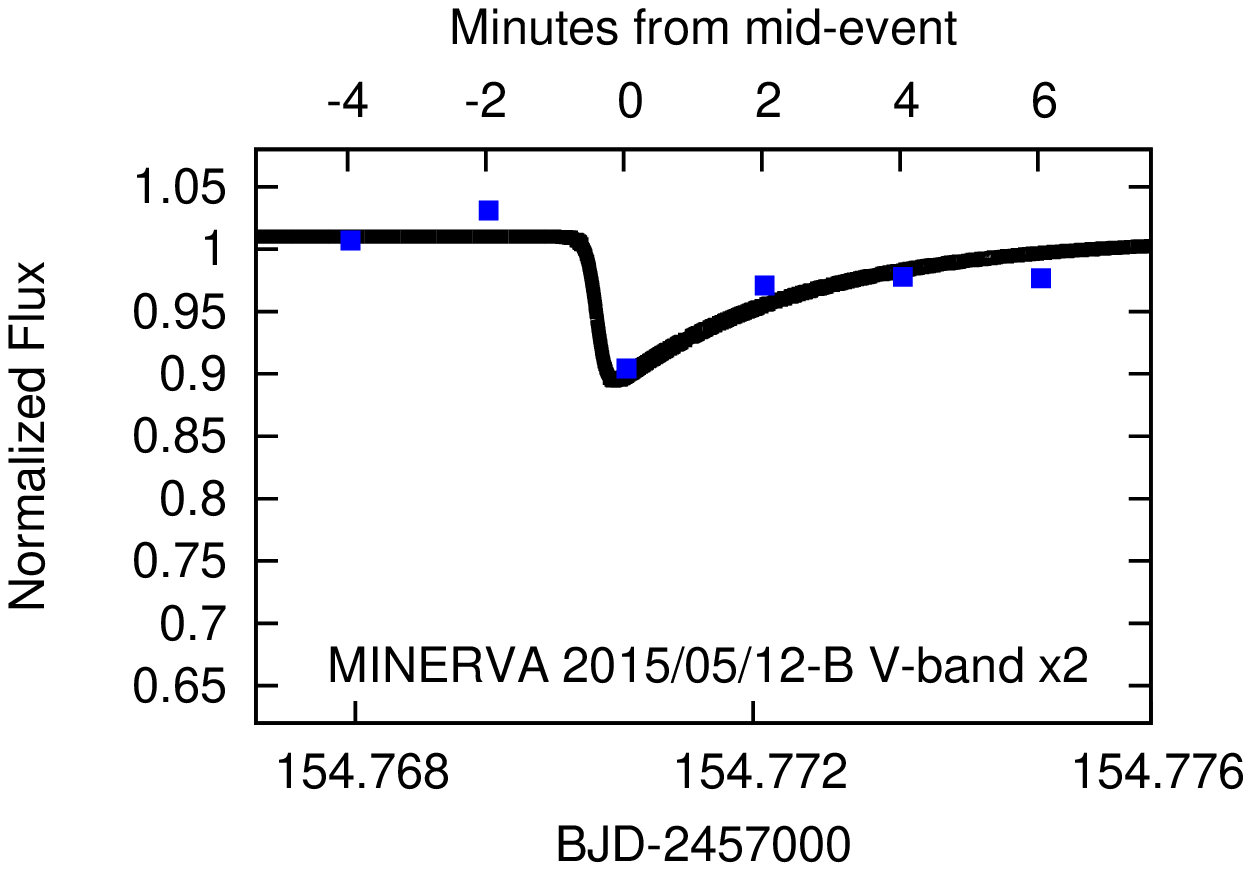} %
\includegraphics[scale=0.33, angle = 270]{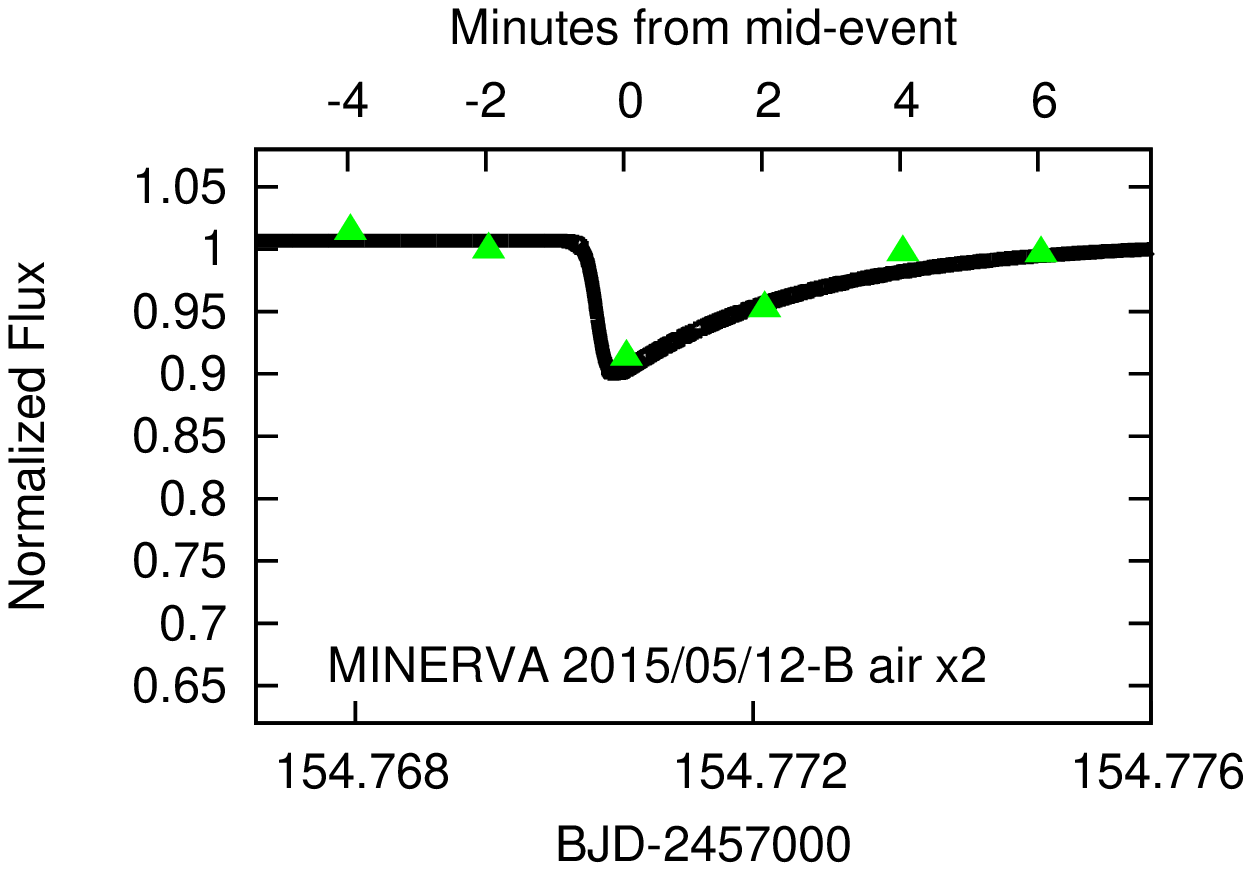} %

\includegraphics[scale=0.33, angle = 270]{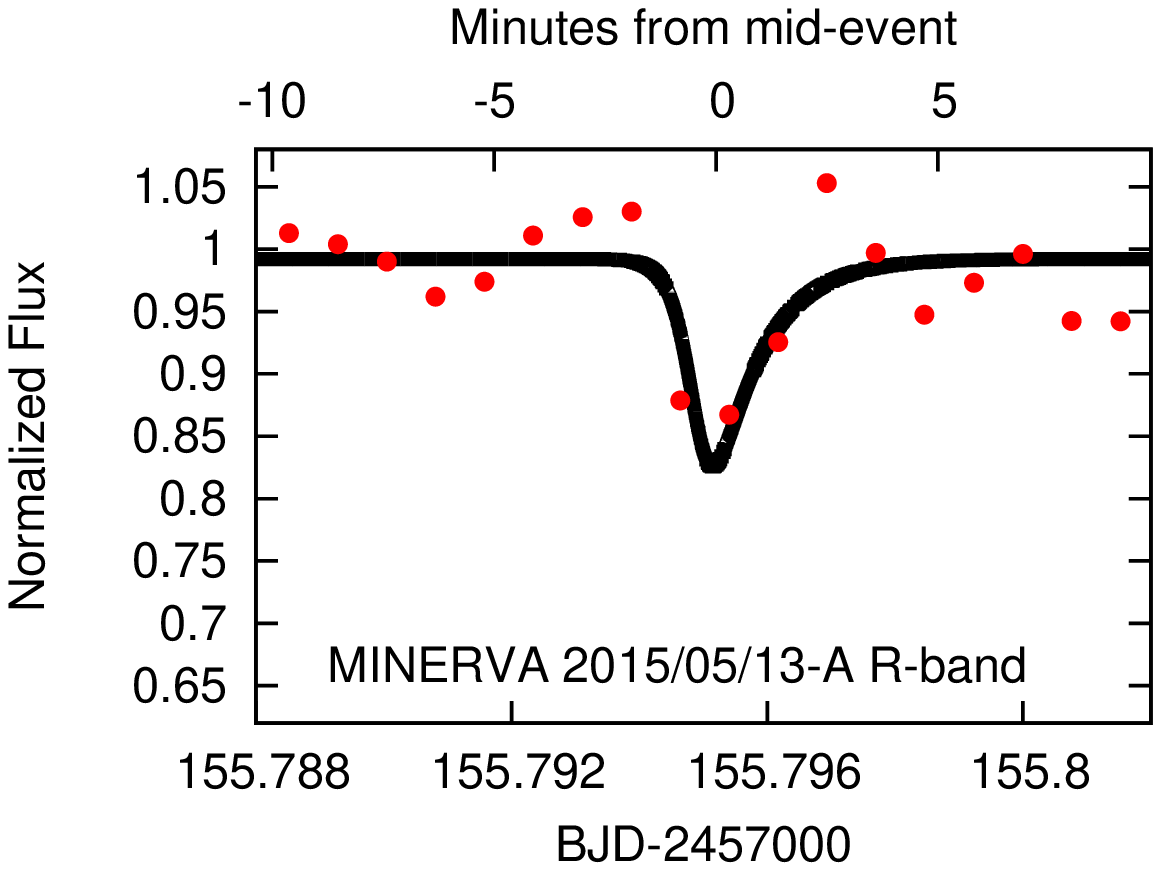} %
\includegraphics[scale=0.33, angle = 270]{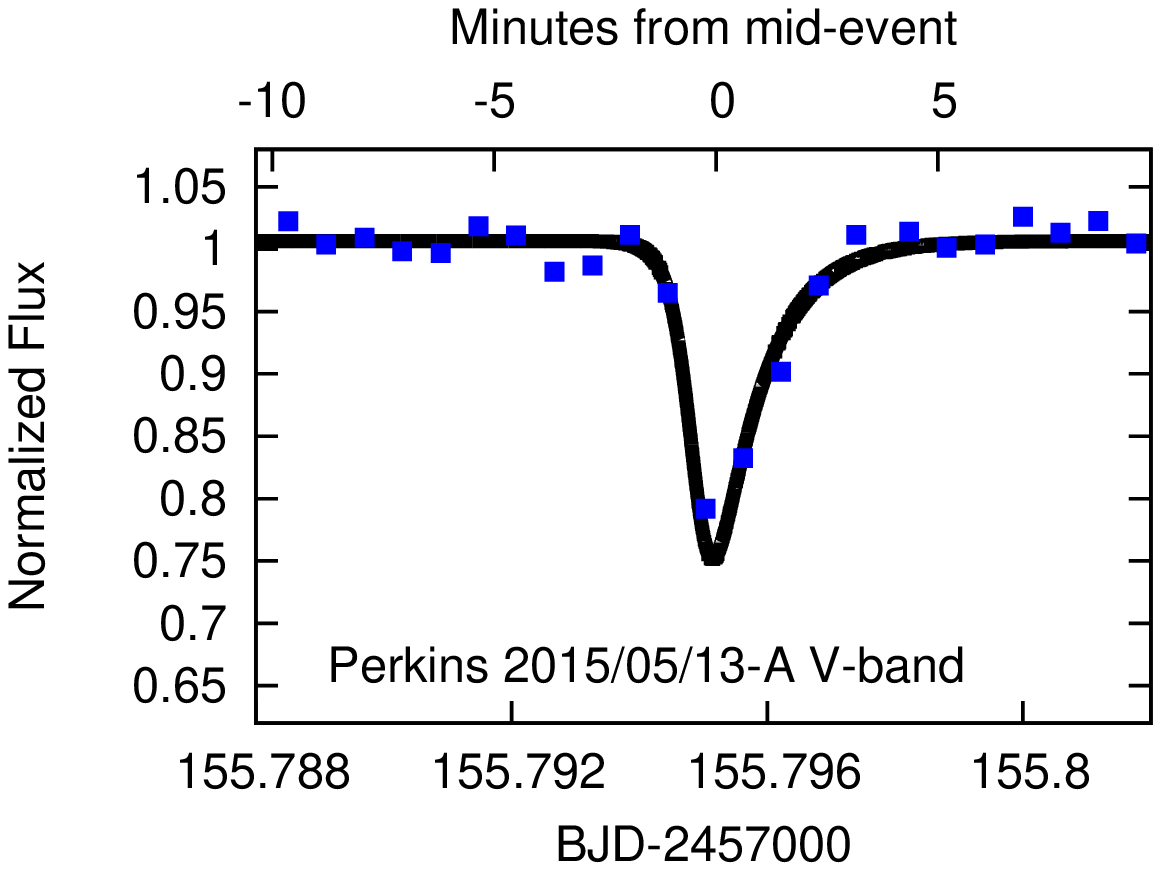} %
\includegraphics[scale=0.33, angle = 270]{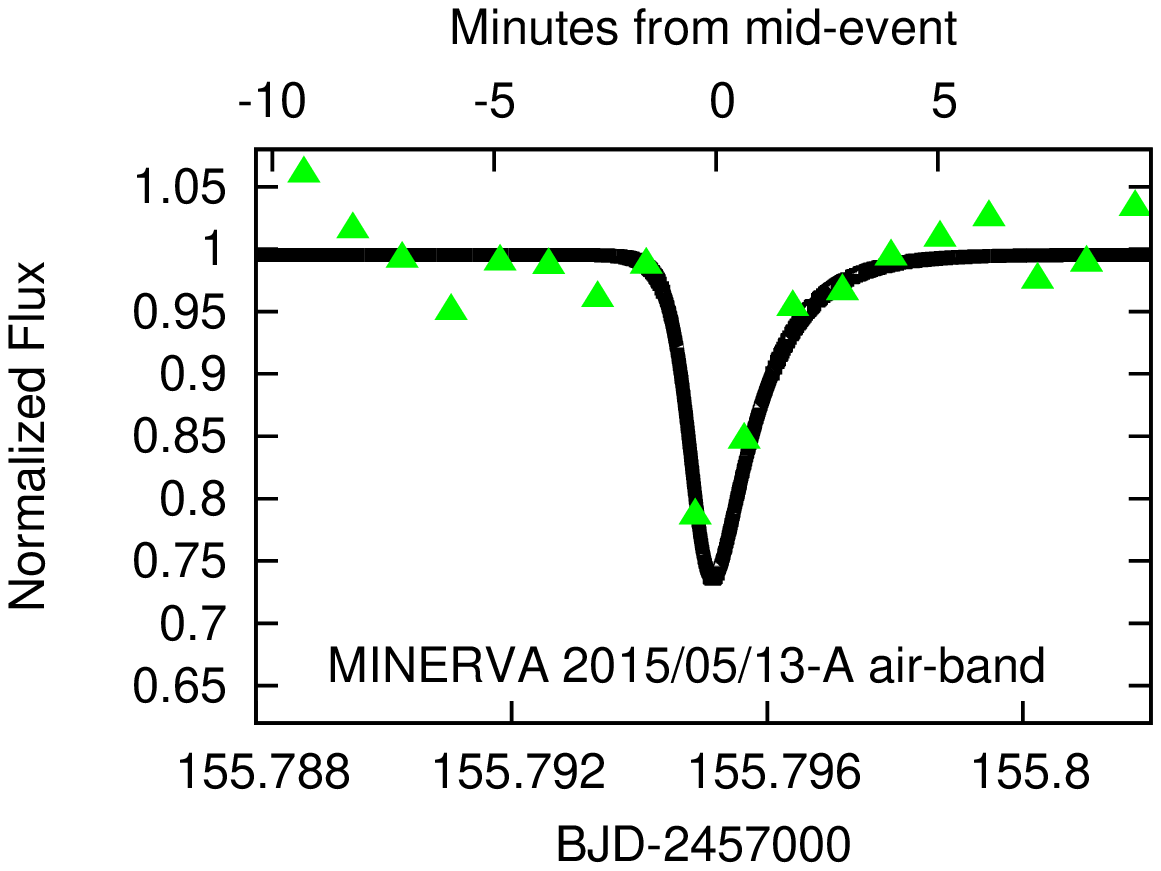} %

\caption[]
	{	Asymmetric hyperbolic secant function fits to various significant flux drops of WD 1145+017,
		observed with the DCT, Perkins, FLWO and MINERVA telescopes.
		Each row of observations in the plot occur at the same time and are from different telescopes, often
		at a different wavelength, as indicated in the panels. 
	}
\label{FigHSFitsJoint}
\end{figure*}

\begin{figure}
 \centering

\includegraphics[scale=0.5, angle = 270]{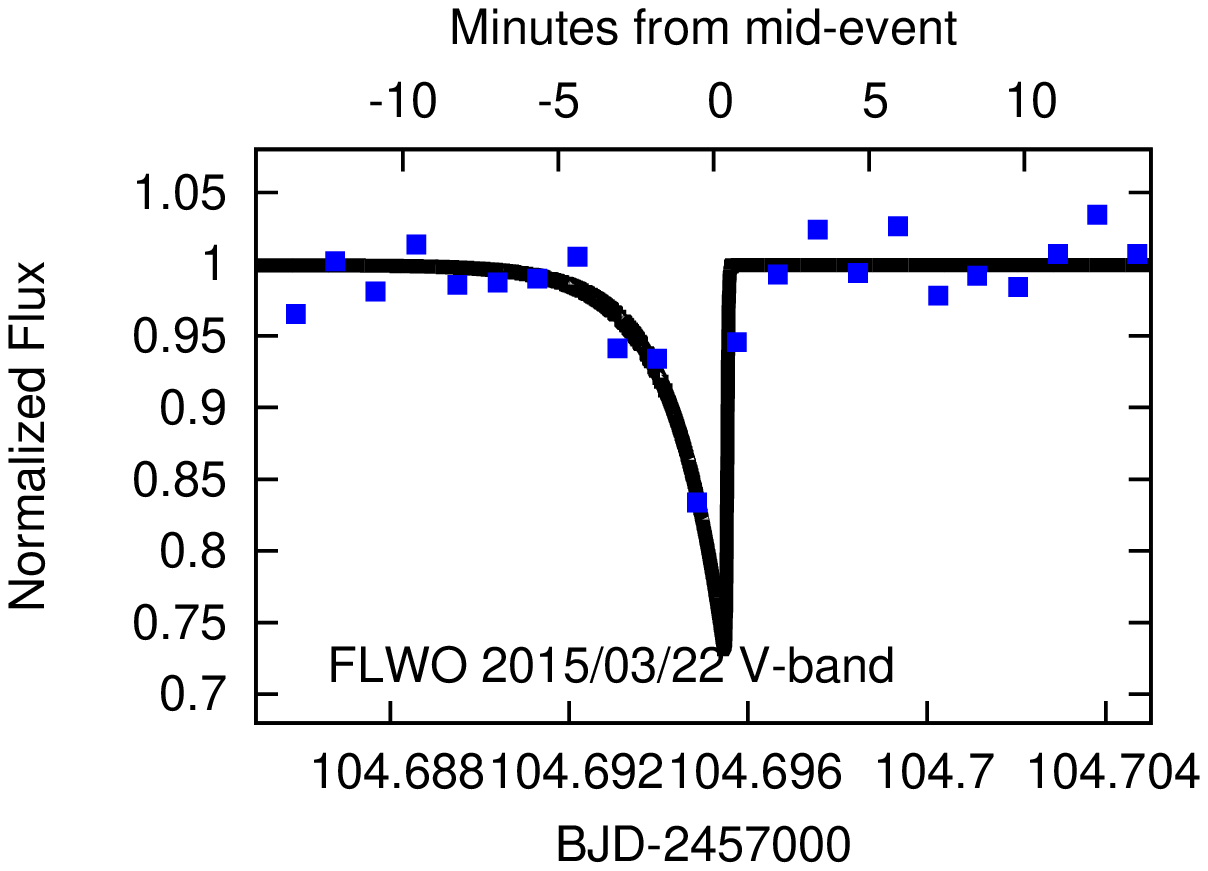} %

\includegraphics[scale=0.5, angle = 270]{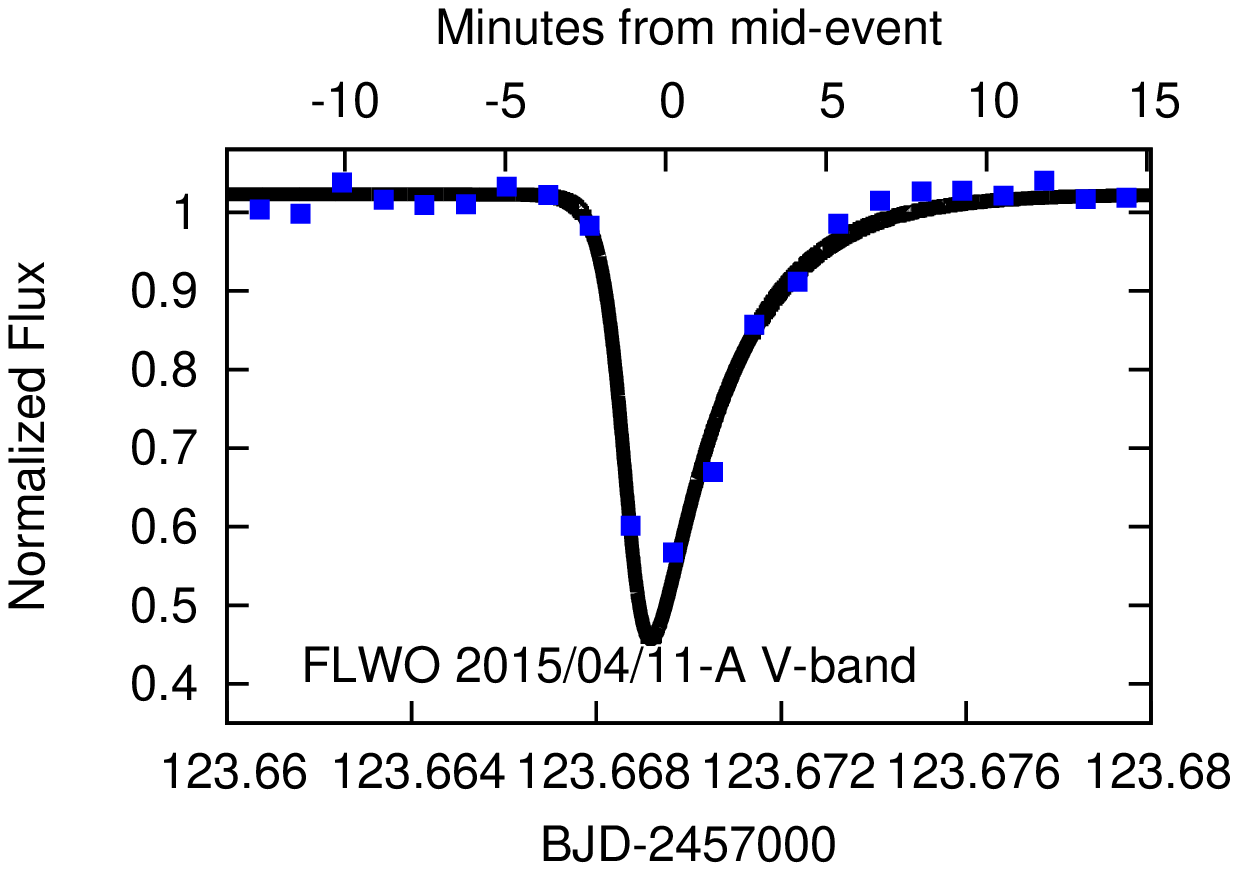} %
\includegraphics[scale=0.5, angle = 270]{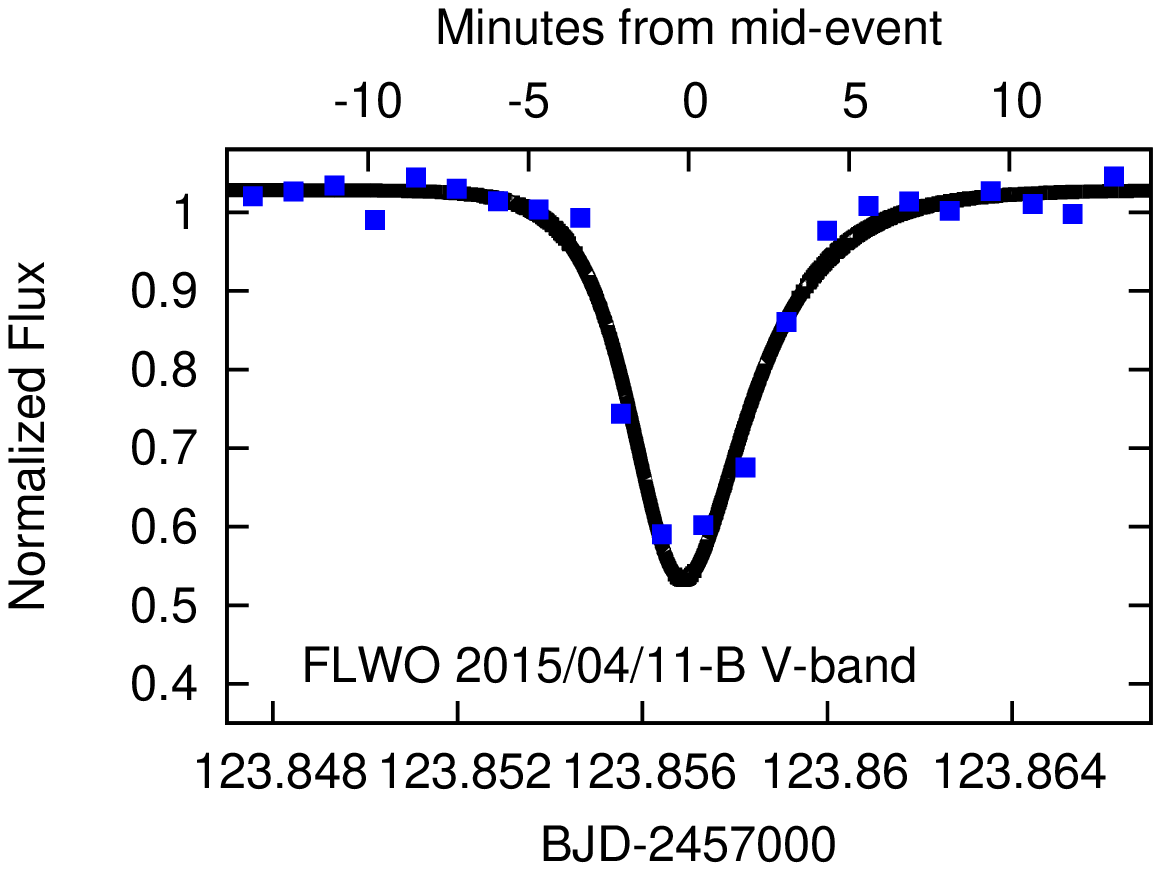} %

\includegraphics[scale=0.5, angle = 270]{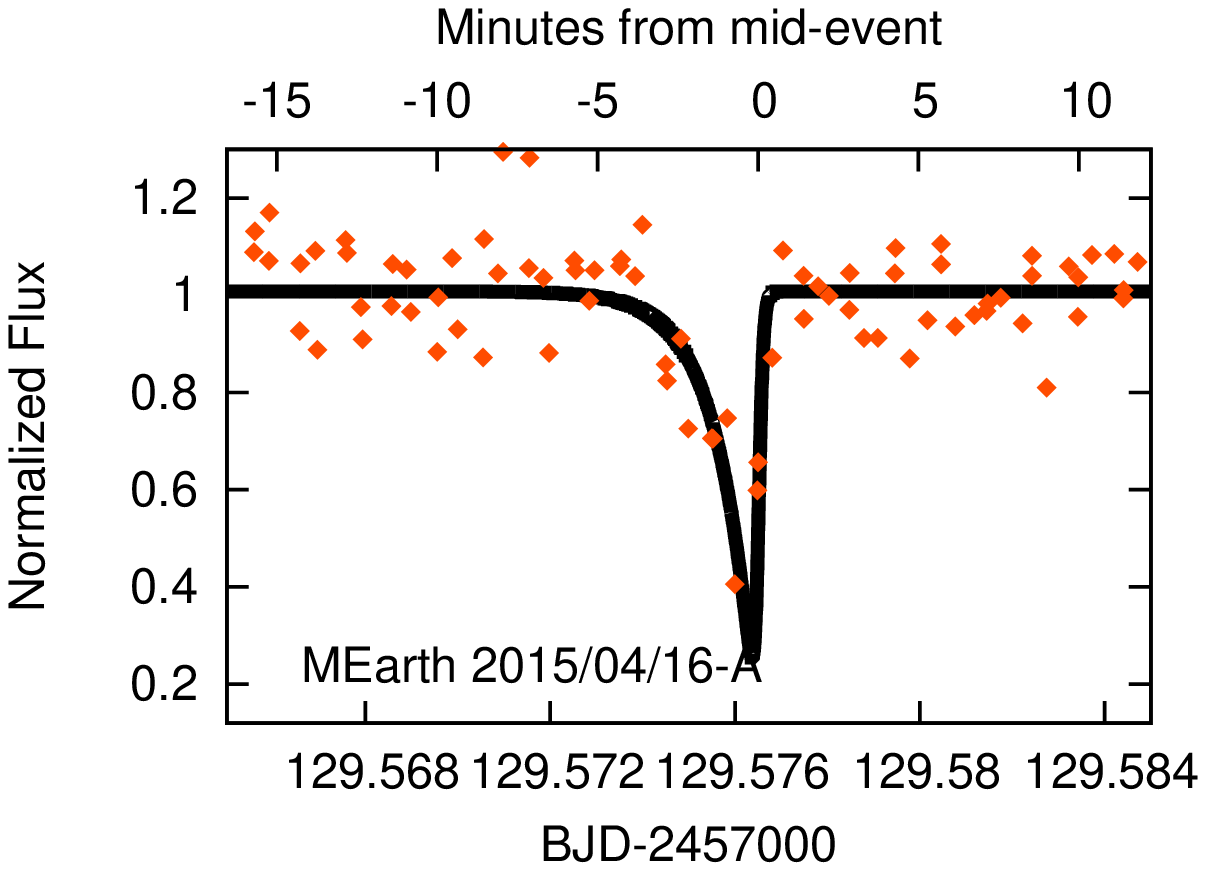} %
\includegraphics[scale=0.5, angle = 270]{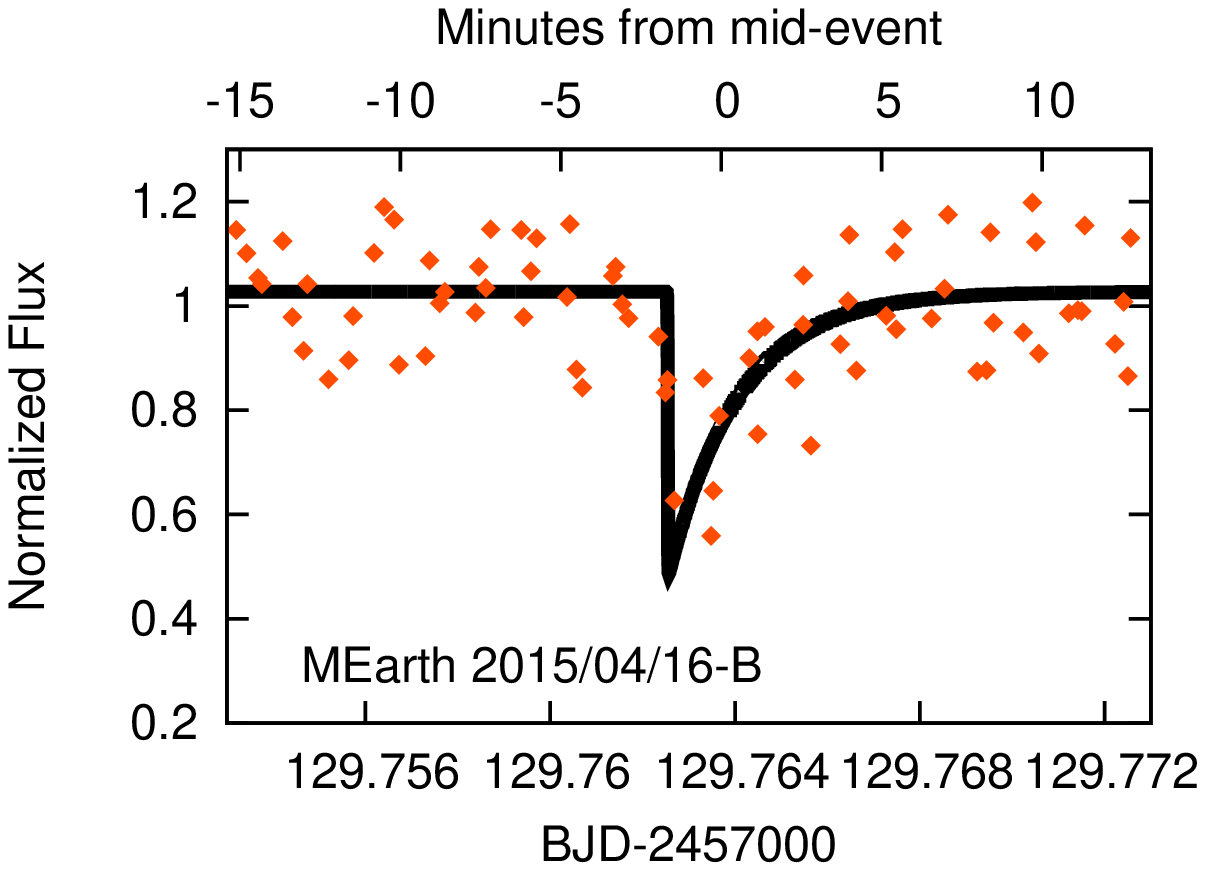} %

\caption[]
	{	Asymmetric hyperbolic secant function fits to the significant flux drops of WD 1145+017 published
		previously by \citet{Vanderburg15}, and observed with the FLWO and MEarth telescopes.
	}
\label{FigHSFitsVanderburg}
\end{figure}

\begin{figure*}
\centering
\includegraphics[scale=0.50, angle = 270]{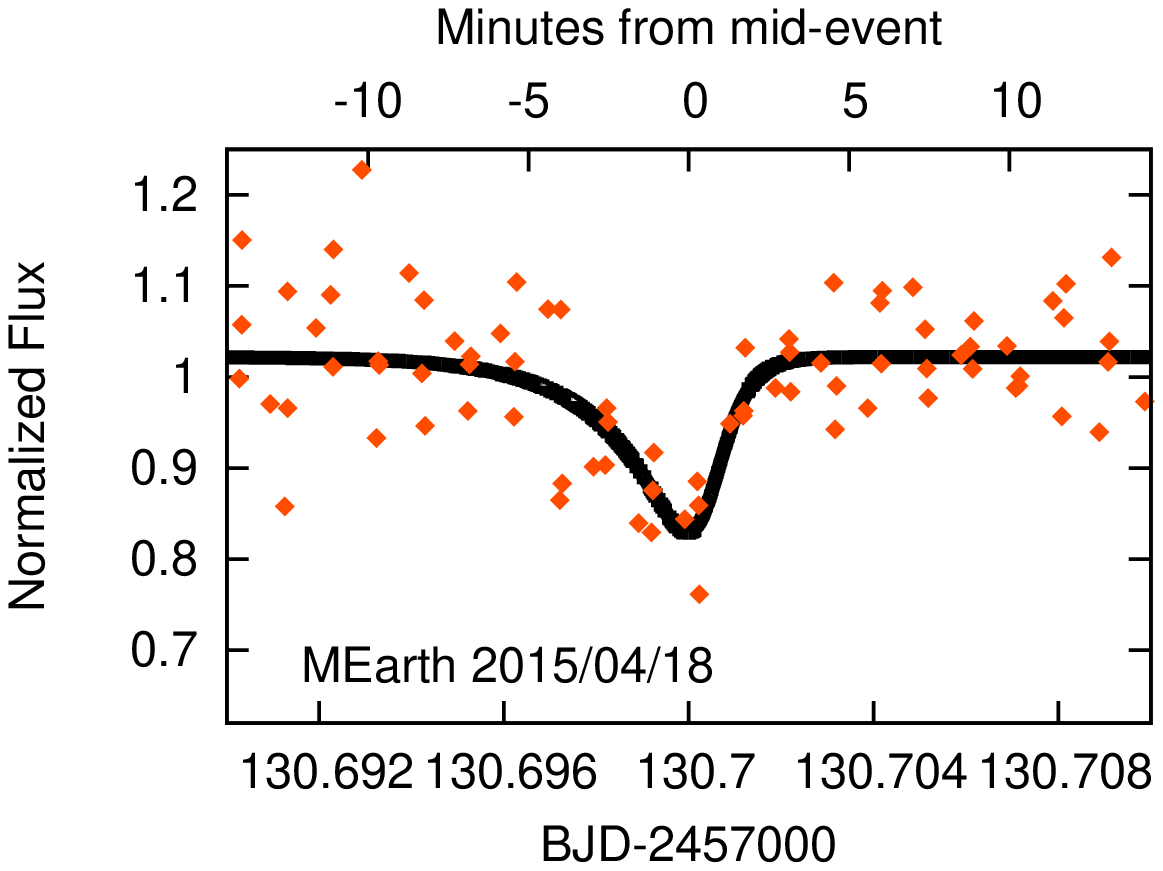} %
\includegraphics[scale=0.50, angle = 270]{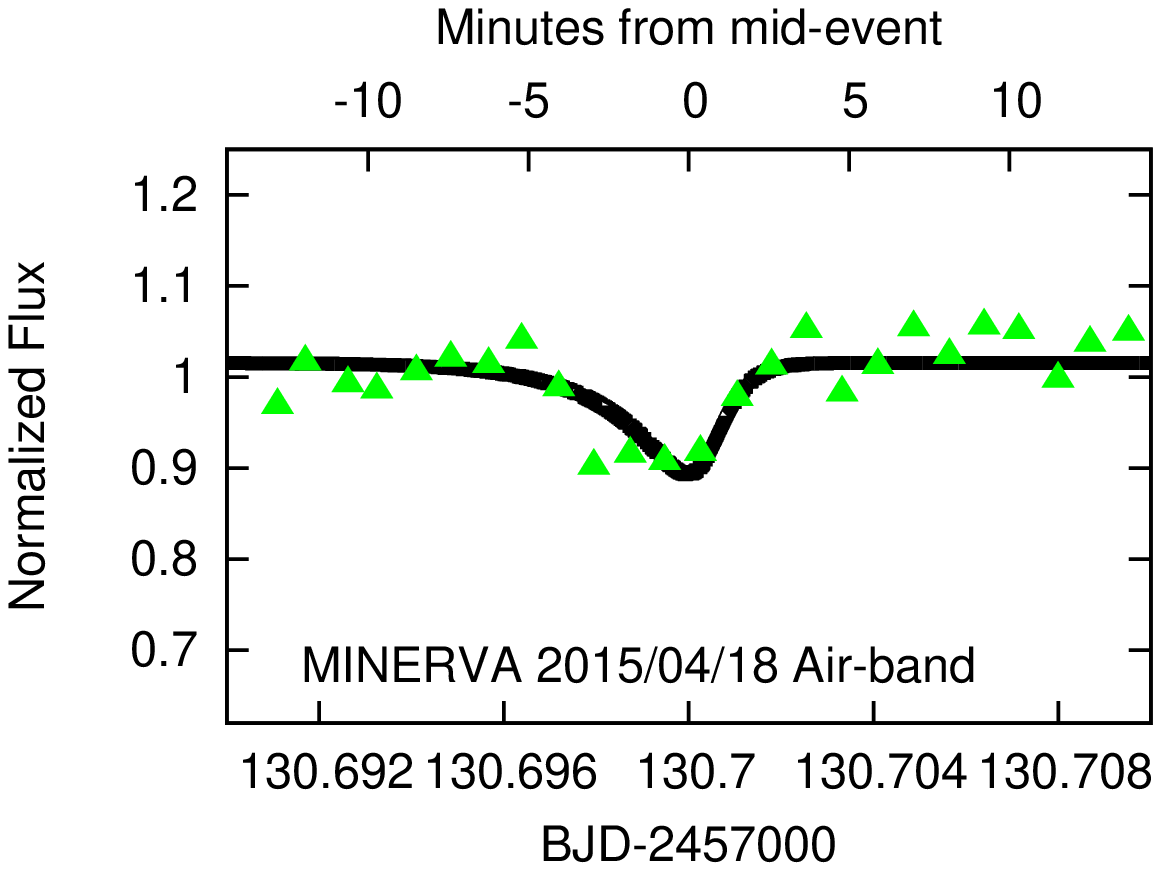} %
\caption[]
	{	Asymmetric hyperbolic secant function fits to the flux drop of WD 1145+017
		observed simultaneously on 2015 April 18 (UTC) with the MEarth and MINERVA/T3 telescopes.
		These data were published previously by \citet{Vanderburg15},
		and the MINERVA/T3 data have been reanalyzed here.
	}
\label{FigHSFitsVanderburgJoint}
\end{figure*}


For our light curves where a significant flux decrement is observed 
we fit the apparent asymmetric transit dips 
with an ``asymmetric hyperbolic secant'', which has been applied
previously to fit the transit profile of the candidate
disintegrating low-mass planet KOI-2700b \citep{Rappaport14}. As the periods of the planetesimals
around WD 1145+017 are not well defined, we replace
the explicit reference to phase with that of the time, $t$, in the asymmetric hyperbolic secant expression of
\citet{Rappaport14}; therefore the profile we fit our transits with has a flux, $F$, at time, $t$, of:
\begin{equation}
F(t) = F_0 - C[ e^{-(t - \tau_0 )/\tau_1} + e^{(t - \tau_0 )/\tau_2}]^{-1}
\label{EquationHS}
\end{equation} 
$F_0$ is the out-of-transit flux, $C$/2 is approximately the transit depth as indicated in equation \ref{EquationDepth}, 
$\tau_0$ is approximately the transit mid-point, and $\tau_1$ and $\tau_2$ are the characteristic durations
of the ingress and egress, respectively.
The minimum of the function has a depth -- which we will refer to as the transit depth, $D$ -- given
by:
\begin{equation}
D = \frac{C \xi^{\xi/(1+\xi)}}{1+\xi}
\label{EquationDepth}
\end{equation}
and occurs at a time $T_{\rm min}$ of:
\begin{equation}
T_{min} = \tau_0 + \frac{\tau_1 \tau_2 \ln(\tau_2/\tau_1)}{\tau_1 + \tau_2 }
\end{equation}
where $\xi$=$\tau_2/\tau_1$.

For those transits that we observe with a single telescope we fit these light curves individually using Equation \ref{EquationHS};
the results are given at the bottom of Table \ref{TableHSFitsJoint} and are presented in Figure \ref{FigHSFitsIndividual}.
For the transits where we are able to obtain multi-telescope coverage, and usually multiwavelength coverage,
we fit $F_0$ \& $C$ individually for each light curve, while fitting $\tau_0$, $\tau_1$ \& $\tau_2$ jointly.
The results for the joint fits are given at the top of Table \ref{TableHSFitsJoint} and are presented in Figure \ref{FigHSFitsJoint}.
We quote $D$ and $T_{\rm min}$, rather than $C$ and $\tau_0$, in Table \ref{TableHSFitsJoint}.
We use Markov Chain Monte Carlo (MCMC) fitting, as described for our purposes in \citet{CrollMCMC}.
When multiple apparent dips are observed in a single evening of observations, we differentiate
between these different transits using a letter label; we label the first transit
of the evening with ``A'' and the next with ``B'' and so-on.
We note that we bin our asymmetric hyperbolic secant model to account for the finite exposure times when
comparing with our photometry during our fitting process.

We also fit the $\sim$40\% occultations of WD 1145+017 that were detected with the FLWO and MEarth telescopes
that were published in \citet{Vanderburg15}; we also confirm two other events that were listed as possible events 
in \citet{Vanderburg15}, including 
an additional $\sim$25\% event observed with the FLWO on 2015 March 22,
and a $\sim$15\% event observed on 2015 April 18 with the MEarth and MINERVA/T3 telescopes.
The MEarth data are reduced and photometry is produced as outlined in \citet{Vanderburg15},
while the FLWO and MINERVA/T3 data are reduced as discussed in Section \ref{SecObs}. 
The MINERVA/T3 data on 2015 April 18 (UTC) was obtained with the altitude-azimuth derotator not functioning,
and therefore suffers from systematics that are most apparent after the drop in flux that we interpret 
as a transit that evening;
we therefore conservatively, artifically scale
up the errors on these points to 10\%, and do not use this transit for transit depth comparisons.
The fits to the \citet{Vanderburg15} photometry 
are presented in Figure \ref{FigHSFitsVanderburg}, and \ref{FigHSFitsVanderburgJoint} and are given in Table \ref{TableHSFitsJoint}.

Lastly, we summarize the ratio of the transit depths, $D$,
we find for our multiwavelength observations in Table \ref{TableRatio}.

\begin{deluxetable*}{cccccc}
\centering
\tablecaption{Joint ``Transit Depth'' Ratio Fits}
\tablehead{
\colhead{Date (UTC) \& Transit \#}	& \colhead{$T_{\rm min}$ (JD-2457000)} & \colhead{$D_{\rm air}/D_R$}	& \colhead{$D_V/D_R$}	&  \colhead{$D_{\rm air}/D_V$}	\\
}
\centering
\startdata
2015/05/09	& \TminMayNineJoint$^{+\TminPlusMayNineJoint}_{-\TminMinusMayNineJoint}$				& \CRatioCAMayNineJoint$^{+\CRatioCAPlusMayNineJoint}_{-\CRatioCAMinusMayNineJoint}$, \CRatioBAMayNineJoint$^{+\CRatioBAPlusMayNineJoint}_{-\CRatioBAMinusMayNineJoint}$ & n/a & n/a \\
2015/05/10	& \TminMayTenJoint$^{+\TminPlusMayTenJoint}_{-\TminMinusMayTenJoint}$				& \CRatioCAMayTenJoint$^{+\CRatioCAPlusMayTenJoint}_{-\CRatioCAMinusMayTenJoint}$	& \CRatioBAMayTenJoint$^{+\CRatioBAPlusMayTenJoint}_{-\CRatioBAMinusMayTenJoint}$	& \CRatioCBMayTenJoint$^{+\CRatioCBPlusMayTenJoint}_{-\CRatioCBMinusMayTenJoint}$\\
2015/05/11-A	& \TminMayElevenSecondJoint$^{+\TminPlusMayElevenSecondJoint}_{-\TminMinusMayElevenSecondJoint}$	& n/a & n/a & \CRatioBAMayElevenSecondJoint$^{+\CRatioBAPlusMayElevenSecondJoint}_{-\CRatioBAMinusMayElevenSecondJoint}$, \CRatioCAMayElevenSecondJoint$^{+\CRatioCAPlusMayElevenSecondJoint}_{-\CRatioCAMinusMayElevenSecondJoint}$\\
2015/05/11-B	& \TminMayElevenJoint$^{+\TminPlusMayElevenJoint}_{-\TminMinusMayElevenJoint}$			& n/a & n/a & \CRatioBAMayElevenJoint$^{+\CRatioBAPlusMayElevenJoint}_{-\CRatioBAMinusMayElevenJoint}$, \CRatioCAMayElevenJoint$^{+\CRatioCAPlusMayElevenJoint}_{-\CRatioCAMinusMayElevenJoint}$\\
2015/05/11-C	& \TminMayElevenThirdJoint$^{+\TminPlusMayElevenThirdJoint}_{-\TminMinusMayElevenThirdJoint}$	& n/a & n/a & \CRatioBAMayElevenThirdJoint$^{+\CRatioBAPlusMayElevenThirdJoint}_{-\CRatioBAMinusMayElevenThirdJoint}$, \CRatioCAMayElevenThirdJoint$^{+\CRatioCAPlusMayElevenThirdJoint}_{-\CRatioCAMinusMayElevenThirdJoint}$\\
2015/05/12-A	& \TminMayTwelveJoint$^{+\TminPlusMayTwelveJoint}_{-\TminMinusMayTwelveJoint}$			& \CRatioCAMayTwelveJoint$^{+\CRatioCAPlusMayTwelveJoint}_{-\CRatioCAMinusMayTwelveJoint}$ 	& \CRatioBAMayTwelveJoint$^{+\CRatioBAPlusMayTwelveJoint}_{-\CRatioBAMinusMayTwelveJoint}$ & \CRatioCBMayTwelveJoint$^{+\CRatioCBPlusMayTwelveJoint}_{-\CRatioCBMinusMayTwelveJoint}$\\
2015/05/12-B	& \TminMayTwelveSecondJoint$^{+\TminPlusMayTwelveSecondJoint}_{-\TminMinusMayTwelveSecondJoint}$	& \CRatioCAMayTwelveSecondJoint$^{+\CRatioCAPlusMayTwelveSecondJoint}_{-\CRatioCAMinusMayTwelveSecondJoint}$ 	& \CRatioBAMayTwelveSecondJoint$^{+\CRatioBAPlusMayTwelveSecondJoint}_{-\CRatioBAMinusMayTwelveSecondJoint}$ & \CRatioCBMayTwelveSecondJoint$^{+\CRatioCBPlusMayTwelveSecondJoint}_{-\CRatioCBMinusMayTwelveSecondJoint}$\\
2015/05/13	& \TminMayThirteenJoint$^{+\TminPlusMayThirteenJoint}_{-\TminMinusMayThirteenJoint}$		& \CRatioCAMayThirteenJoint$^{+\CRatioCAPlusMayThirteenJoint}_{-\CRatioCAMinusMayThirteenJoint}$ 	& \CRatioBAMayThirteenJoint$^{+\CRatioBAPlusMayThirteenJoint}_{-\CRatioBAMinusMayThirteenJoint}$ & \CRatioCBMayThirteenJoint$^{+\CRatioCBPlusMayThirteenJoint}_{-\CRatioCBMinusMayThirteenJoint}$\\
\hline
Weighted Mean	& n/a	& \TransitDepthAirOverR \ $\pm$ \TransitDepthAirOverRError	&  \TransitDepthVOverR \ $\pm$ \TransitDepthVOverRError	& \TransitDepthAirOverV \ $\pm$ \TransitDepthAirOverRError	\\
\enddata
\label{TableRatio}
\end{deluxetable*}

\subsection{The Frequency of Significant Transits}

In Table \ref{TableHSFitsJoint} we present \NumSignificantTransits \ transits with depths, $D$, greater than $\sim$10\%
of the stellar flux from our 2015 May photometry.
The number of non-overlapping hours of observations from MINERVA, FLWO, the DCT or the Perkins telescope during the month of 2015 May
that we would be 
sensitive\footnote{To be conservative we exclude $\sim$10 minutes at the start and end of each night of observations,
and exclude data with significant cloud cover, or that is otherwise unreliable.}
to these \NumSignificantTransits \ significant transits is $\sim$\NumSignificantHours \ $\rm hr$;
therefore during our observations
the frequency of 10\% transit dips is $\sim$\FrequencySignificantTransits \ per hour, or
$\sim$\HoursPerSignificantTransits \ $\rm hr$ per significant event.

\subsection{Transit Duration Changes}
\label{SecTransitDuration}

\begin{figure}
\includegraphics[scale=0.615, angle = 270]{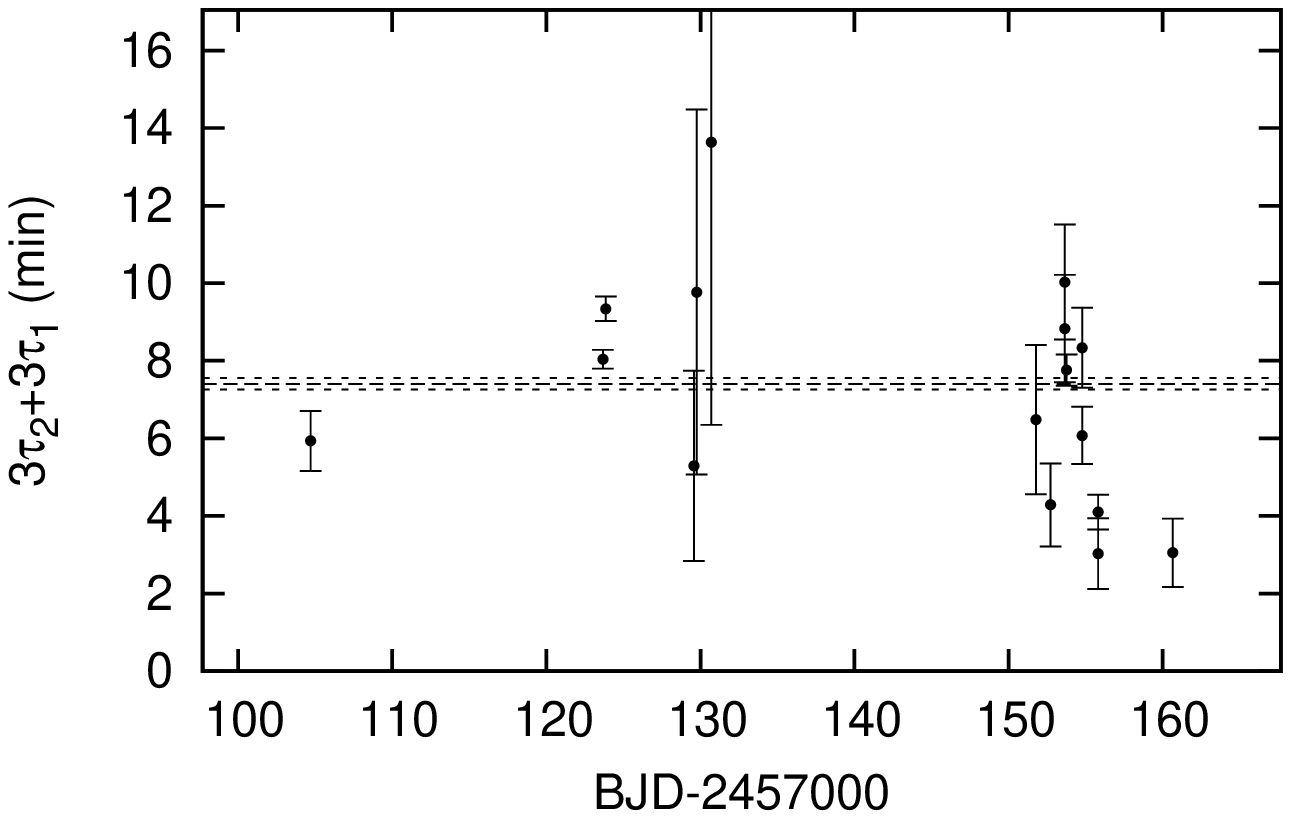} %
\includegraphics[scale=0.615, angle = 270]{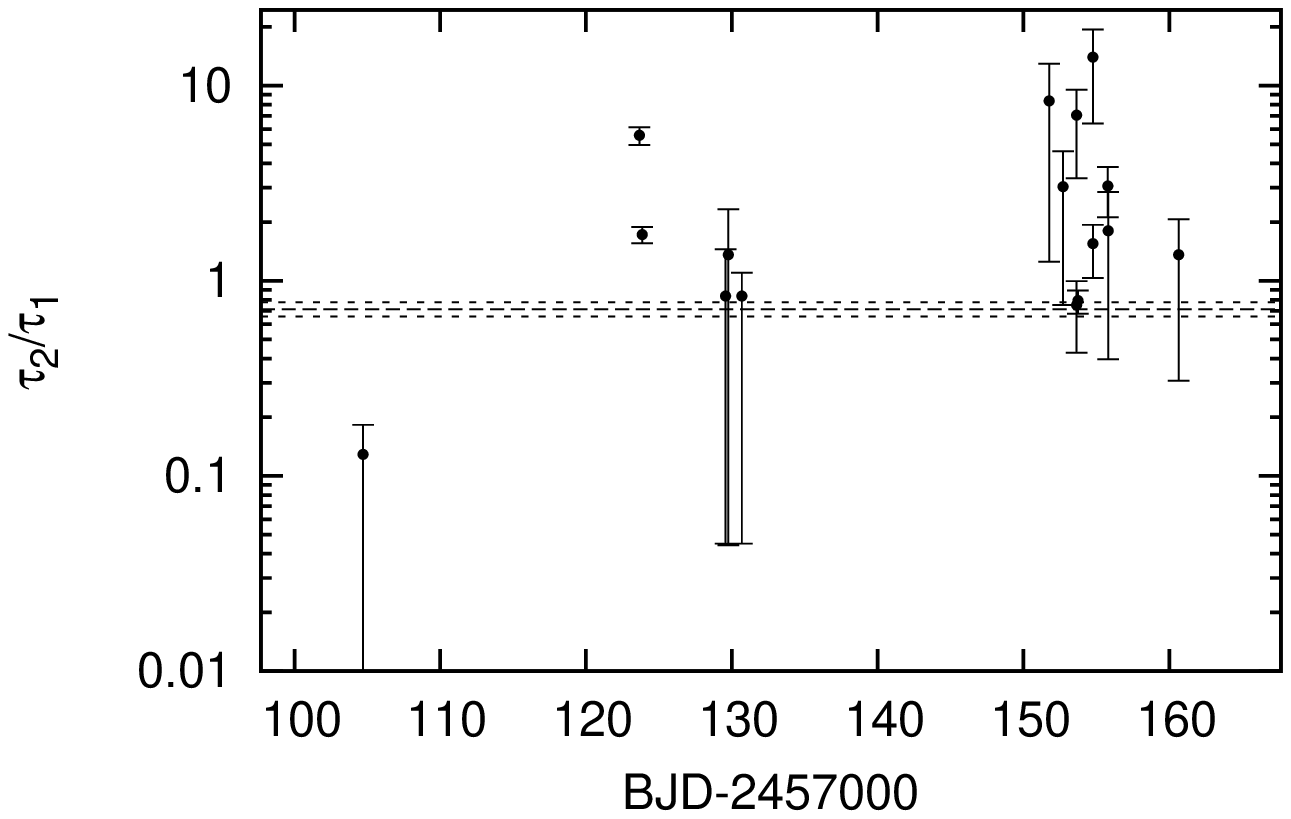} %
\caption[]
	{	Constraints on the transit duration (3$\tau_1$+3$\tau_2$; top panel) and the ratio
		of the egress to ingress times ($\tau_2$/$\tau_1$; bottom panel);
		in both panels the horizontal dashed line shows the weighted mean
		of the transit duration (top) and egress to ingress times (bottom), while
		the dotted line shows the 1$\sigma$ uncertainty in these values.
		The transit duration and 		
		the ratio of the egress to ingress times display evidence that they may not be constant from eclipse to eclipse.
	}
\label{FigWeather}
\end{figure}

We approximate the transit duration by 
\TransitDurationUse $\times$$(\tau_1 + \tau_2)$; this captures $\sim$94\% of the area of the asymmetric hyperbolic secant curve, and appears to qualitatively
match the approximate transit duration as indicated by visually inspecting Figures \ref{FigHSFitsIndividual} - \ref{FigHSFitsVanderburgJoint}. 
The weighted mean of the transit durations we measure 
is:
\TransitDurationUse $\times$$(\tau_1 + \tau_2)$ = \TransitDurationWeightedMean \ $\pm$ \TransitDurationWeightedError \ $\rm min$.
In comparison 
the crossing time of an object with a 4.5 - 4.9 $\rm hr$ period in a circular orbit
around WD 1145+017 should be $\sim$1 $\rm min$.

If we assume that all the ground-based eclipses in Table \ref{TableHSFitsJoint}
have the same transit duration, the reduced $\chi^2$ of this model is \TransitDurationReducedChiSquared. 
Therefore, there may be evidence that the transit duration is not constant for the transits we observe. 
For instance, the 2015/05/11 ``A'' transit has an eclipse duration of
\TransitDurationUse $\times$$(\tau_1 + \tau_2)$ = \TransitDurationMayElevenSecondJoint$^{+\TransitDurationPlusMayElevenSecondJoint}_{-\TransitDurationMinusMayElevenSecondJoint}$ $\rm min$. 
We display the transit duration of the eclipses in the top panel of Figure \ref{FigWeather}.

 Our mean transit duration also informs us on the size of the occulting region transiting in front of the white dwarf - in this case
the size, or the length, of the candidate planetesimal and the cometary tail streaming behind it. Using the typical equation for the
transit duration (equation 14 of \citealt{Winn10}) and assuming 
a circular, edge-on orbit with a $\sim$4.5 hour period, 
the stellar radius and mass ($R_*$ = 1.4 $R_{\oplus}$; $M_*$ = 0.6 $M_{\odot}$) quoted in 
\citet{Vanderburg15}, 
and a planetesimal
mass much less than the stellar mass, a 
transit duration of \TransitDurationUse $\times$$(\tau_1 + \tau_2)$ $\sim$ \TransitDurationWeightedMean \ $\rm min$
is produced by an occulting region of size $R_{o}$ $\sim$ \ROccultREarth \ $R_{\oplus}$.

\subsection{Transit Profiles and a Limit on the Variability of the Ratio of Egress to Ingress times}

The weighted mean of the ratio of the egress to ingress times for all the ground-based 
transits in Table \ref{TableHSFitsJoint} is 
$\tau_2$/$\tau_1$ = \TauRatioWeightedMean \ $\pm$ \TauRatioWeightedError.
A constant egress to ingress time ratio model fits our eclipses with a 
reduced $\chi^2$ of \TauRatioReducedChiSquared.
Therefore there may be evidence of variations in the ratio of egress to ingress times.
Although the weighted mean of the ratio of the egress to ingress times is just below unity, 
for most transits the transit egress lasts significantly longer than the ingress.
For a few of our ground-based transits
the ingress seems to last marginally longer than the egress, and the errors on the ratio are smaller for these transits,
leading to the weighted mean being near unity.
In comparison, the median egress and ingress times from our ground-based transits are
$\tau_1$ = \TauOneMedian \ $\rm min$ and
$\tau_2$ = \TauTwoMedian \ $\rm min$, respectively, and the 
median ratio of the egress to ingress times is 
$\tau_2$/$\tau_1$ = \TauRatioMedian.

We display the ratio of the egress to ingress times in the bottom panel of Figure \ref{FigWeather}.
That the ingress is longer than the egress for at least some of our transits suggests
the possibility of a leading cometary tail in addition to a trailing cometary tail
for at least one or more of the planetesimals.

Our ground-based photometry also has sufficient precision that we are able to inspect the transit profiles of all the
new eclipses we present in Table \ref{TableHSFitsJoint}; the transits appear to be well fit by our asymmetric,
hyperbolic secant (Equation \ref{EquationHS}), suggesting that the transit profile is indeed very different than that of
a solid transiting planet without a cometary tail passing in front of its host star.




\subsection{A Limit on Single Size Grains in the Cometary Tails Trailing the Planetesimals around WD 1145+017}

\begin{figure}
\includegraphics[scale=0.42, angle = 0]{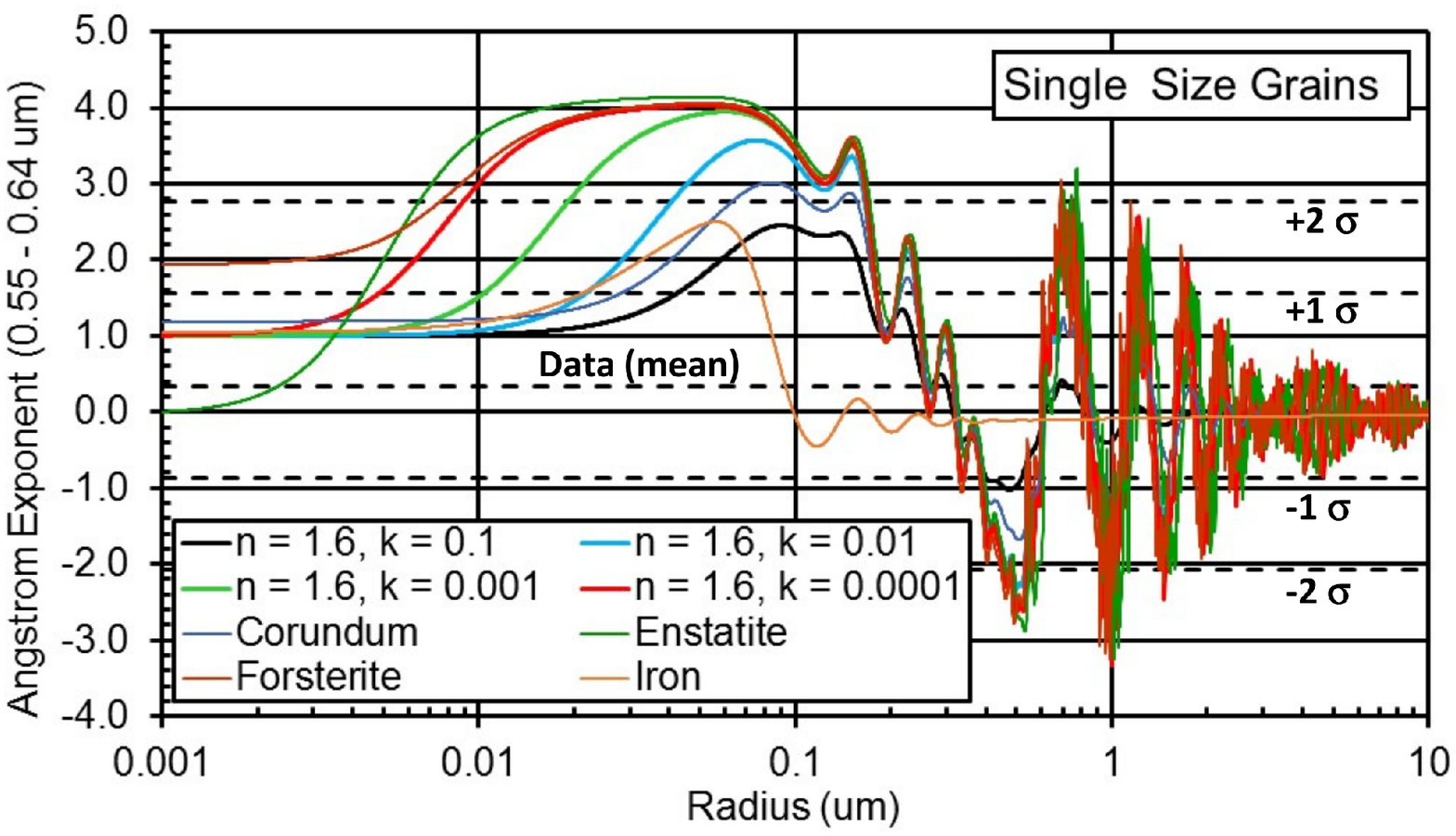} %
\caption[]
	{	Plot of the \r{A}ngstr\"{o}m exponent for spherical grains of
		a given radius for various materials, and for an 
		index of refraction of n = 1.6 and for various imaginary components of
		the index of refraction, $k$.
		The horizontal dashed lines give the 1 and 2$\sigma$ limits on the ratio 
		of the transit depths, $D$, between our V- and R-band observations.
		Our limits on the \r{A}ngstr\"{o}m exponent
		allow us to state that the radius of single sized grains in the dusty tails streaming behind these planetesimals
		must be 
		$\sim$\MicronSingleSizeLimit \ ${\rm \mu m}$ or larger, or $\sim$\MicronSingleSizeLowLimit \ ${\rm \mu m}$ or smaller, 
		with 2$\sigma$ confidence.
	}
\label{FigAngstrom}
\end{figure}

As the 
extinction efficiency generally increases with wavelength until the wavelength is comparable
to the particle circumference \citep{HansenTravis74}, one can
deduce the size of small dust grains from the wavelength dependence
of their extinction.
Although the wavelength, $\lambda$, differences in our current study between the V ($\lambda$$\sim$0.55 ${\rm \mu m}$)
and R-bands ($\lambda$$\sim$0.64 ${\rm \mu m}$) are small, the lack of wavelength-dependent
transit depth differences in Table \ref{TableRatio} allows us to rule out
small dust grains. As the MINERVA ``air'' filter throughput is particularly wide, spanning near-ultraviolet 
to near-infrared wavelengths,
we do not believe the ratio of our V or R to ``air''-band observations will be particularly constraining 
on the particle sizes trailing the planetesimals 
around this white dwarf; we therefore do not use the ``air''-band observations to attempt to place a particle size limit.

 We employ the methodology we have already developed in \citet{Croll14} to place a limit on the size of particles trailing
the planetesimals orbiting WD 1145+017. We
employ the \r{A}ngstr\"{o}m exponent, $\alpha(a,\lambda_1,\lambda_2)$, a measure of
the dependence of extinction on
wavelength, defined as:
\begin{equation}
\alpha(a,\lambda_1,\lambda_2) \equiv - \frac{ \log{ [ \sigma_{ext}(a,\lambda_2) / \sigma_{ext}(a,\lambda_1) ] } }{ \log{ ( \lambda_2 / \lambda_1 ) } }
\label{eqn:WavelengthDependence01}
\end{equation}
where $\lambda_1$ and $\lambda_2$ are the two wavelengths of interest, and $a$ is the grain radius.
The ratio of the transit depths, $D_{\lambda_2}/D_{\lambda_1}$, is approximately the ratio of the extinctions at these wavelengths:
$\sigma_{\rm ext}(a,\lambda_2) / \sigma_{\rm ext}(a,\lambda_1)$.
Therefore the ratio of the transit depths between the V and the R-bands from Table \ref{TableRatio} 
of $D_V$/$D_R$ = \TransitDepthVOverR \ $\pm$ \TransitDepthVOverRError \ results in a ratio on the 
\r{A}ngstr\"{o}m exponent of $\alpha(a,0.55 \mu m,0.64 \mu m)$ = \AngstromVOverR \ with ranges 
\AngstromVOverROneSigmaLow \ to \AngstromVOverROneSigmaHigh \ (1$\sigma$) and 
\AngstromVOverRTwoSigmaLow \ to \AngstromVOverRTwoSigmaHigh \ (2$\sigma$).

We compare this limit to a Mie scattering calculation using the methodology discussed in \citet{Croll14}. We present
the results, adapted to this white dwarf host, in Figure \ref{FigAngstrom}
assuming there is a single particle size in all the cometary tails in this system.
We compare
to hypothetical materials with a given index of refraction, $n$, and a complex index of refraction, $k$, as well
as a number of materials that have previously been suggested to make-up
the grains trailing disintegrating planets \citep{Rappaport12,Budaj13,Croll14}; these materials
include pure iron,
forsterite (Mg$_{2}$SiO$_{4}$; a silicate from the olivine family),
enstatite (MgSiO$_3$; a pyroxene without iron),
and corundum (Al$_2$O$_3$; a crystalline form of aluminium oxide).
Using our transit depth ratio,
materials with a typical index of refraction ($n$$\sim$1.6)
and a low complex index of refraction ($k$$<$0.01; enstatite and forsterite satisfy these constraints at these wavelengths) 
single size particles must be at least $\sim$0.15 ${\rm \mu m}$ 
or larger, or $\sim$0.04 ${\rm \mu m}$ or smaller, with 2$\sigma$ confidence.
For corundum particles ($n$$\sim$1.6 and $k$$<$0.04 at these wavelengths), or
other materials with similar indices of refraction,
the limit on single size particles are 
$\sim$0.15 ${\rm \mu m}$ 
or larger, or $\sim$0.06 ${\rm \mu m}$ or smaller, with 2$\sigma$ confidence.
We cannot place a limit 
on pure iron particles, or on other materials with a high complex index of refraction ($k$$>$0.1).

 These limits prompt the question
of whether such small iron, corundum, enstatite, and forsterite particles
could survive for long enough to create the observed transits in the
$\sim$\TEquilibriumFourPi \ $K$ environment (assuming the dust particles reradiate isotropically)
at these short orbital periods around this white dwarf.
The expected time for grains to travel the length of the cometary tail in this system
(from equation 6 of \citealt{Rappaport12}, using 
the size of the occulting region, $R_o$ = \ROccultREarth \ $R_{\oplus}$ from Section \ref{SecTransitDuration})
is $\sim$\TravelTime $\times$10$^{3}$ s.
\citet{Kimura02} presents the sublimation lifetimes of various grains at various solar insolation levels. 
The stellar insolation of a dust grain in a $\sim$4.5 hour period around WD 1145+017 is equivalent to
an orbit of 12 $R_{\odot}$ around our Sun (ignoring the difference in the shape of the stellar spectra);
at these distances an amorphous olivine particle of size $a$ $\sim$ 0.15 ${\rm \mu m}$ 
survives for $\sim$$100$ s, while similar size crystalline olivines, and pyroxenes survives for several 
orders of magnitude longer.
Iron, which has a vapor pressure $\sim$50 times greater than that for olivines \citep{PerezBeckerChiang13}, 
and non-crystalline
forsterite seem unlikely
to survive for the travel time required to create the observed transit durations.
Orthoclase, albite, and fayalite were also 
mentioned by \citet{Vanderburg15} as possible materials that might make up the cometary
tails of these planetesimals; as they have similar or higher vapour pressures than iron,
it seems unlikely that small particles of these materials could survive for long enough 
to create the observed transit durations.
Crystalline forsterite, enstatite and corundum of $\sim$\MicronSingleSizeLimit \ ${\rm \mu m}$ or larger should survive for long enough without sublimating
to travel the length of the occulting region.
Generally sublimation timescales, $t_{\rm sub}$, scale with the radius of the particle \citep{Lebreton15},
and therefore a related question is whether particles smaller than $\sim$\MicronSingleSizeLimit \ ${\rm \mu m}$ are likely to survive the stellar insolation levels
in this system for the required travel time. Even very small pyroxenes particles, such as enstatite, should
be able to survive these stellar insolation levels; it is less clear whether very small crystalline olivines, such as forsterite,
will be able to survive without sublimating.
Given these arguments, henceforth we quote our 
limit on the radius of single-size particles in this system of 
$\sim$\MicronSingleSizeLimit \ ${\rm \mu m}$ or larger, or $\sim$\MicronSingleSizeLowLimit \ ${\rm \mu m}$ or smaller,
with 2$\sigma$ confidence, which applies to crystalline forsterite, enstatite, corundum and materials with similar properties.

\subsection{The low level variability of WD 1145+017}
\label{SecLLV}

\begin{figure*}
\includegraphics[scale=0.80, angle = 270]{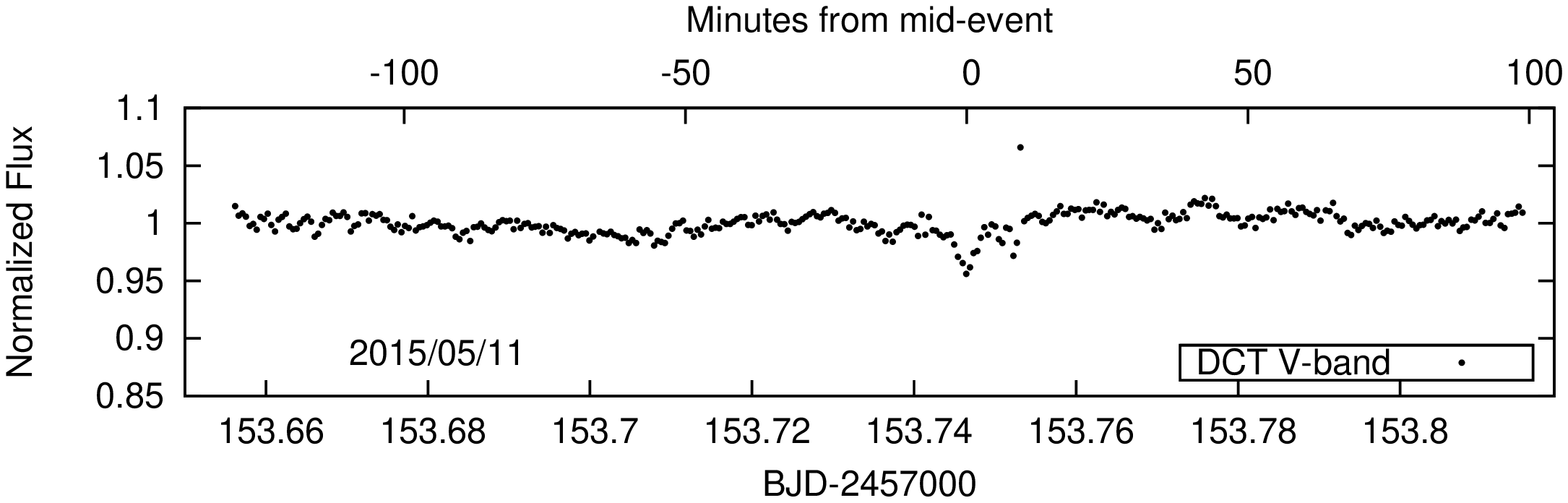}
\includegraphics[scale=0.80, angle = 270]{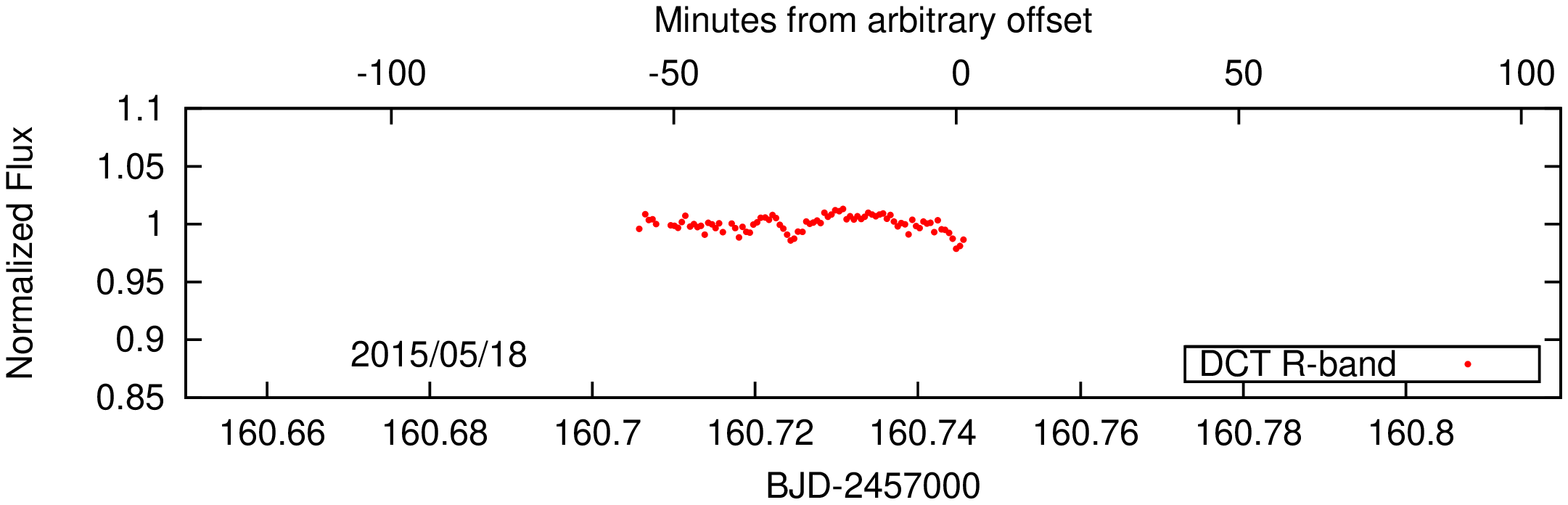}
\caption[]
	{	DCT/LMI observations of WD 1145+017 on UTC 2015 May 11 in the V-band (top) and 
		on 2015 May 18 in the R-band (bottom).
		The observed low level variability is unlikely to be due to pulsations, and is likely
		due to dusty material passing in front of the white dwarf and scattering light
		out of the line of sight.
	}
\label{FigLLV}
\end{figure*}

Our photometry of WD 1145+017 also displays low level variability; this low level variability 
is best displayed in our 2015 May 11 and 2015 May 18
DCT/LMI photometry of WD 1145+017 (Figure \ref{FigLLV}).
The variability we observe is not present in similarly faint
reference stars or with other blue reference stars in our field. On 2015 May 11 the
variability we observe in our DCT light curve is generally consistent with the FLWO and MINERVA observations taken 
simultaneously\footnote{On 2015 May 18 the MINERVA data obtained simultaneously is not of sufficient precision to 
make the statement that it is generally consistent (or inconsistent) with the DCT data obtained simultaneously.}.
For our 2015 May 11 photometry, even after excluding the data around the significant flux decrement at 
BJD-2457000 $\sim$ \TminMayElevenThirdJoint \ (we exclude data from BJD-2457000 = 153.735 to 153.750), 
the reduced $\chi^2$ of a flat light curve is $\sim$\ChiSquaredReducedFlatLine, suggesting a flat light curve
is an extremely poor fit to the data.


We do not attribute this low level variability to pulsation, but instead due to dusty particles
passing in front of the star along our line of sight, either due to the debris disk,
or that have been ejected
beyond the Roche lobe of one of the candidate planetesimals.
The reason we do not attribute this variability to pulsation, is that a 
white dwarf of the effective temperature ($T_{\rm eff}$=15,900 $\pm$ 500 K), surface gravity (log g $\sim$ 8.0)
and helium abundance (H/He $<$ $10^{-4.5}$)				
of WD 1145+017 \citep{Vanderburg15}
is not believed to pulsate; a white dwarf with these characteristics has not been observed to pulsate previously,
and is not near a known white dwarf instability strip \citep{VanGrootel15}.

\subsection{Transit-Timing Analysis}

\begin{figure*}
\includegraphics[scale=1.05, angle = 270]{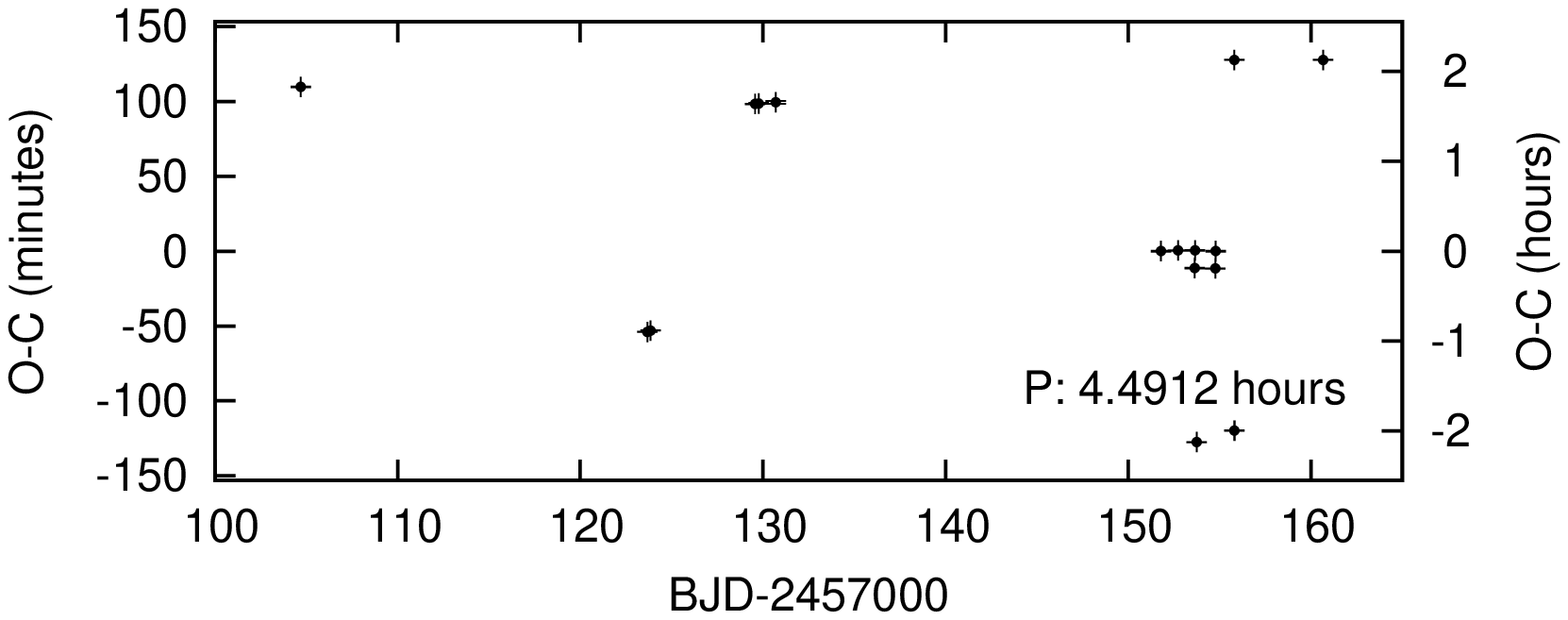} %
\includegraphics[scale=1.05, angle = 270]{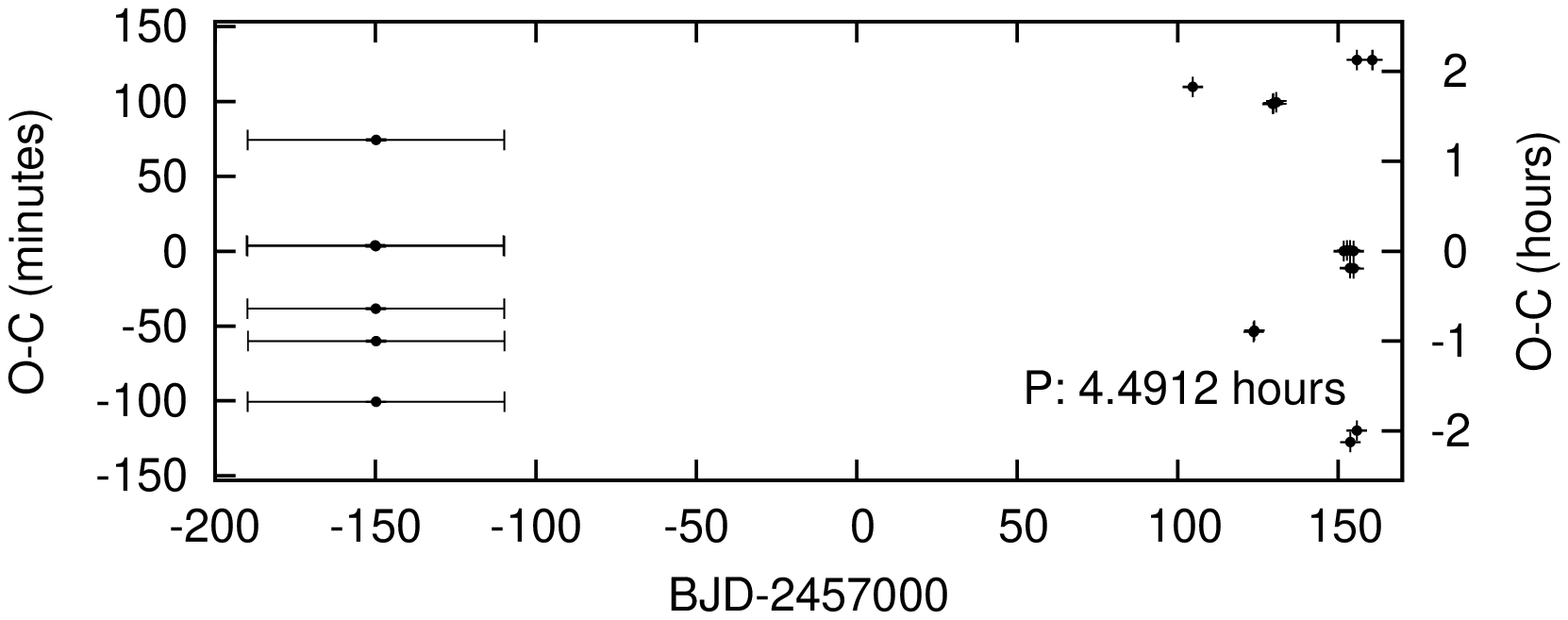} %
\caption[]
	{	Transit-timing analysis phased to a period, $P$=4.4914 $\rm hr$, 
		using the mid-transit times, $T_{\rm min}$, and associated errors from Table \ref{TableHSFitsJoint}.
		The top panel features only the ground-based transit detections, while the bottom panel includes the 
		six K2 detections 
		from \citet{Vanderburg15}.
		The x-error bars in the bottom panel signify that those data-points are averages over the $\sim$80 days of K2 data.
		Multiple separate groups of transits appear to phase
		with a $\sim$4.5 $\rm hr$ period, that is near the ``A'' period of  \citet{Vanderburg15}.
	}
\label{FigTTV}
\end{figure*}

We perform a transit timing analysis using the $T_{\rm min}$ values from Table \ref{TableHSFitsJoint}.
The $T_{\rm min}$ values do not phase well
with any one period and ephemeris, supporting the conclusion of \citet{Vanderburg15}
that there is more than one planetesimal transiting in front of the stellar host
in this system. We notice that a number of our transits
phase up with a $\sim$4.5 hour period and display these results in Figure \ref{FigTTV}.
Potentially up to three pairs of transits, one group of three transits,
and one group of four transits, phase up with a $\sim$4.5 hour period,
but with different ephemerides.
The group of four transits are the transits identified in Table \ref{TableHSFitsJoint} as 
2015/05/09, 2015/05/10, 2015/05/11-B, and 2015/05/12-B; these four transits phase up with a period
$P$=4.4912 $\pm$ 0.0004 $\rm hr$. We note that
this is near the ``A'' period ($P$=4.49888 $\pm$ 0.00007 $\rm hr$) from \citet{Vanderburg15}, but
the errors indicate that these two periods are inconsistent with one another with strong confidence.
The other pairs of transits that appear to phase up with a $\sim$4.5 hour period include
the pairs of ground-based transits observed on 2015/04/11 (A \& B), 
the 2015/05/11-A and 2015/05/12-A transits, and the 2015/05/13-A and 2015/05/18 transits,
while the group of three transits
are the 2015/04/17 (A \& B) and 2015/04/18 transits.

For the group of four transits (2015/05/09, 2015/05/10, 2015/05/11-B, and 2015/05/12-B), 
and the associated $P$=4.4912 $\pm$ 0.0004 $\rm hr$ orbit, on 
the subsequent night of observations (2015/05/13 UTC) and on several other occasions,
we have photometry that overlaps with a predicted transit for this period and ephemeris; 
no obvious deep transits
($>$10\% of the stellar flux) are observed. Similarly, the ground-based photometry of \citet{Vanderburg15} on 2015/04/17 displayed a pair of
$\sim$40\% deep transits separated by $\sim$4.5 hours that were followed by $\sim$15\% transits on the following night.
Given the variability that we observe in the 
depths and shapes for the transits on 2015/05/09, 2015/05/10, 2015/05/11-B, and 2015/05/12-B,
the lack of transits on subsequent nights, and the similar deep transits followed by much shallower transits displayed
in the \citet{Vanderburg15} ground-based photometry,
this suggests that the dust tail trailing these candidate planetesimals evolves rapidly.

On the evenings of 2015/05/11, and 2015/05/12
the 2015/05/11-B, and 2015/05/12-B transits come accompanied by another event that occurs approximately $\sim$12
minutes earlier; this event may occur on 2015/05/10 as well. 

We have attempted to phase our ground-based eclipses with the periods and ephemerides of the 
A - F periods from \citet{Vanderburg15} and include these in the bottom panel of Figure \ref{FigTTV};
our ground-based times are not obviously coincident with the predicted transit
times from these periods and ephemerides, and therefore we cannot provide evidence in favour of the six specific periods 
and ephemerides
given by \citet{Vanderburg15}.
Arguably, this could have been foreseen as the durations of the events in the K2 photometry are generally inconsistent 
with the sharp, short-duration events that have been observed from the ground \citep{Vanderburg15}.
Our ground-based transits also do not phase with the predicted ephemerides from the 
ground-based MEarth and FLWO transits and the $\sim$4.5 hour period observed by \citet{Vanderburg15}.

We have also performed a blind period search to determine
if there are any other compelling
periods for which a large fraction of the ground-based transit times phase up with a given period.
To perform this seach we step through in frequency space in small frequency increments
from periods of a few hours to a few days, and for each of the $T_{\rm min}$ values we predict future and past
ephemerides using this tested period. We then determine the number of other
$T_{\rm min}$ values that are close to the predicted ephemerides (we allow the $T_{\rm min}$ values to differ 
from the predicted ephemerides by at most
2\% from the integer number of cycles of the tested period).
No period other than the $\sim$4.5 hour period was particularly compelling.

Although period evolution (e.g. orbital decay) could reduce the number of planetesimals needed
to explain the number and timing of the observed ground-based transits, nonetheless it appears
that multiple planetesimals would still be required.
Therefore, the timing of the eclipses we observe
supports multiple planetesimals orbiting in close-period orbits around this white dwarf, but the exact
number and periods of these bodies are unclear.

\subsection{Analysis of the Persistence of the 6 Periodic Transits in the K2 Photometry}

\begin{figure}
\includegraphics[scale=0.59, angle = 0]{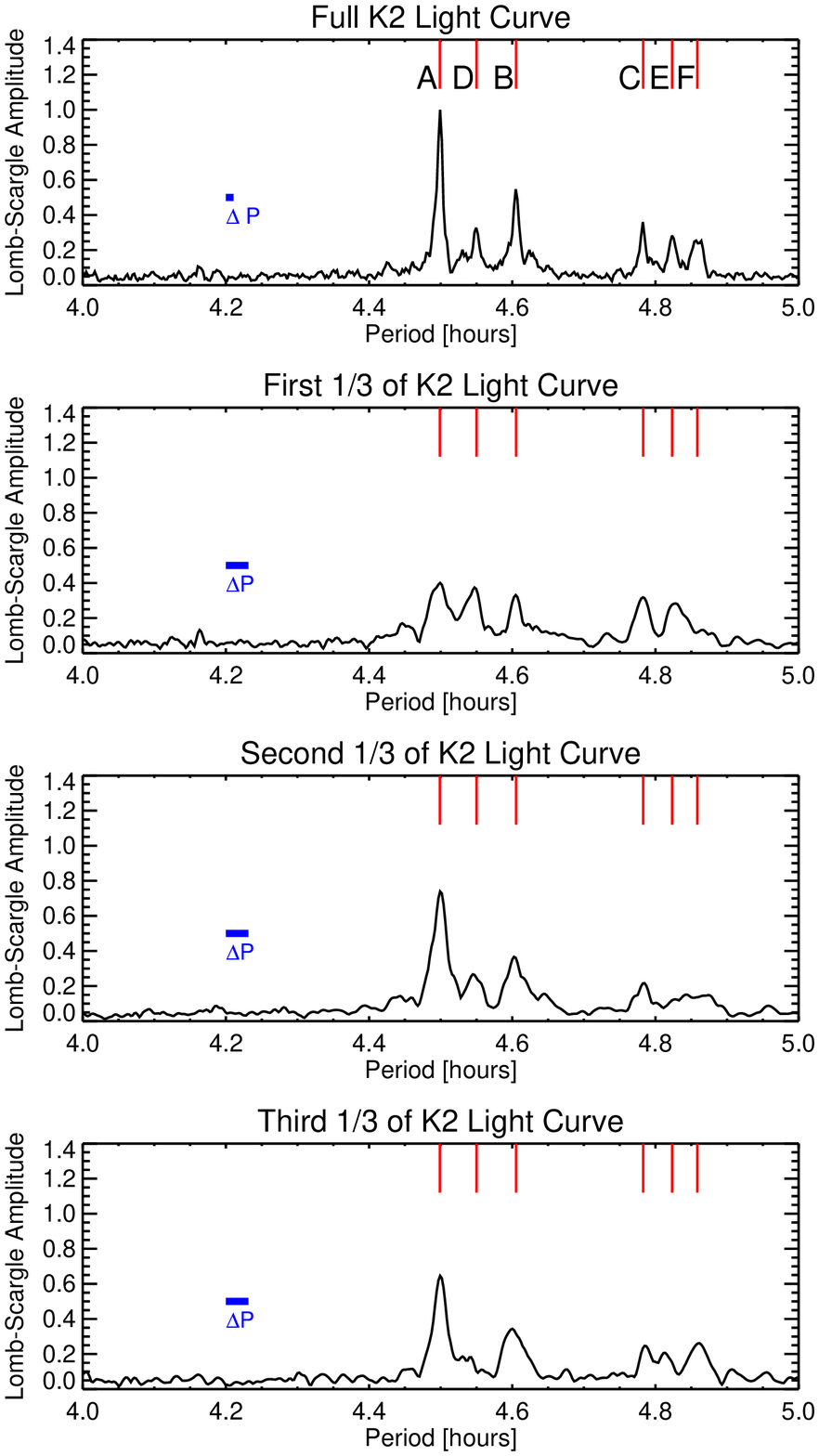} %
\caption[]
	{	Summed Lomb Scargle (LS) periodogram analysis of the K2 photometry of WD 1145+017 (top panel),
		and the K2 photometry split in three equal $\sim$26.6 ${\rm d}$ sections (bottom three panels).
		The vertical red dashes denote the A - F periods of \citet{Vanderburg15}.
		The horizontal blue line marked ``$\Delta P$'' denotes the frequency resolution
		of the LS periodogram given by $\Delta P$ = 1/$T$, where $T$ is the duration of the photometry
		(80 ${\rm d}$ in the top panel, and $\sim$26.6 ${\rm d}$ in the bottom panels).
	}
\label{FigFTs}
\end{figure}

\begin{figure*}
\includegraphics[scale=0.60, angle = 0]{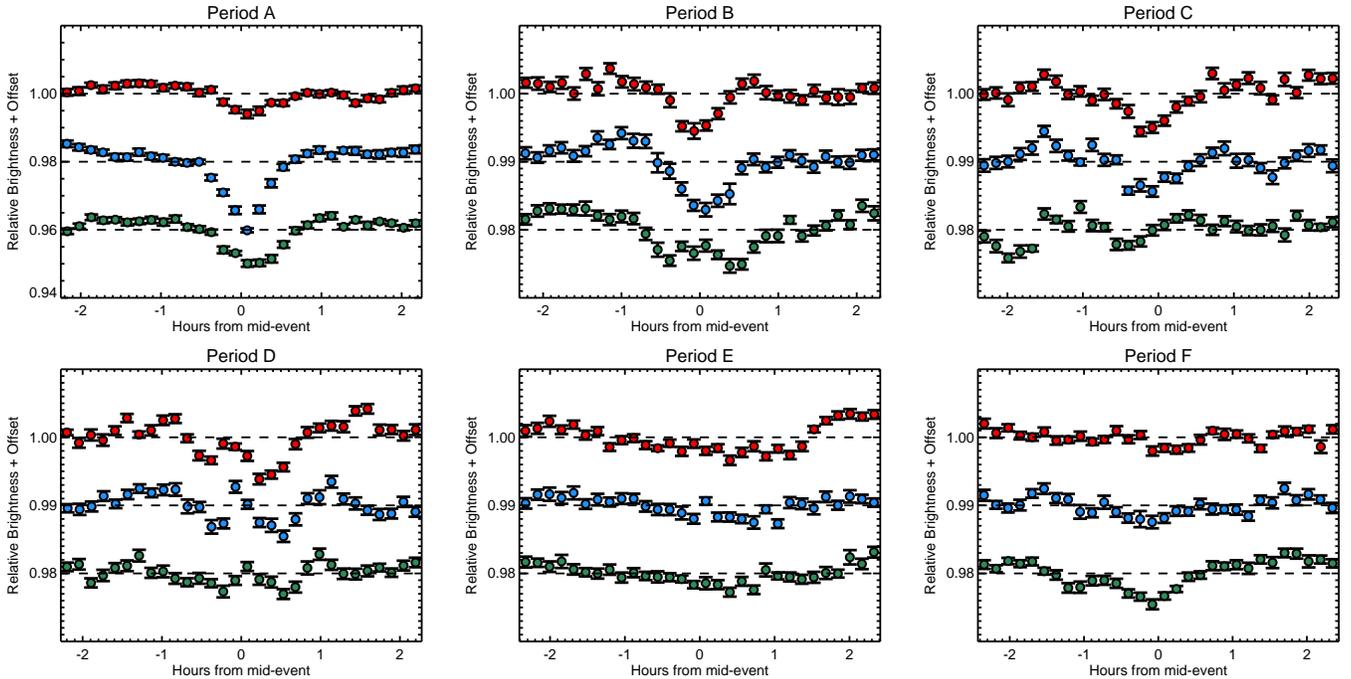} %
\caption[]
	{	K2 photometry of WD 1145+017 folded on the \citet{Vanderburg15} 
		A - F periods as indicated at the top of each panel.
		The red circles represent the first third ($\sim$26.6 ${\rm d}$) of the K2 photometry,
		while the second and third $\sim$26.6 ${\rm d}$ stretches of K2 photometry are represented by the blue and green circles,
		respectively.
		The data is binned every $\sim$0.033 in phase, and 
		the various light curves are vertically offset for clarity.
	}
\label{FigK2}
\end{figure*}

 Given the apparent rapid evolution of the depths and profiles of the transits in the ground-based photometry of WD 1145+017, 
we reanalyzed
the K2 photometry of this system to determine if the 6 claimed transiting bodies in this system 
persist in duration and depth throughout the 80 days of K2 photometry. 
To do this we split the K2 photometry into three equal sections of $\sim$26.6 ${\rm days}$; 
three equal 26.6 ${\rm day}$ sections were chosen as these sections 
were long enough in duration to allow for sufficient statistical accuracy, and short enough
to allow for the evolution of these K2 signals to be investigated.
Reduction of the photometry was performed as discussed in \citet{Vanderburg15}.
On each third of the K2 photometry we perform 
a harmonic-summed Lomb Scargle (LS) periodogram \citep{Lomb76,Scargle82,Ransom02},
where the amplitudes from the first two harmonics are added to the fundamental frequency in the period range from 4 - 5 ${\rm hours}$.
We display these results, compared to the original LS periodogram signal from all the K2 photometry,
in Figure \ref{FigFTs}. We also phase the K2 photometry to these periods, and present the phase binned transit signals
in Figure \ref{FigK2}.
The transit dips at the original \citet{Vanderburg15} A ($\sim$4.499 ${\rm d}$) 
and B ($\sim$4.605 ${\rm d}$) periods are present in all three thirds of the K2 photometry,
although they appear to vary in depth.
The situation is less clear for
the transits dips at the C ($\sim$4.783 ${\rm d}$), D ($\sim$4.550 ${\rm d}$),
E ($\sim$4.823 ${\rm d}$) and F ($\sim$4.858 ${\rm d}$) periods. 
For the C, D and F periods, in addition to varying in depth, it is unclear if these signals exist in all
three thirds of the K2 photometry; for the C and D periods it is unclear if the signal exists in the last
third of K2 photometry, while the F period is not clearly present in the first and second third of K2 photometry.
Although there appears to be a slight decrement at the transit mid-point
in all three thirds of the K2 photometry for the E period, the statistical significance of the E period detections in
each third of K2 photometry are not overwhelming.
Therefore, one possibility is that this analysis indicates that the transits at the A - F periods simply evolve in depth
over the 80 ${\rm d}$ of K2 photometry; another possibility is that the transits at the C - F
periods may not start to transit or may cease to transit for up to or more than $\sim$26 ${\rm d}$
of the 80 ${\rm d}$ of K2 photometry.

\section{Discussion \& Conclusions}

We have presented multiwavelength, multi-telescope, ground-based photometry of the white dwarf 
WD 1145+017 that revealed \NumSignificantTransits \ significant dips in flux of more than 10\% of the stellar flux,
and up to $\sim$30\%. 
During our 2015 May observations we observe a transit with a depth greater than 10\% of the stellar flux on average every
$\sim$\HoursPerSignificantTransits \ $\rm hr$ of observations.
Through fits to the transits that we observe, we confirm 
that the transit egress timescale is usually longer than the ingress timescale, and that the transit duration is longer
than expected for a solid body at these short periods. All these lines of evidence support the conclusion
of \citet{Vanderburg15} that WD 1145+017
is likely orbited by multiple, low-mass planets/planetesimals in short-period orbits, likely with dusty cometary tails trailing behind
them.

The exact number of planets/planetesimals orbiting WD 1145+017 and the periods of these objects are unclear.
Given the substantial number of transit events that we observed, and that have been previously observed by \citet{Vanderburg15},
it seems likely that there are a number of planetesimals orbiting WD 1145+017. A number of our ground-based
transits phase up well with a $\sim$4.5 hour period, however these events 
are best-fit by drastically different ephemerides.
Four of our ground-based transit times are consistent
with a constant ephemeris and a $\sim$4.5 hour period, but this ephemeris is not obviously
consistent with the ephemerides of 
our other ground-based transits, the ground-based transit ephemerides of \citet{Vanderburg15},
or the K2 periods and ephemerides of \citet{Vanderburg15}.
This suggests that there are likely multiple objects in this system, and a number of these objects
might have $\sim$4.5 hour orbital periods.

 We have also reanalyzed the K2 photometry of this system to determine if the signals for the 
six claimed transiting objects persist throughout the K2 photometry. For four of the six claimed
signals, the K2 photometry is consistent with that either the depths vary to nearly undetectable levels 
over the 80 days of K2 photometry, or that the signals may not transit or cease to transit during
a significant fraction of the 80 days of K2 photometry. This suggests that the amount of material
in the cometary tails trailing these candidate planetesimals may evolve rapidly, or that we may be observing
collisions, tidal break-up, or gravitational interactions that causes the orbits of these planetesimals to rapidly evolve.

That we are unable to establish the rough number and the exact periods of the candidate planetesimals
orbiting WD 1145+017 has implications on the mass of these objects.
The suggestion that the planetesimals orbiting WD 1145+017 might be approximately Ceres-mass ($\sim$1.6 $\times$ 10$^{-4}$ $M_{\oplus}$)
or less, came from an N-body simulation assuming six stable orbits with periods from $\sim$4.5 - 4.9 $\rm hr$ --
such a collection of short-period objects with higher masses would quickly become unstable.
Since we are unable to confirm the strict periodicity suggested
by the K2 photometry of this system, strict stability is not required by our observations and
higher mass objects may be possible, including even planetary-mass objects; such an orbital configuration is arguably unlikely, 
as a number of short-period, planet-mass objects 
would likely become unstable after
a few million orbits or less \citep{Vanderburg15}, meaning that we would have to be observing this system during a unique epoch
in its history.

 The mechanism for the mass loss leading to the cometary tails that are believed to
be trailing the candidate planetesimals in this system is also unclear. 
Our 2$\sigma$ limit that the radius of single-size particles in the cometary tails streaming
behind planetesimals in this system must be
$\sim$\MicronSingleSizeLimit \ ${\rm \mu m}$ or larger, or $\sim$\MicronSingleSizeLowLimit \ ${\rm \mu m}$ or smaller, 
is consistent
with a variety of scenarios. If the objects in this system have a mass more
typically associated with planets than planetesimals, then a Parker thermal wind may be required
to lift material and escape the relatively strong surface gravity, similar to the other disintegrating planetary-mass candidates
that have been presented thus far \citep{PerezBeckerChiang13}. 
If the objects in this system are planetesimal-mass objects -- a more likely scenario since the orbits of such objects could 
be stable for a few million orbits
or more -- then in this lower surface gravity regime, the dusty material may escape the planetesimal via a number of mechanisms. 
For the first, analogous to comets in our own solar system \citep{Cowan79},
vapourization of volatiles from the planetesimal's surface would drive dusty material from the surface.
The second possibility
is that the high temperature on the surface of the planetesimal would cause sublimation of rocky material into metal vapours.
Under these high temperatures the thermal speed of metal vapours would exceed the escape speed of 
the planetesimal; at altitude these vapours could condense into dusty material forming the observed cometary tails.
Lastly, two other very different mechanisms may explain the candidate planetesimals and their cometary tails: collisions
with other planetesimals in the system or with the debris disk, and tidal disruption of these plantesimals.
For the collision scenario, if there are a number of planetesimals with short-period orbits that are 
embedded within, or nearby, the observed debris disk, collisions could lead to material trailing behind the planetesimals that would
quickly shear to form tails \citep{Veras14}; shear would also lead to cometary tails in a tidally disrupted body.
Numerical simulations will have to be performed to determine whether such scenarios are consistent
with the rapid night-to-night variability observed in both the transit depth and shape, including that deep transits 
are followed a night later by significantly shallower transits\footnote{This statement assumes that these transits are associated
with an object with a $\sim$4.5 hour period, that we have identified as the most likely
for at least some of the planetesimals we have observed.}.


 Also, our highest precision photometry, obtained with the DCT, displays low amplitude variations. These variations are not believed
to be due to pulsations from the white dwarf. Instead, this variability more likely suggests that dusty material consistently
passes in front of the white dwarf. This observed material could be either from the detected debris disk in this system,
or could be from material that has been ejected beyond the Roche lobe of the candidate planetesimals; either possibility could be consistent
with the idea that these planetesimals are analogous to the Jupiter Ring-Moon system \citep{Burns04}. If these lower amplitude events are periodic, due
to for instance a number of smaller planetesimals in the system, then these lower amplitude variations could be responsible for the C - F periods
from the K2 photometry \citep{Vanderburg15}.

That we are unable to 
determine the number of candidate planetesimals,
and with the rapid transit-to-transit depth or profile evolution that we observe, suggests the possibility
that rather than observing a planet or asteroid that has been disrupted, 
we may be observing a planet or asteroid in the midst of being tidal disrupted. Tidal disruption events 
have been previously suggested to endure for as short as a few years \citep{Debes12,XuJura14}.
Therefore, follow-up observations over the next few years will determine whether 
the transit frequency, depths and profiles are consistent with previous observing seasons, and whether
there are a consistent number of orbiting objects in the system.

Lastly, that we observe a significant transit dip (greater than 10\% of the stellar flux) on average every
$\sim$\HoursPerSignificantTransits \ $\rm hr$ of observations, indicates that WD 1145+017 is a very favourable candidate
for follow-up observations with larger telescopes. If this frequency of transit dips persists into the future 
then it is likely that a single night of observations 
of WD 1145+017 with larger ground-based or space-based telescopes will detect significant transit events,
and reveal further information about this fascinating and confounding system.

\acknowledgements

These results made use of Lowell Observatory's Discovery Channel Telescope.
Lowell operates the DCT in partnership with Boston University, Northern Arizona University, the University
of Maryland, and the University of Toledo. Partial support of the DCT was provided by Discovery
Communications. LMI was built by Lowell Observatory using funds from the National Science Foundation
(AST-1005313).

MINERVA is a collaboration among the Harvard-Smithsonian Center for Astrophysics, the Pennsylvania
State University, University of Montana, and University of New South Wales. MINERVA is made possible
by generous contributions from its collaborating institutions and Mt. Cuba Astronomical Foundation,
the David and Lucile Packard Foundation, the National Aeronautics and Space Administration, and the
Australian Research Council.

J.A.J is supported by a generous grant from the David and Lucile Packard Foundation.
The Center for Exoplanets and Habitable Worlds is supported by the
Pennsylvania State University, the Eberly College of Science, and the
Pennsylvania Space Grant Consortium.


 We thank Dan Feldman and Connor Robinson for their assistance in observing this object with the Discovery Channel Telescope,
Zachary Hall for his assistance with observing this object with the Perkins Telescope,
Brian Taylor for his assistance in preparing observations of this object on both the Perkins and Discovery Channel Telescope,
and Gilles Fontaine and Patrick Dufour for helpful conversations on white dwarf variability and pulsations.


\begin{thebibliography}{}

\bibitem[Bochinski et al.(2015)]{Bochinski15} Bochinski, J.J. et al. 2015, \apjl, 800, L21
\bibitem[Budaj (2013)]{Budaj13} Budaj, J. 2013, \aap, 557, A72
\bibitem[Burns et al.(2004)]{Burns04} Burns, J. A., Simonelli, D. P., Showalter, M. R., et al. 2004, in Jupiter: The Planet, Satellites and Magnetosphere, ed. F. Bagenal et al., Chap. 11, 241
\bibitem[Carter et al.(2011)]{Carter11} Carter, J.A. et al. 2011, \apj, 730, 82
\bibitem[Cowan \& A'Hearn(1979)]{Cowan79} Cowan, J.J. \& A'Hearn, M.F. 1979, Moon and the Planets, 21, 155
\bibitem[Croll et al.(2014)]{Croll14} Croll, B. et al. 2014, \apj, 786, 100
\bibitem[Croll(2006)]{CrollMCMC} Croll, B. 2006, \pasp, 118, 1351
\bibitem[Croll et al.(2015)]{Croll15} Croll, B. et al. 2015, \apj, 802, 1
\bibitem[Debes \& Sigurdsson (2002)]{DebesSigurdsson02} Debes, J.H. \& Sigurdsson, S. 2002, \apj, 572, 556
\bibitem[Debes et al. (2012)]{Debes12} Debes, J.H., Walsh, K.J. \& Stark, C. 2012, \apj, 747, 148
\bibitem[Dupuis et al. (1993)]{Dupuis93} Dupuis, J., Fontaine, G. \& Wesemael, F. 1993, \apjs, 87, 345
\bibitem[Eastman et al. (2010)]{Eastman10} Eastman, J., Siverd, R., \& Gaudi, B. S. 2010, \pasp, 122, 935
\bibitem[Hansen \& Travis~(1974)]{HansenTravis74} Hansen, J.E. \& Travis, L.D. 1974, Space Science Reviews, 16, 527
\bibitem[Howell et al. (2014)]{Howell14} Howell, S.B. et al. 2014, PASP, 126, 398
\bibitem[Irwin et al. (2015)]{Irwin15} Irwin, J.M. et al. 2015, 18th Cambridge Workshop on Cool Stars, Stellar Systems, and the Sun, arXiv:astro-ph/1409.0891 
\bibitem[Janes et al. (2004)]{Janes04} Janes, K.A. et al. 2004, BAAS, 36, 672	
\bibitem[Jewitt et al. (2010)]{Jewitt10} Jewitt, D. et al. 2010, Nature, 7317, 817
\bibitem[Jewitt et al. (2013)]{Jewitt13} Jewitt, D. et al. 2013, \apjl, 778, L21
\bibitem[Jura (2003)]{Jura03} Jura, M. et al. 2003, \apjl, 584, L91
\bibitem[Kimura et al. (2002)]{Kimura02} Kimura, H. et al. 2002, Icarus, 159, 529
\bibitem[Koester et al. (2014)]{Koester14} Koester, D. et al. 2014, \aap, 566, A34
\bibitem[Lebreton et al. (2015)]{Lebreton15} Leberton, J. et al. 2015, \aap, 555, A146
\bibitem[Lomb (1976)]{Lomb76} Lomb, N.~R. 1976, Ap\&SS, 39, 447
\bibitem[Massey et al. (2013)]{Massey13} Massey, P. et al. 2013, AAS, 221, \#345.02	
\bibitem[Michikami et al. (2008)]{Michikami08} Michikami, T. et al. 2008, Earth Planets Space, 60, 13
\bibitem[Nutzmann \& Charbonneau (2008)]{NutzmannCharbonneau08} Nutzmann, P. \& Charbonneau, D. 2008, \pasp, 120, 317
\bibitem[Perez-Becker \& Chiang (2013)]{PerezBeckerChiang13} Perez-Becker, D. \& Chiang, E. 2013, \mnras, 433, 2294
\bibitem[Raddi et al. (2015)]{Raddi15} Raddi, R. et al. 2015, \mnras, 450, 2083
\bibitem[Ransom et al. (2002)]{Ransom02} Ransom, S.M., Eikenberry, S.S. \& Middleditch, J. 2002, \aj, 124, 1788
\bibitem[Rappaport et al. (2012)]{Rappaport12} Rappaport, S. et al. 2012, \apj, 752, 1
\bibitem[Rappaport et al. (2014)]{Rappaport14} Rappaport, S. et al. 2014, \apj, 2014, 784, 40
\bibitem[Sanchis-Ojeda et al. (2015)]{Sanchis15} Sanchis-Ojeda, R. et al. 2015, \apj, submitted, arXiv:astro-ph/1504.04379
\bibitem[Scargle (1982)]{Scargle82} Scargle, J.D. 1982, \apj, 263, 835
\bibitem[Swift et al.(2015)]{Swift15} Swift, J.J. et al. 2015, Journal of Astronomical Telescopes, Instruments, and Systems, 1, 2
\bibitem[Szentgyorgyi et al.(2005)]{Szentgyorgyi05} Szentgyorgyi, A. H., et al. 2005, BAAS, 37, 1339
\bibitem[Vanderburg et al.(2015)]{Vanderburg15} Vanderburg, A. et al. 2015, Nature, accepted
\bibitem[Van Grootel et al.(2015)]{VanGrootel15} Van Grootel, V., Fontaine, G., Brassard, P. \& Dupret, M.-A. 2015, \aap, 575, A125
\bibitem[Veras et al.(2014)]{Veras14} Veras, D. et al. 2014, \mnras, 445, 2244
\bibitem[Winn (2010)]{Winn10} Winn, J.N. 2010, arXiv:astro-ph/1001.2010
\bibitem[Xu \& Jura (2014)]{XuJura14} Xu, S. \& Jura, M. 2014, \apjl, 792, L39
\bibitem[Zuckerman et al.(2003)]{Zuckerman03} Zuckerman, B. et al. 2003, \apj, 596, 477
\bibitem[Zuckerman et al.(2007)]{Zuckerman07} Zuckerman, B. et al. 2007, \apj, 671, 872
\bibitem[Zuckerman et al.(2010)]{Zuckerman10} Zuckerman, B. et al. 2010, \apj, 722, 725

\end{thebibliography}
\end{document}